\documentclass[prd, aps,nofootinbib, preprintnumbers, showpacs, superscriptaddress,twocolumn]{revtex4}
\usepackage{latexsym}
\usepackage{amssymb}
\usepackage{amsfonts}
\usepackage{amsmath}
\usepackage[dvips]{graphicx}\usepackage{color}

\usepackage{ulem}

\renewcommand{\emph}[1]{{\it #1}}

\newcommand{\be}{\begin{equation}}  
\newcommand{\ee}{\end{equation}}
\newcommand{\bea}{\begin{eqnarray}}           
\newcommand{\eea}{\end{eqnarray}} 
\newcommand{\beqn}{\begin{eqnarray*}}
\newcommand{\eeqn}{\end{eqnarray*}}
\newcommand{\ba}{\begin{align}}
\newcommand{\ea}{\end{align}}

\def\vp{v_{\rm pole}}
\def\de{\partial}
\def\lm{{\ell m}}

\def\ii{{\rm i}}
\def\l{{\ell }}

\def\y{{\bar{y}}}

\def\J{{\cal J}}

\def\k{\hat{\hat{k}}}

\begin{document}


\title{Improved resummation of post-Newtonian multipolar waveforms from
  circularized compact binaries.}

  \author{Thibault \surname{Damour}}
  \affiliation{Institut des Hautes Etudes Scientifiques, 91440 Bures-sur-Yvette, France}
  \affiliation{ICRANet, 65122 Pescara, Italy}

  \author{Bala R.~\surname{Iyer}}
  \affiliation{Institut des Hautes Etudes Scientifiques, 91440 Bures-sur-Yvette, France}
  \affiliation{Raman Research Insitute, Bangalore 560 080, India}
  \author{Alessandro \surname{Nagar}}
  \affiliation{Institut des Hautes Etudes Scientifiques, 91440 Bures-sur-Yvette, France}
  \affiliation{ICRANet, 65122 Pescara, Italy}
  \affiliation{INFN, Sezione di Torino, Via Pietro Giuria 1, Torino, Italy}

  \date{\today}
  
\begin{abstract}
We improve and generalize a resummation method of post-Newtonian multipolar
waveforms from circular (nonspinning) compact binaries 
introduced in Refs.~\cite{Damour:2007xr,Damour:2007yf}. One of
the characteristic features of this resummation method is to replace the usual
{\it additive} decomposition of the standard post-Newtonian approach by a 
{\it multiplicative} decomposition of the complex multipolar waveform $h_{\lm}$
into several (physically motivated) factors: (i) the ``Newtonian'' 
waveform, (ii) a relativistic correction coming from an ``effective source'', 
(iii) leading-order tail effects linked to propagation on a Schwarzschild
background, (iv) a residual tail dephasing, and (v) residual relativistic 
amplitude corrections $f_{\lm}$. We explore here a new route for resumming
$f_{\lm}$ based on replacing it by its $\ell$-th root: $\rho_{\lm}=f_{\lm}^{1/\ell}$.
In the extreme-mass-ratio case, this resummation procedure results in 
a much better agreement between analytical and numerical 
waveforms than when using standard post-Newtonian approximants. 
We then show that our best approximants behave in a robust and continuous
manner as we ``deform'' them by increasing the symmetric mass ratio
$\nu\equiv m_1 m_2/(m_1+m_2)^2$ from $0$ (extreme-mass-ratio case) to $1/4$
(equal-mass case).
The present paper also completes our knowledge of the first post-Newtonian
corrections to multipole moments by computing
ready-to-use explicit expressions for the first post-Newtonian
contributions to the odd-parity (current) multipoles. 
\end{abstract}

  
  \maketitle
  

\section{Introduction}
\label{intro}
One of the prime targets for the currently operating network of laser
interferometer gravitational wave (GW) detectors is the inspiral and merger
of binary black hole systems.
To detect and interpret the GW signals from such systems one will need
accurate templates to match theoretically computed signals to the 
noisy output of the detectors.
The prime analytical framework allowing one to compute (within General
Relativity) the GW signal emitted by a comparable mass binary system in 
the mildly relativistic regime\footnote{Our notation is: $M\equiv m_1+m_2$,
  $\mu\equiv m_1 m_2/M$, $\nu\equiv\mu/M=m_1 m_2/(m_1+m_2)^2$,
  $\Omega\equiv\text{orbital frequency}$, $v\equiv (GM\Omega)^{1/3}$, 
$x\equiv v^2/c^2\equiv (GM\Omega/c^3)^{2/3}$.
We shall generally use $x$ as PN ordering parameter, and often use 
(without warning) units where either $c=1$ or 
$G=1$. We recall that a term $x^n\sim v^{2n}/c^{2n}$ is said to belong to
the n-PN approximation.}  $x \sim (v/c)^2\sim GM/(c^2 R)\ll 1$ 
is the post-Newtonian (PN) approximation scheme 
(see Ref.~\cite{Blanchet:2002av} for a review).
This raises the issue of the convergence
of the PN expansion, or, in practical
terms, of the largest value of the PN-expansion parameter $x$ 
for which the currently known PN expansions yield accurate enough
GW templates.
Note that, when speaking of
``convergence'' in this paper we shall not have in mind the mathematical
question of whether the full PN expansion of, say, the (Newton-normalized) 
GW radiation flux,
$\hat{F}^{\rm Taylor}(x)=\sum_{n=0}^{+\infty} f_n(\nu;\log x)x^n$ is a 
mathematically point-wise
convergent series (for some fixed $x$ belonging to some range) as
$n\to+\infty$, but the more practical question of how small is the numerical 
difference (say in the supremum, $L_{\infty}$, norm) between the currently known truncated PN
expansions, say, $\hat{F}_N^{\rm Taylor}\equiv \text{Taylor}_N \hat{F}(x)=\sum_{n=0}^N
f_n(\nu;\log x)x^n$, for $N=3$ (3PN approximation), and the ``exact'' 
flux $\hat{F}^{\rm  Exact}(x)$ in some physically relevant 
interval $0<x<x_{\rm max}$, where $x_{\rm max}$ is equal or close to the value
corresponding to the Last Stable (circular) Orbit (LSO).
We shall then consider that
some resummation method, which transforms $\hat{F}_N^{\rm Taylor}(x)$ into $\hat{F}_N^{\rm
Resummed}(x)$ (say for $N=3$) is, {\it effective} if
$\text{sup}_{x<x^{\rm max}}\left|\hat{F}_N^{\rm Resummed}(x)-\hat{F}^{\rm Exact}(x)\right|$
 is significantly smaller than
$\text{sup}_{x<x^{\rm max}}\left|\hat{F}_N^{\rm Taylor}(x)-\hat{F}^{\rm Exact}(x)\right|$ 
when $x_{\rm max}$ corresponds to the LSO 
(i.e., $x_{\rm max}=1/6$ in the extreme-mass-ratio limit $\nu\to0$).

It was pointed out by Cutler et al.~\cite{Cutler:1993vq} 
and Poisson~\cite{Poisson:1995vs} that the convergence (in the sense just
explained) of the PN series
is rather poor, especially near the LSO 
(i.e. when $x\simeq 1/6$) in the extreme-mass-ratio case that they considered.  
It was then suggested by Damour, Iyer, Sathyaprakash~\cite{Damour:1997ub}, 
to use {\it resummation methods} to extend the numerical  validity of the
PN expansions (at least) up to the LSO. They used several resummation
techniques, and in particular Pad\'e approximants. New resummation
methods, aimed at extending the validity of suitably resummed PN results
beyond the LSO, and up to the merger, were later introduced in 
the ``Effective-One-Body'' (EOB) approach 
and used to estimate the complete GW signal emitted by inspiralling,
plunging, merging and ringing binary black hole 
systems~\cite{Buonanno:1998gg,Buonanno:2000ef}. The EOB method has been
recently improved, notably by the introduction of a new, {\it resummed},
$3^{+2}$PN accurate\footnote{The notation $3^{+2}$PN refers to a ``hybrid'' 
expression which incorporates both the comparable-mass ($\nu\neq0$) 3PN terms and the
extreme-mass ratio ($\nu=0$) 4PN and 5PN terms. See below for the precise
definition of the ``hybridization'' procedure we use here.} waveform for 
the $\ell=m=2$ case~\cite{Damour:2007xr,Damour:2007yf}. 
For several, comparable-mass cases, such a waveform (married to the EOB dynamics) 
has been shown to agree remarkably well, both in phase and in modulus, with 
Numerical Relativity data (see~\cite{Pretorius:2007nq} for review of binary
black hole numerical simulations).
For instance, Ref.~\cite{Damour:2008te} found a phase difference smaller 
than $\pm 0.025$ radians with Jena data all over the inspiral and plunge 
up to merger, while Ref.~\cite{Damour:2007yf} found a remarkable amplitude
agreement with published Caltech-Cornell data over the inspiral
and part of the plunge. Let us note in this respect that the use of a
theoretically less accurate waveform (Newtonian-accurate multipolar waveform)
still allows for a rather small phase difference, but leads to significantly
larger differences in the modulus~\cite{Buonanno:2007pf}.

The main aim of this paper is to further improve the type of resummed
multipolar waveform introduced in~\cite{Damour:2007xr,Damour:2007yf} for
circularized (nonspinning) compact binaries.
More precisely, we shall achieve here two goals: (i) on the one hand, we
shall generalize the resummed $\ell=m=2$ waveform
of~\cite{Damour:2007xr,Damour:2007yf} to higher multipoles by using the most
accurate currently known PN-expanded 
results~\cite{Kidder:2007rt,Berti:2007fi,Blanchet:2008je}
as well as the higher PN terms which are known in the test-mass
limit~\cite{Tagoshi:1994sm,Tanaka:1997dj}, and (ii), on the other hand, we shall introduce a 
{\it new resummation procedure} which consists in considering a new
theoretical quantity, denoted below as $\rho_{\lm}(x)$, which enters the
$(\ell,m)$ waveform (together with other building blocks, see below) only
through its $\ell$-th power: $h_{\ell m}\propto \left(\rho_{\lm}(x)\right)^{\ell}$.
In this paper we shall primarily use the small-mass-ratio limit ($\nu\to 0$), in which
one knows both high PN expansions of $\rho_{\lm}(x)$~\cite{Tagoshi:1994sm,Tanaka:1997dj}
and the ``exact'' value of $\rho_{\lm}(x)$ from numerical studies of test particles
around black holes~\cite{Cutler:1993vq,Poisson:1995vs,Yunes:2008tw}, to study
the quality of the convergence of $\rho_{\lm}^{\rm Taylor}(x)$.
Then we shall explore the robustness and consistency of our new approximants
in the comparable-mass case.

\begin{figure*}[t]
\begin{center}
(a)\includegraphics[width=82 mm]{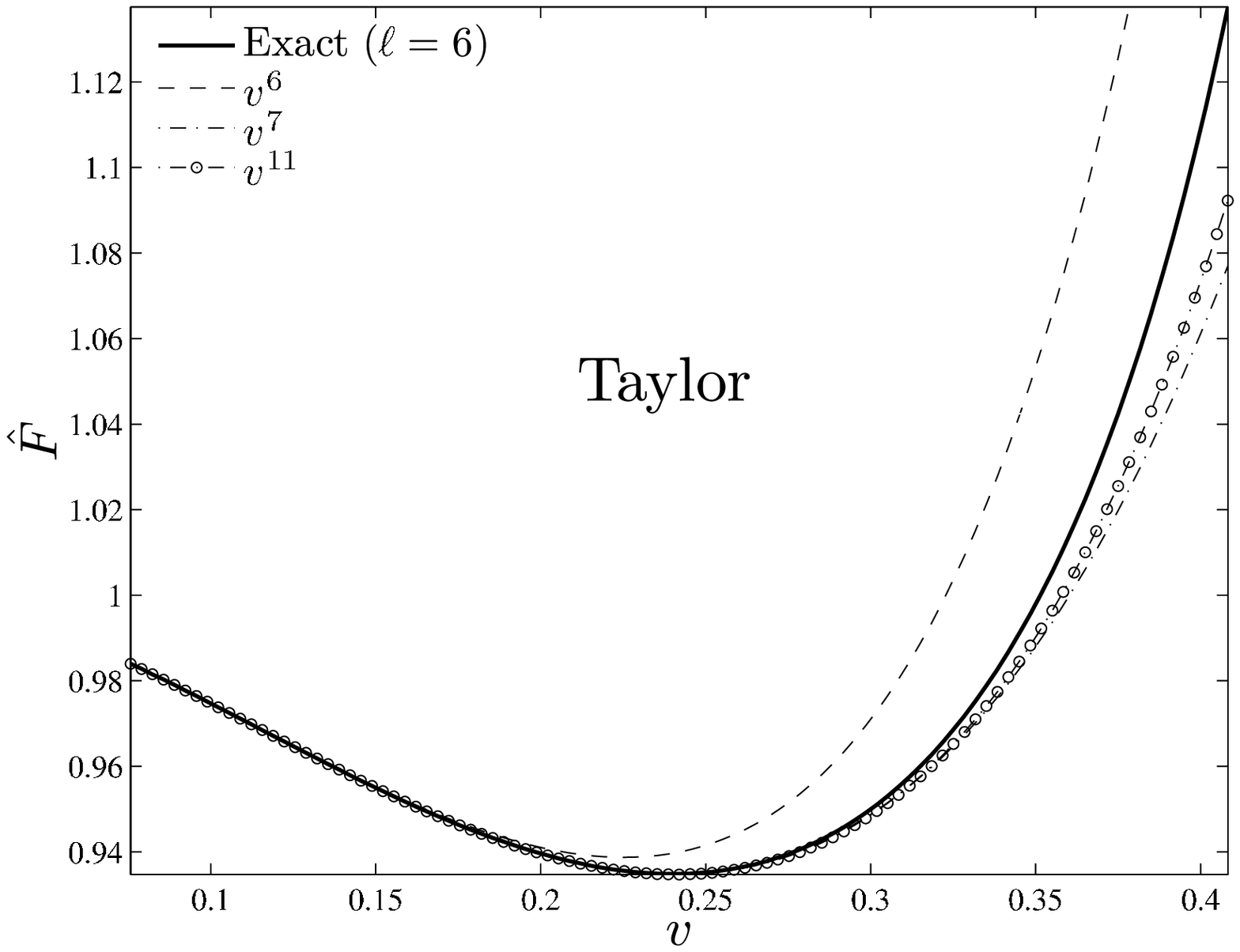}
\hspace{3 mm}
(b)\includegraphics[width=82 mm]{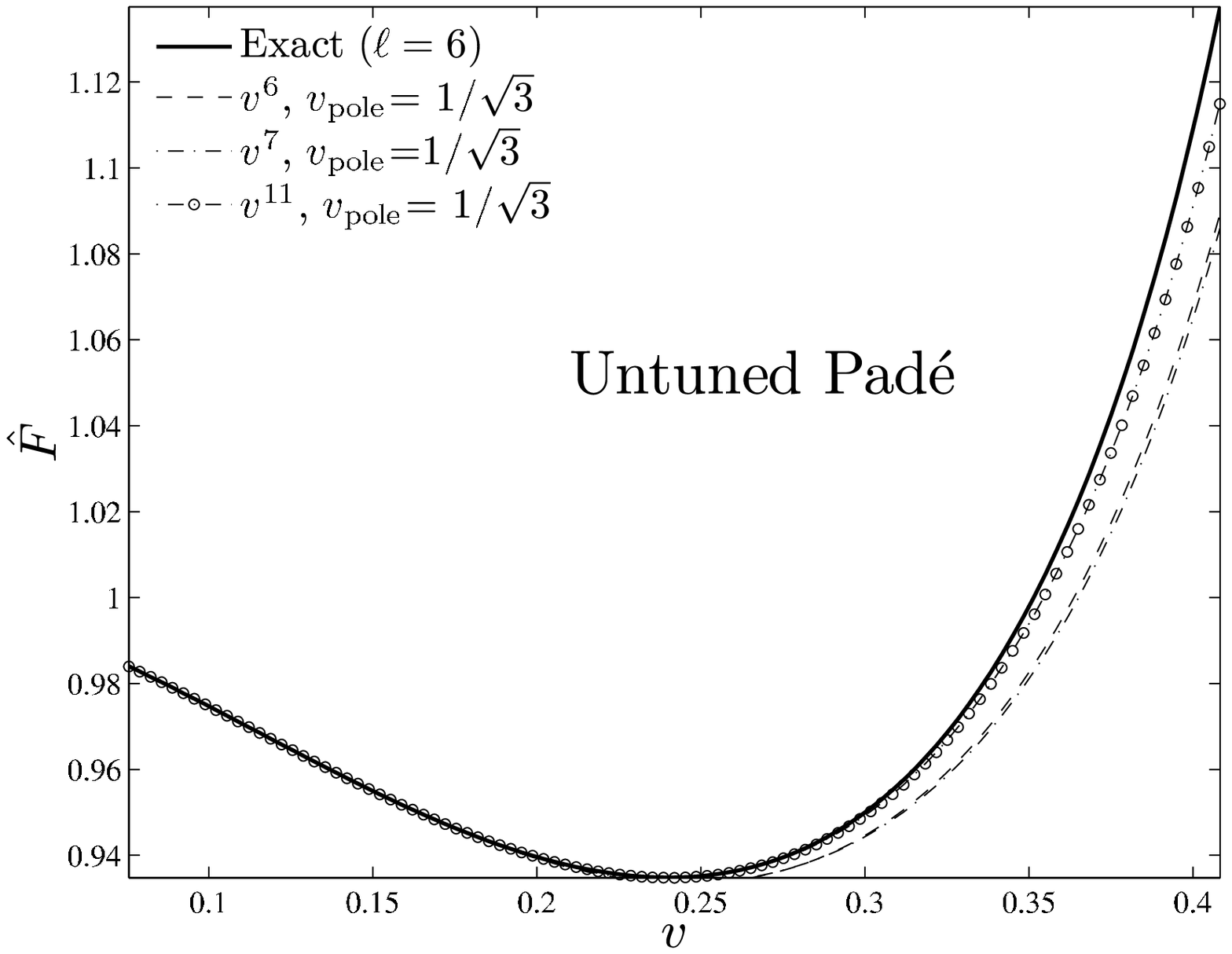}\\
\vspace{5mm}
(c)\includegraphics[width=82 mm]{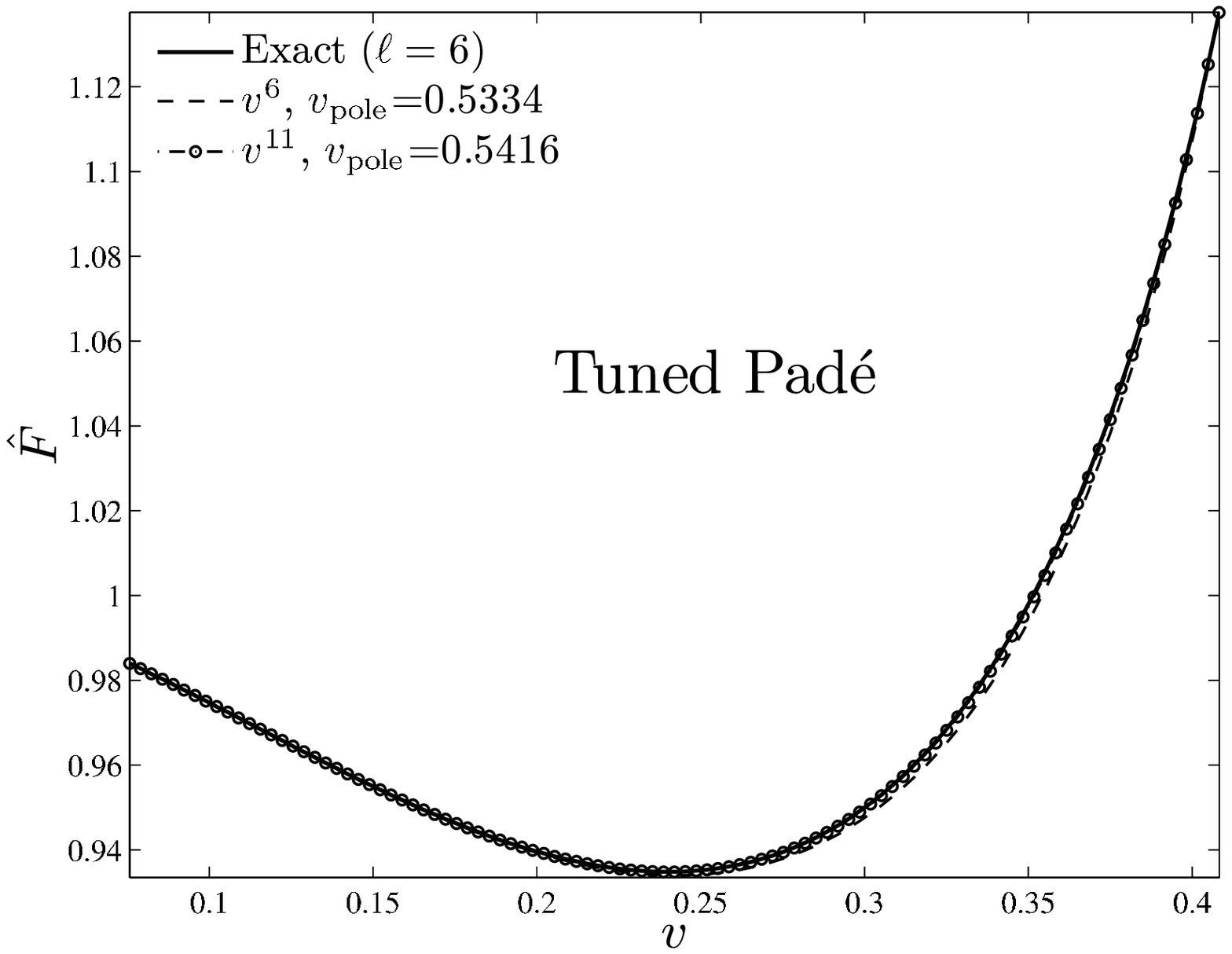}
\hspace{3 mm}
(d)\includegraphics[width=82 mm]{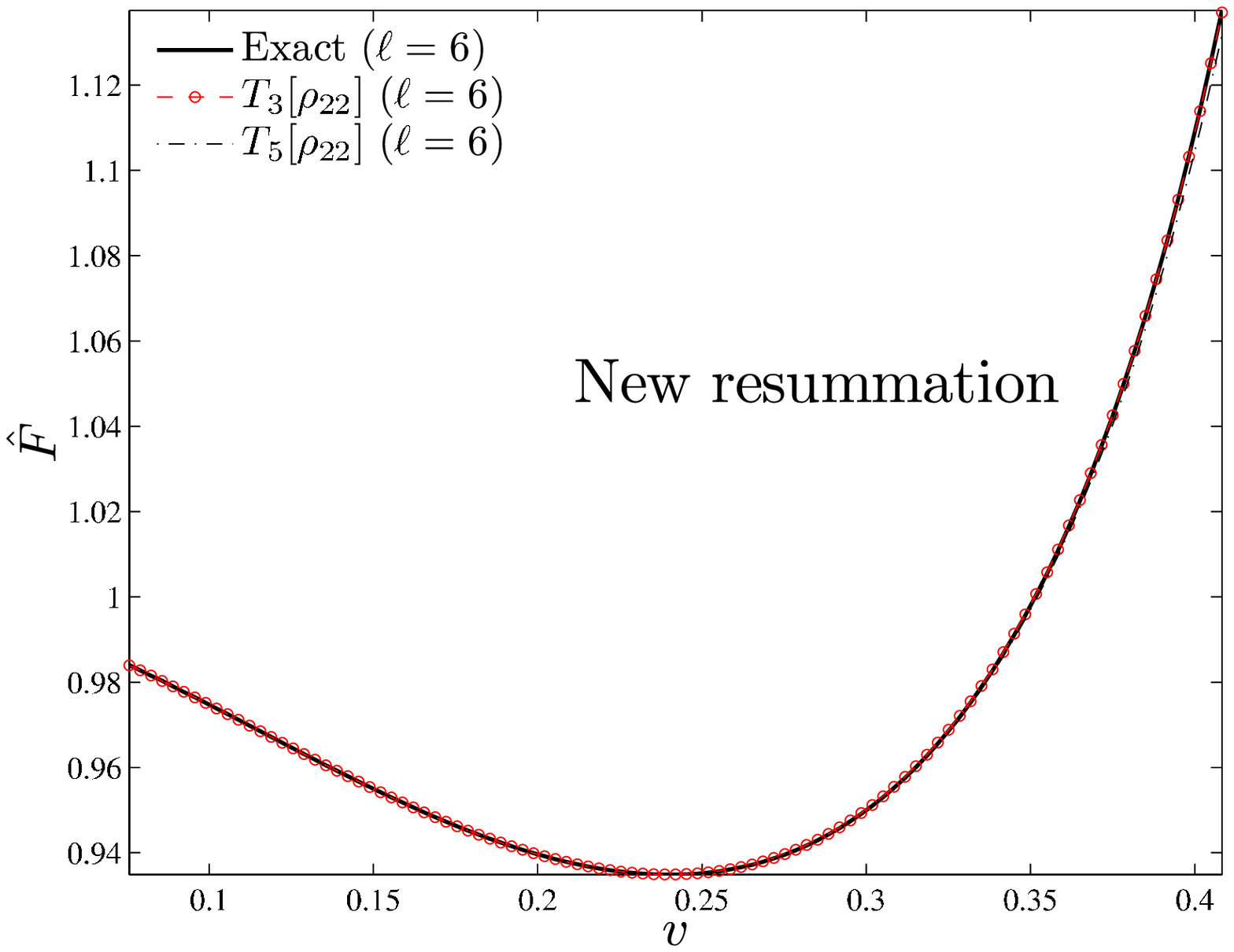}
\vspace{2mm}
\caption{ \label{fig:vpole} Extreme-mass-ratio limit ($\nu=0$).
Comparing various resummations of the (Newton-normalized) gravitational wave energy flux: 
(a) standard Taylor expansion; (b) Pad\'e resummation 
as proposed in Ref.~\cite{Damour:1997ub} with $\vp=1/\sqrt{3}$; (c) Pad\'e 
resummation {\it flexing} $\vp$ according to the discussion of Sec.~II of 
Ref.~\cite{Damour:2007yf}; (d) new resummation technique based on the 
$\rho_{\lm}$ functions discussed in this paper.}
  \end{center}
\end{figure*}

Though we leave to later sections the precise definition of the various
building blocks of our new, resummed waveform, let us already sketch here
its structure. The basic idea is to write the $(\ell,m)$ multipolar waveform
emitted by a circular\footnote{In this paper, we focus on the waveform
emitted by exactly circular orbits (see, however, footnote 12 below). 
We leave to future work the study of ``non-quasi-circular'' corrections 
that must be introduced in the realistic
case of inspiralling and plunging orbits (such corrections have already been
introduced in the EOB approach, see~\cite{Damour:2007vq,Damour:2008te}).}
nonspinning compact binary as the product of several factors, namely
\begin{align}
\label{eq:hlm}
h_{\lm}^{(\epsilon)}&=\dfrac{GM\nu}{c^2 R} n_{\lm}^{(\epsilon)} c_{\l+\epsilon}(\nu)
x^{(\ell+\epsilon)/2}Y^{\ell-\epsilon,-m}\left(\dfrac{\pi}{2},\Phi\right)\nonumber\\
                  &\times \hat{S}_{\rm  eff}^{(\epsilon)}T_{\ell m} e^{\ii\delta_{\lm}} \rho_{\lm}^\ell.
\end{align}
Here $\epsilon=0$ for ``even-parity'' (mass-generated) multipoles ($\ell+m$ even), and
$\epsilon=1$ for ``odd-parity'' (current-generated) ones ($\ell+m$ odd); $n_{\lm}^{(\epsilon)}$ and
$c_{\l+\epsilon}(\nu)$ are numerical coefficients;
$\hat{S}^{(\epsilon)}_{\rm eff}$ is a $\mu$-normalized effective source 
(whose definition comes from the EOB formalism); $T_{\ell m}$ is a resummed
version~\cite{Damour:2007xr,Damour:2007yf} of an infinite number of
``leading logarithms'' entering the {\it tail effects}~\cite{Blanchet:1992br,Blanchet:1997jj};
$\delta_{\lm}$ is a supplementary phase (which corrects the phase effects not
included in the {\it complex} tail factor $T_{\lm}$), and, finally,
$\left(\rho_{\lm}\right)^\ell$ denotes the $\ell$-th power of the quantity
$\rho_{\lm}$ which is the new building block introduced and studied in this
paper. [In previous papers~\cite{Damour:2007xr,Damour:2007yf} the quantity
$\left(\rho_{\lm}\right)^\ell$ was denoted as $f_{\lm}$.]

We shall discuss in quantitative details below the various facts showing that
the new ingredient $\rho_{\lm}(x)$ is a useful quantity to consider (mainly
because its PN expansion has better convergence properties than the
straightforward PN expansion of $h_{\lm}$ itself). In this introductory
section, we shall whet the appetite of the reader by comparing the performance
of our new resummed method, to some of the previously considered PN-based
methods.
For definiteness, we shall do this initial comparison at the level of the total energy
flux, say $F$, which is related to the individual waveforms via
\begin{align}
\label{eq:flux}
F^{(\ell_{\rm max})} = &\sum_{\ell =2}^{\ell_{\rm max}}\sum_{m=1}^{\ell} F_{\lm}=
                       \dfrac{2}{16\pi G}\sum_{\ell =2}^{\ell_{\rm max}}\sum_{m=1}^{\ell}|R\dot{h}_{\lm}|^2\nonumber\\
&=\dfrac{2}{16\pi G}\sum_{\ell =2}^{\ell_{\rm max}}\sum_{m=1}^{m=\ell} (m\Omega)^2|Rh_{\lm}|^2.
\end{align}
Note that $F_{\ell m}=F_{\ell|m|}$ denotes the sum of two equal contributions
corresponding to $+m$ and $-m$ ($m\neq0$ as $F_{\ell0}$ vanishes for circular orbits).
 This explains the explicit factor two in the
last two equations above, which relate $F_{\lm}$ to $h_{\lm}$.
It is convenient to consider the total flux $F$ for continuity with previous 
studies of the ``convergence'' of PN-expansions that focussed 
on $F$~\cite{Cutler:1993vq,Poisson:1995vs,Damour:1997ub,Yunes:2008tw}
and because of its physical importance as a measure of the radiation reaction
that acts on inspiralling binaries.
To be fully precise, we shall consider here the (rather accurate) approximation
$F^{(6)}$ obtained by truncating the sum over $\ell$ in Eq.~\eqref{eq:flux}
beyond $\ell=6$, and we normalize the result onto the ``Newtonian'' (i.e.,
quadrupolar) result $F_{22}^N=32/5(\mu/M)^2 x^5$. In other words, we consider here the
quantity $\hat{F}\equiv F^{(6)}/F_{22}^N$.
\begin{table}[t]
\caption{\label{tab:table1} Errors in the flux of the two (untuned or tuned) 
  Pad\'e resummation procedures.  
  From left to right, the columns report: the PN-order; the difference between the
  resummed and the exact flux,  $\Delta\hat{F}=\hat{F}^{\rm
  Resummed}-\hat{F}^{\rm Exact}$, at the LSO,
  and the $L_{\infty}$ norm  of $\Delta\hat{F}$, $||\Delta\hat{F}||_\infty$
  (computed over the interval $0<v<v_{\rm LSO}$), for $\vp=1/\sqrt{3}$;
   the {\it flexed} value of $v_{\rm pole}$ used here;
   $\hat{\Delta}F$ at the LSO and the corresponding $L_{\infty}$ norm 
  (over the same interval) for the flexed value of $\vp$.}
\begin{center}
  \begin{ruledtabular}
  \begin{tabular}{ccccccc}
    PN-order   & $\Delta\hat{F}^{1/\sqrt{3}}_{\rm LSO}$ 
               & $||\Delta\hat{F}||_\infty^{1/\sqrt{3}}$
               & $v_{\rm pole}$
               & $\Delta\hat{F}^{\vp}_{\rm LSO}$ 
               & $||\Delta\hat{F}||_\infty^{\vp}$  \\ 
    \hline \hline
 3   ($v^6$)    &   -0.048 &  0.048 & $0.5334$  &  $7.06\times 10^{-5}$ & 0.00426   \\
 3.5 ($v^7$)    &   -0.051 &  0.051 & $0.5425$  &  $5.50\times 10^{-5}$ & 0.00429   \\  
 5.5 ($v^{11}$) &   -0.022 &  0.022 & $0.5416$  &  $2.52\times 10^{-5}$ & 0.000854 
  \end{tabular}
\end{ruledtabular}
\end{center}
\end{table}%

Fig.~\ref{fig:vpole} compares and contrasts four different ways of using the
same PN information about the total Newton-normalized GW flux function
$\hat{F}(v)=F(v)/F_{22}(v)$
(i.e., the same finite set of coefficients $\{f_k(\log x); 1\leq k\leq n \}$ of
the $n$-PN expansion of the Newton-normalized 
flux $\text{Taylor}_n(\hat{F}(x))=\sum_{k=0}^nf_k(\log x) x^k$ of the GW flux).
As in many previous works, we use the extreme mass ratio limit $\nu\to 0$ as a
laboratory for devising and testing resummation procedures. Indeed, in that
case, the quasi-circular adiabatic description of inspiralling binaries is
expected to hold up to the LSO ($x_{\rm LSO}(\nu=0)=1/6$) and one can compare
PN-based analytical results~\cite{Poisson:1993vp,Tagoshi:1994sm,Tanaka:1997dj}
to numerical estimates of the GW flux, based on the numerical integration of some
(Regge-Wheeler-Zerilli or Teukolsky) wave equation~\cite{Cutler:1993vq,Yunes:2008tw}.

Panel (a) of the figure recalls the results of
Refs.~\cite{Cutler:1993vq,Poisson:1995vs}, namely the rather poor convergence
of the standard {\it Taylor approximants} of $\hat{F}(x)$ in the full interval
$0<x<x_{\rm LSO}$ where one might hope to tap the PN information. For
clarity, we selected only three Taylor approximants: 3PN ($v^6$), 3.5PN $(v^7)$
and 5.5PN ($v^{11}$). These three values suffice (by contrast with the other
panels) to illustrate the rather large scatter among Taylor approximants, and
the fact that, near the LSO, the convergence towards the exact value (solid
line) is rather slow, and non monotonic. 
[See Fig.~1 in Ref.~\cite{Poisson:1995vs} and Fig.~3 of Ref.~\cite{Damour:1997ub}
for fuller illustrations of the scattered and non monotonic way in which
successive Taylor expansions approach the numerical result.]
\begin{figure}[t]
   \begin{center}
      \includegraphics[width=85 mm]{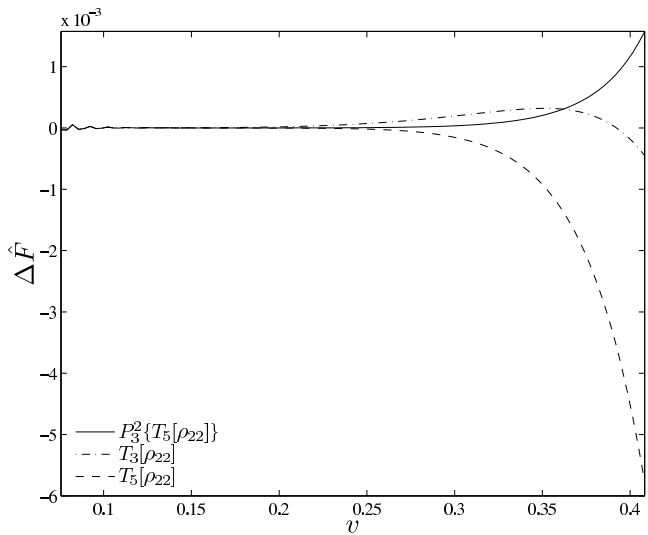}
      \caption{ \label{fig:error} Extreme-mass-ratio limit ($\nu=0$). 
	Complement to panel (d) of Fig.~\ref{fig:vpole}.
	Difference between the resummed and exact energy flux, 
	for different approaches to the resummation of the $\rho_{22}$ 
	function. See text for explanations.}
  \end{center}
\end{figure}

On the other hand, panel (b) recalls the results of~\cite{Damour:1997ub}, namely the
significantly better (and monotonic) way in which successive 
{\it Pad\'e approximants} approach (in $L_{\infty}$ norm on the full interval
$0<x<x_{\rm LSO}$) the numerical result. Ref.~\cite{Damour:1997ub} 
also showed that the observationally relevant overlaps (of both the ``faithfulness'' and
the ``effectualness'' types) between analytical and numerical adiabatic
signals were systematically better for Pad\'e approximants than for Taylor
ones. Note that our present panel (b) is slightly different from the
corresponding results in panel {\it (b)} of Fig.~3 in~\cite{Damour:1997ub} (in
particular, the present panel (b) exhibits a better ``convergence'' of the
$v^{11}$ curve). This difference is due to the new treatment of the
logarithmic terms $\propto \log x$. Instead of factoring them out in front
as proposed in~\cite{Damour:1997ub}, we consider them here
(following~\cite{Damour:2007yf}) as being part of the ``Taylor coefficients''
$f_n(\log x)$ when Pad\'eing the flux function. Note also that panel (b)
follows Ref.~\cite{Damour:1997ub} in introducing a pole in the resummed flux
$\hat{F}(v)$ located at the value $\vp^{(\nu=0)}=1/\sqrt{3}$. 

By contrast, panel (c) of the figure illustrates the remarkable improvement in the
($L_{\infty}$) closeness 
between $\hat{F}^{\text{Pad\'e-resummed}}(v)$ and $\hat{F}^{\rm Exact}(v)$
obtained, as recently suggested by Damour and Nagar~\cite{Damour:2007yf}
(following ideas originally introduced in Ref.~\cite{Damour:2002vi}),
by suitably flexing the value of $\vp$.
As proposed in Ref.~\cite{Damour:2007yf}, $\vp$ is tuned until
the difference between the resummed and the exact 
flux at the LSO is zero (or at least smaller than $10^{-4}$).
The resulting closeness between the exact and tuned-resummed fluxes is so
good (compared to the previous panels, where the differences were
clearly visible) that we need to complement panel (c) of Fig.~\ref{fig:vpole}
with Table~\ref{tab:table1}. This table compares in a quantitative way
the result of the ``untuned'' Pad\'e resummation ($\vp=1/\sqrt{3}$) 
of Ref.~\cite{Damour:1997ub} to the result of the ``$\vp$-tuned'' Pad\'e
resummation of Ref.~\cite{Damour:2007yf}. Defining the function
$\Delta\hat{F}(v;\vp)=\hat{F}^\text{Resummed}(v;\vp)-\hat{F}^\text{Exact}(v)$
measuring the difference between a resummed and the exact energy flux,
Table~\ref{tab:table1} lists both the values of $\Delta\hat{F}$ 
at $v=v_{\rm  LSO}$ and its $L_{\infty}$ norm on the interval $0<v<v_{\rm LSO}$
for both the untuned and tuned cases.
Note, in particular, how the $\vp$-flexing approach permits to reduce the $L_{\infty}$ 
norm over this interval by more than an order of magnitude with respect to
the untuned case. Note that the closeness between the tuned flux and the 
exact one is remarkably good ($4.3\times 10^{-3}$) already at the 3PN level. 

Finally, panel (d) of Fig.~\ref{fig:vpole} illustrates the even more
remarkable improvement in the closeness between
$\hat{F}^{\text{New-resummed}}$ and $\hat{F}^{\rm Exact}$ obtained by
 means of the new resummation procedure proposed in this paper.
More precisely, panel (d) plots two examples of fluxes obtained from our
new $\rho_{\lm}$-representation, Eq.~\eqref{eq:hlm}, for the individual
multipolar waveforms $h_{\lm}$ in the sum Eq.~\eqref{eq:flux}.
These two examples differ in the choice of approximants for the $\ell=m=2$
partial wave. One example uses for $\rho_{22}$ its 3PN Taylor expansion,
$T_3[\rho_{22}]$, while the other one uses its 5PN Taylor expansion,
$T_5[\rho_{22}]$. All the other partial waves are given by their maximum known
Taylor expansion. Note that the fact that we use here for the $\rho_{\lm}$'s
some straightforward Taylor expansions does not mean that our new procedure 
is not a resummation technique. Indeed, the defining resummation features of
our procedure have four sources: (i) the factorization of the PN
corrections to the waveforms into four different blocks, namely 
$\hat{S}_{\rm  eff}^{(\epsilon)}$, $T_{\lm}$, $e^{\ii \delta_{\lm}}$ and
$\rho_{\lm}^{\ell}$ in Eq.~\eqref{eq:hlm}; (ii) the fact 
the $\hat{S}_{\rm  eff}^{(\epsilon)}$ is by itself a resummed source whose
PN expansion would contain an infinite number of terms; (iii) the fact that
the tail factor is a closed form expression (see Eq.~\eqref{eq:tail_factor} below) whose
PN expansion also contains an infinite number of terms and (iv) the fact
that we have replaced the Taylor expansion of $f_{\lm}\equiv \rho_{\lm}^\ell$
by that of its $\ell$-th root, namely $\rho_{\lm}$.

Even more so than in the $\vp$-tuned case of panel (c), the closeness between
analytical and exact results exhibited by the ``new-resummed'' case of panel (d)
is so good that it is undistinguishable by eye. We therefore complement panel
(d) by displaying in Fig.~\ref{fig:error} the residual differences
$\Delta\hat{F}(v)=\hat{F}^{\rm New-resummed}(v)-\hat{F}^{\rm Exact}(v)$.
We included in Fig.~\ref{fig:error} a third curve corresponding to the case
where we further resum our ``new-resummed'' flux by using for 
(the 5PN accurate) $\rho_{22}$ its near-diagonal $(2,3)$ Pad\'e approximant, 
say $P^2_3\left\{T_5[\rho_{22}(x)]\right\}$, instead of its Taylor expansion.
[The other $\rho_{\lm}$'s being still used in Taylor expanded form].
Note that the difference $\Delta\hat{F}$ at the LSO is $\approx 4.5\times 10^{-4}$
when using $T_3[\rho_{22}]$, is $-5.7\times 10^{-3}$ when using  $T_5[\rho_{22}]$
and is $+1.6\times 10^{-3}$ when using $P^2_3\left\{T_5[\rho_{22}(x)]\right\}$.
Note that these numbers are in the same ball park than the $v^{11}$-accurate
$\vp$-tuned result ($8.5\times 10^{-4}$) quoted in Table~\ref{tab:table1}.
Discarding the very small difference corresponding to the 3PN-accurate $T_3$ case
as being probably accidental we conclude that using the normal,
near-diagonal\footnote{We will explore other Pad\'e approximants of $\rho_{22}$ below.}
Pad\'e resummation of {\it only} the leading multipolar contribution $\rho_{22}$
has the effect of significantly improving the agreement with the exact result
(compare the two 5PN-accurate curves, $T_5[\rho_{22}]$ and $P^2_3\left\{T_5[\rho_{22}(x)]\right\}$,
in Fig.~\ref{fig:error}). We therefore expect that Pad\'eing some of the
higher multipoles will further improve the agreement between the energy fluxes.
Note also, in passing, that the new resummation procedure explored here
is more ``predictive'' than the $\vp$-flexing technique in that it does
not need to rely on the knowledge of the exact answer. We will also show below
that it is ``robust'' under the deformation brought about by increasing the 
symmetric mass ratio $\nu$ from $0$ up to its maximal value $1/4$.

This paper is organized as follows: in Sec.~\ref{building_blocks} we
explicitly define the building blocks entering the resummation of the
gravitational waveform. Section~\ref{results_nu0} analyzes the performance
of this resummation procedure in the extreme-mass-ratio ($\nu\to0$) case,
while Sec.~\ref{sec:eqmass} considers the comparable mass case ($\neq 0$).
We present some concluding remarks in Sec.~\ref{conclusions}.
The paper is completed by three Appendices which  collect many useful 
formulas that complete the discussion of the main text.
In particular, Appendix A provides the first explicit, ready-to-use expression
for the $1$-PN corrections to the Symmetric-Trace-Free (STF) 
current-multipole moments and to the corresponding 
spherical-harmonics odd-parity multipoles for arbitrary $\ell$ and $m$.

\section{Defining the building blocks of the resummation 
         of the multipolar gravitational waveform}
\label{building_blocks}

Let us now explicitly define each of the building blocks of our
new resummed waveform, Eq.~\eqref{eq:hlm}. Note that our methodology
differs from the PN-methodology in a basic way. The PN approach consists
in writing any relativistic quantity as a {\it sum} of various contributions
starting with the so-called Newtonian approximation. In other words,
a PN-expanded multipolar waveform has the structure $h_{\lm}=h_{\lm}^N
+h_{\lm}^{1\rm PN} + h_{\lm}^{1.5\text{PN}}+\dots$. By contrast to this
{\it additive} approach we will use here, as advocated 
in Refs.~\cite{Damour:2007yf,Damour:2007xr}, 
a {\it multiplicative} approach in which any relativistic quantity is
decomposed as a {\it product} of  various 
contributions\footnote{This multiplicative approach can be naturally applied
to the multipolar waveform $h_{\lm}$ which is a {\it complex number}.}. 
One of the factors of this multiplicative decomposition will be the 
Newtonian waveform, $h_{\lm}^N$. Some of the other factors are chosen 
so as to best capture part of the essential physics contained 
in the waveform. The remaining 
factors will then resum the subleading effects that have not been included
in the previous ones.
First of all, it is convenient to introduce the following notation
\begin{equation}
\label{hlm_expanded}
h_{\lm} = h_{\lm}^{(N,\epsilon)} \hat{h}_{\lm}^{(\epsilon)},
\end{equation}
where $h_{\lm}^{(N,\epsilon)}$ represents the Newtonian contribution 
and $\hat{h}_{\lm}^{(\epsilon)}$ the product of all the other factors
in our multiplicative decomposition. As all these other factors represent
resummed version of PN corrections (of the type $1+{\cal O}(x)$) their
product will also have the structure $\hat{h^{(\epsilon)}}_{\lm}=1+{\cal O}(x)$.

\subsection{The Newtonian multipolar waveform}
\label{sbsc:newt}

Though all the work in this paper will focus on the resummation of the
PN-correcting factor $\hat{h}_{\lm}^{(\epsilon)}$, let us, for completeness,
recall the well-known~\cite{Thorne:1980ru,Kidder:2007rt,Blanchet:2008je} structure of the Newtonian
multipolar waveform\footnote{We mostly follow here the 
conventions of Refs.~\cite{Thorne:1980ru,Kidder:2007rt}, 
except that we take into account some of the 
simplifications used in~\cite{Blanchet:2008je}. Note the presence of a factor
$1/\sqrt{2}$ in the relation between the $(\ell,m)$ Newtonian waveform and the
corresponding $(\ell,m)$ radiative multipoles: 
$R h_{\lm}^{\rm even}=(G/\sqrt{2})U_{\lm}=(G/\sqrt{2})I_{\lm}^{(\ell)}$ for
mass multipoles and 
$R h_{\lm}^{\rm odd}=-\ii(G/\sqrt{2})V_{\lm}=-\ii(G/\sqrt{2})S_{\lm}^{(\ell)}$ for current multipoles.}, 
here considered  for the adiabatic
circular case. The Newtonian contribution for circular orbits is, for given
$(\ell,m)$, a function of $x\equiv(GM\Omega/c^3)^{2/3}$ and the two mass
ratios $X_1=m_1/M$ and $X_2=m_2/M$\footnote{Note that: $X_1+X_2=1$,
$X_1X_2=\nu$ where $\nu$ is the symmetric mass ratio $\nu\equiv m_1 m_2/(m_1+m_2)^2$, 
while $X_1-X_2=\text{sign}(m_1-m_2)\sqrt{1-4\nu}$, where the sign
depends whether $m_1>m_2$ or the reverse.}
\begin{align}
\label{eq:hNdef}
h_{\lm}^{(N,\epsilon)}&=\dfrac{GM\nu}{c^2 R}n_{\lm}^{(\epsilon)} c_{\l+\epsilon}(\nu)
x^{(\ell+\epsilon)/2}Y^{\ell-\epsilon,-m}\left(\dfrac{\pi}{2},\Phi\right).
\end{align}
Here, $\epsilon$ denotes the parity of the multipolar 
waveform, i.e., even ($\epsilon=0$) for mass-generated multipoles and odd
($\epsilon=1$) for  current-generated ones. In the circular case,
$\epsilon$ is equal to the parity of the sum $\ell+m$:  
$\epsilon=\pi(\ell+m)$. In other words $\epsilon=0$ 
when $\ell+m$ is even, and $\epsilon=1$ when $\ell+m$ is odd.
The $Y^{\lm}(\theta,\phi)$ are the usual 
scalar spherical harmonics\footnote{We use the $Y_{\lm}$'s defined in
Eqs.~(2.7) and (2.8) of Ref.~\cite{Thorne:1980ru}, or equivalently by
the $s=0$ case of Eqs.~(4) and (5) of Ref.~\cite{Kidder:2007rt}.}
while
\begin{align}
\label{eq:newtnorm}
n^{(0)}_{\lm}  & = (\ii m)^{\ell} \dfrac{8\pi}{(2\ell +
  1)!!}\sqrt{\dfrac{(\ell+1)(\ell+2)}{\ell(\ell-1)}}, \\
n^{(1)}_{\lm} & = -(\ii m)^\ell \dfrac{16\pi\ii}{(2\ell+1)!!}\sqrt{\dfrac{(2\ell+1)(\ell+2)(\ell^2-m^2)}{(2\ell-1)(\ell+1)\ell(\ell-1)} },
\end{align} 
are ($\nu$-independent) numerical coefficients. Finally, the $\nu$-dependent
coefficients $c_{\ell+\epsilon}(\nu)$ (such that
\hbox{$|c_{\ell+\epsilon}(\nu=0)|=1$}), 
can be expressed in terms of $\nu$ (as in Ref.~\cite{Kidder:2007rt,Blanchet:2008je}), 
although it is more conveniently
written in terms of the two mass ratios $X_1$ and $X_2$ 
in the form\footnote{When expressing 
$c_{\ell+\epsilon}(\nu)$ as an explicit function of 
$\nu$, as in Ref.~\cite{Kidder:2007rt}, 
it is useful to note that $c_{\ell+\epsilon}(\nu)$ vanishes in the equal mass
case when $\ell+\epsilon$ is odd, which is equivalent (given that
$\epsilon=\pi(\ell+m)$ for circular orbits), to $m$ being odd. 
In such cases one can factor 
out, as in Ref.~\cite{Kidder:2007rt,Blanchet:2008je},
from $c_{\ell+\epsilon}(\nu)$ a factor 
$\Delta\equiv X_1-X_2=\text{sign}(m_1-m_2)\sqrt{1-4\nu}$.
The  $c_\ell(\nu)$ in this paper are the same as the $s_\ell(\nu)$ 
in~\cite{Blanchet:2008je}.
Note however that in Appendix~A of Ref.~\cite{Kidder:2007rt},
in the second line above Eq.~(A7), the definition of 
$d_{\ell}$ should include a supplementary factor
$m/\delta m=1/(X_1-X_2)$ on the right hand side.}
\begin{align}
\label{eq:cl}
c_{\ell+\epsilon}(\nu) & 
= X_2^{\ell+\epsilon-1}+(-)^{\ell+\epsilon}X_1^{\ell+\epsilon-1}\nonumber\\
&=X_2^{\ell+\epsilon-1}+(-)^m X_1^{\ell+\epsilon-1}.
\end{align}
In the second form of the equation we have used the fact that, 
as $\epsilon=\pi(\ell+m)$, $\pi(\ell+\epsilon)=\pi(m)$.

\subsection{The first three factors in the multiplicative decomposition 
            of the PN fractional correction $\hat{h}_{\lm}^{(\epsilon)}$ }
\label{sbsc:first_three}

Let us recall that, in the comparable mass case ($\nu\neq 0$), 
the $\hat{h}_{\lm}^{(\epsilon)}$ PN-correction can be computed within 
the perturbative 
multipolar-post-Minkowskian (MPM) formalism~\cite{Blanchet:1992br,Blanchet:1997jj,Blanchet:2002av},
while in the test-mass limit ($\nu\to 0$) it can be obtained by black hole
perturbation theory~\cite{Poisson:1994yf,Tagoshi:1994sm,Tanaka:1997dj,lrr-2003-6}.

The final result is that $\hat{h}_{\lm}^{(\epsilon)}$ is given by a PN
expansion of the form $\hat{h}_{\lm}=1 + h_1 x + h_{1.5} x^{3/2} + \dots $.
For comparable-mass circularized compact binaries, the partial wave which
is known with the highest PN accuracy is the leading even-parity quadrupolar wave
$\hat{h}_{22}$, which is known to fractional 3PN accuracy
~\cite{Damour:2007yf,Kidder:2007rt,Arun:2004ff,Blanchet:2008je}.
Note that Ref.~\cite{Blanchet:2008je} provides $h_{\lm}$ half a PN 
order more accurately than Ref.~\cite{Kidder:2007rt} for multipolar 
orders $(\ell,m)=(2,1),(3,3),(3,2),(3,1),
(4,3),(4,1),(5,5),(5,3)$ and $(5,1)$. 
This information is fully employed in this work.  
In the extreme-mass-ratio case, the partial waves are known with 
even higher PN accuracy. For instance, $\hat{h}_{22}$ is known to 
5.5PN~\cite{Tagoshi:1994sm,Tanaka:1997dj} and other 
multipoles to accuracies consistent with 5.5PN GW flux. As explained later, 
this information is also appropriately exploited in our construction.

As indicated above, the resummation method 
we shall use here consists in: (i) decomposing the PN-correction 
factor $\hat{h}^{(\epsilon)}_{\lm}=1 + h_1 x + h_{1.5} x^{3/2} + \dots$ 
into the {\it product of four factors}, each of which has a similar PN expansion,
$1+{\cal O}(x)$, namely
\begin{equation}
\label{eq:hhat}
\hat{h}_{\lm}^{(\epsilon)} = \hat{S}_{\rm  eff}^{(\epsilon)}T_{\ell m} e^{\ii\delta_{\lm}} \rho_{\lm}^\ell,
\end{equation}
and then (ii), resumming separately each factor.

The choice of these various factors is based on our physical intuition of
the main physical effects entering the final waveform.
The first factor is motivated by thinking about the form of the equation
satisfied by each partial wave in the (circular) test-mass limit: indeed in this 
limit $\hat{h}_{\lm}$ is the asymptotic value (at spatial infinity) of
a solution of a (frequency-domain) wave equation of 
the Regge-Wheeler-Zerilli type (see e.g. Ref.~\cite{Nagar:2005ea}). The source
term appearing on the right-hand side 
of this equation is a linear combination of 
terms linear in the stress-energy tensor $T_{\mu\nu}$ of a test-particle
of mass $\mu$ moving around a black hole of mass $M$. As the
Effective-One-Body method has shown that the dynamics of comparable-mass
black holes can be mapped onto the dynamics of an effective particle 
of mass $\mu$ moving in some effective metric (which reduces to the
Schwarzschild metric of mass $M$ when $\nu\to 0$), it is natural to introduce 
(both when $\nu\to 0$ and $\nu\neq 0$) effective source terms in the 
partial waves made up from the important dynamical characteristics of 
the EOB dynamics, namely the effective EOB Hamiltonian $H_{\rm eff}$ and
the EOB angular momentum ${\cal J}$.
This motivates us to define as first factor in $\hat{h}_{\lm}$ an effective
source term $S^{(\epsilon)}_{\rm eff}$ proportional either to $H_{\rm eff}$
or ${\cal J}$. Note that this idea of factoring $H_{\rm eff}$ or $\J$ from the wave
amplitude is similar to the suggestion of Ref.~\cite{Damour:1997ub} of factoring 
out a pole in the energy flux.
Indeed, the analytical continuation in $x$ of the flux function $F(x)$
below the LSO inherits, in the $\nu\to 0$ limit, a simple pole from the
fact that $F(x)$ is proportional to the square\footnote{Note that 
in the $\nu\to0$ limit both $H_{\rm eff}$ and $\J$ have a square-root 
singularity $\propto 1/\sqrt{1-3x}$ at the light-ring. 
See e.g. Eqs.~\eqref{Hnu0}-\eqref{jnu0} below.} of the energy of the
``rotating source'' (see discussion p.~893 of~\cite{Damour:1997ub}).

Our second factor is motivated by thinking about the structure 
of the ``transfer'' function relating (in the comparable mass case) 
the far-zone GW amplitude $h_{\lm}$ to the near zone one.
If we keep, in the full Einstein equations considered outside the 
binary system, only the terms coupling the instantaneous ``monopolar''
Arnowitt-Deser-Misner (ADM) 
mass of the system, $M_{\rm ADM}=M+\text{binding energy}=H_{\rm real}$,
to the multipolar wave amplitude, we get (in the circular approximation
and in the Fourier domain)
a Schr\"odinger-type equation, for each multipole order $\ell$, 
containing a potential $V_\ell(r)$ whose
leading behavior as $r\to\infty$ is dominated by two effects: (i) the
$\ell(\ell+1)/r^2$ centrifugal barrier, and (ii) a more slowly decreasing term
$\sim -4M_{\rm ADM}\omega^2/r$ coming from the coupling to a curved
(Schwarzschild-like) ``monopolar'' background metric.
One can solve this leading-order equation 
by means of Coulomb wave functions. When doing this, it is found that 
each asymptotic partial wave is related to its corresponding near-zone
expression by a certain ``tail'' factor $T_{\lm}$. 
It can be checked that, in the comparable mass case, this tail factor 
represents the resummation of the infinite number of 
leading logarithms (see Eqs.~(7)-(9) in~\cite{Damour:2007yf}) 
that appear when computing asymptotic multipolar waves in the MPM 
formalism~\cite{Blanchet:1992br,Blanchet:1997jj,Blanchet:2002av}.
Having so factorized two of the main physical effects entering
$\hat{h}_{\lm}$, we define the two other factors as the 
phase, $e^{\ii\delta_{\lm}}$ and  the modulus, $f_{\lm}$ 
of the  remaining quotiented Newton-normalized waveform.
In this subsection we discuss in detail the first three factors, 
postponing to the following subsection the last one, namely the modulus $f_{\lm}$.

Let us start by discussing the structure of the $\hat{S}^{(\epsilon)}_{\rm eff}$
and $T_{\lm}$ factors. In the even-parity case (corresponding to mass
moments), since  the leading order
source of gravitational radiation is given by the energy density, 
it is natural to define
\begin{align}
\label{eq:source_even}
\hat{S}^{(0)}_{\rm eff}(x) &=\hat{H}_{\rm eff}(x)\qquad \ell+m \quad \text{even}.
\end{align}
Here, $\hat{H}_{\rm eff}$ is the {\it effective} EOB Hamiltonian (per unit
$\mu$ mass), that we shall restrict here along the sequence of EOB 
circular orbits. 
When $\nu\to 0$, $\hat{H}_{\rm eff}$ reduces to the usual conserved energy of
a test-mass $\mu$ in a Schwarzschild background of mass $M$ (see Eq.~\eqref{Hnu0} below).

The explicit expression of $\hat{H}_{\rm eff}$, along
circular orbits, as a function of the frequency parameter $x$, cannot
be written in closed form~\cite{Buonanno:1998gg,Buonanno:2000ef}. 
However, it can be written in parametric form in terms of the EOB 
inverse radius parameter\footnote{As usual in EOB work, we use dimensionless variables, 
notably  $r= Rc^2/GM$, where $R$ is the EOB Schwarzschild-like radial
coordinate.} $u=1/r$. More precisely, we have
\begin{equation}
\label{eq:Heff}
\hat{H}_{\rm eff}=\dfrac{H_{\rm eff}}{\mu}=\sqrt{A(u)(1+j^2 u^2)} \quad
\text{(circular orbits)},
\end{equation}
where $u=1/r$, and where  $A(u)(\equiv -g_{00}^{\rm effective}(r))$ 
is the crucial EOB radial potential and 
$j={\cal J}/(\mu G M)$ is the (dimensionless) angular momentum 
along circular orbits.
We recall that the PN expansion of $A(u)$ has the form
\begin{align}
\label{eq:Au}
A^{\text{Taylor}}(u) = &1-2u+2\nu u^3 + \left(\dfrac{94}{3}-\dfrac{41}{32}\pi^2\right)\nu u^4 +a_5\nu u^5 \nonumber\\
&+ {\cal O}(\nu u^6),
\end{align}
where the $u^4$ term corresponds to 3PN contributions to the EOB
dynamics~\cite{Damour:2000we} and where
we have parametrized the presence of yet uncalculated 4PN
(and higher) contributions to $A(u)$ by adding a term $+a_5(\nu)u^5$ 
with the simple form $a_5(\nu)=a_5\nu$. As in previous EOB work, 
we shall not use the Taylor-expanded function $A^{\text{Taylor}}(u)$,
but replace it by a suitably Pad\'e resummed function $A(u)$.

The circular orbits in the EOB formalism are determined by the condition
$\de_u\left\{A(u)[1+j^2u^2]\right\}=0$, which leads to the following
parametric representation of the squared angular momentum:
\begin{equation}
j^2(u)=-\dfrac{A'(u)}{(u^2 A(u))'}\quad\text{(circular orbits)},
\end{equation}
where the prime denotes  $d/du$. Inserting this $u$-parametric representation of 
$j^2$ in Eq.~\eqref{eq:Heff} defines the $u$-parametric representation of the
effective Hamiltonian $\hat{H}_{\rm eff}(u)$.
We can then obtain (at least numerically) $\hat{H}_{\rm eff}$ as a function of $x$
by eliminating $u$ between $\hat{H}_{\rm eff}(u)$ and the corresponding
$u$-parametric representation of the frequency parameter $x=(GM\Omega/c^3)^{2/3}$
obtained by the angular Hamilton equation of motion in the circular case 
\begin{equation}
\label{eq:Omega}
M\Omega(u) = \dfrac{1}{\mu}\dfrac{\de H_{\rm real}}{\de j}=\dfrac{M A(u)j(u) u^2}{H_{\rm real}\hat{H}_{\rm eff}},
\end{equation}
where $H_{\rm real}$ denotes the real EOB Hamiltonian
\begin{equation}
\label{eq:real_hamiltonian}
H_{\rm real} = M \sqrt{ 1 + 2\nu\left( \hat{H}_{\rm eff} - 1\right)}.
\end{equation}

While in the even-parity case we shall factor out $\hat{H}_{\rm eff}(x)$ as 
a ``source term'', in 
the odd-parity one we explored two, equally motivated, possibilities. 
The first one consists simply in still factoring $\hat{H}_{\rm eff}(x)$;
i.e., in defining 
\begin{equation}
\label{eq:source_odd_H}
\hat{S}^{(1,H)}_{\rm eff} =\hat{H}_{\rm eff}(x)\qquad  \ell+m \quad \text{odd}.
\end{equation}
The second one consists in factoring the angular momentum ${\cal J}$.
Indeed, the angular momentum density $\epsilon_{ijk}x^j \tau^{0k}$
enters as a factor in the (odd-parity) current moments, and ${\cal J}$ 
occurs (in the small-$\nu$ limit) as a 
factor in the source of the Regge-Wheeler-Zerilli odd-parity multipoles.
This leads us to define as second possibility
\begin{equation}
\label{eq:source_odd_J}
\hat{S}^{(1,J)}_{\rm eff} =\hat{j}(x)\equiv x^{1/2}j(x) \qquad  \ell+m \quad \text{odd},
\end{equation}
where $\hat{j}$ denotes what can be called the ``Newton-normalized'' 
angular momentum, namely the ratio $\hat{j}(x)=j(x)/j_N(x)$
with $j_N(x)=1/\sqrt{x}$. 
[This Newtonian normalization being such that $\hat{j}(x)=1+{\cal O}(x)$.]
We will compare below the performances of these two possible choices.
Note that the PN expansions of these two possible sources start as 
\begin{align}
\label{H_expansion}
\hat{H}_{\rm eff}(x) &=1-\dfrac{1}{2} x +{\cal O}(x^2), \\
\label{j_expansion}
\hat{j}(x)           &=1+\left(\dfrac{3}{2}+\dfrac{\nu}{6}\right)x + {\cal O}(x^2) .
\end{align}
The second building block in our factorized decomposition is the ``tail
factor'' $T_{\lm}$ (introduced in Refs.~\cite{Damour:2007xr,Damour:2007yf}).
As mentioned above, $T_{\lm}$ is a resummed version of an infinite numbers of ``leading
logarithms'' entering the transfer function between the near-zone multipolar
wave and the far-zone one, due to {\it tail effects} linked to its propagation
in a Schwarzschild background of 
mass $M_{\rm ADM}=H_{\rm real}$. Its explicit expression
reads
\begin{equation}
\label{eq:tail_factor}
T_{\lm} = \dfrac{\Gamma(\ell+1-2\ii\hat{\hat{k}})}{\Gamma(\ell
  +1)}e^{\pi\hat{\hat{k}}}e^{2\ii\hat{\hat{k}}\log(2 k r_0)} ,
\end{equation}
where $r_0=2GM$ and $\k\equiv G H_{\rm real} m\Omega$
and $k\equiv m\Omega$.
Note that  $\k$ differs from $k$ by a rescaling involving 
the {\it real} (rather than the {\it effective}) 
EOB Hamiltonian, Eq.~\eqref{eq:real_hamiltonian}.

The tail factor $T_{\lm}$ is a complex number which already takes into 
account some of the dephasing of the partial waves as they propagate
out from the near zone to infinity. However, as the tail factor only takes 
into account the leading logarithms, one needs to correct it by a complementary 
dephasing term, $e^{\ii\delta_{\lm}}$,  
linked to subleading logarithms and other effects.
This subleading phase correction can be computed as being the phase
$\delta_{\lm}$ of the
complex ratio between the PN-expanded $\hat{h}_{\lm}^{(\epsilon)}$ and the 
above defined source and tail factors. In the comparable-mass case
($\nu\neq0$), the 3PN $\delta_{22}$ phase correction to the leading quadrupolar
wave was computed in Ref.~\cite{Damour:2007yf} (see also
Ref.~\cite{Damour:2007xr} for the $\nu=0$ limit). 
For the subleading partial waves, 
we computed the other $\delta_{\lm}$'s to the highest 
possible PN-accuracy by starting from the currently known 
3PN-accurate $\nu$-dependent waveform~\cite{Blanchet:2008je}.
Our explicit results read 
\begin{align}
\delta_{22} &=\dfrac{7}{3}y^{3/2} + \dfrac{428\pi}{105}y^3 - 24 \nu \y^{5/2},\\
\delta_{21} &=\dfrac{2}{3}y^{3/2}-\dfrac{493\nu}{42}\y^{5/2}, \\
\delta_{33} &= \dfrac{13}{10}y^{3/2}-\dfrac{80897}{2430}\nu \y^{5/2}, \\
\delta_{32} &=\dfrac{10+33\nu}{15(1-3\nu)}y^{3/2},\\
\delta_{31} &=\dfrac{13}{30}y^{3/2}-\dfrac{17\nu}{10}\y^{5/2},\\
\delta_{44} &=\dfrac{112+219\nu}{120(1-3\nu)}y^{3/2},\\
\delta_{43} &= \dfrac{486+4961\nu}{810(1-2\nu)}y^{3/2},\\
\delta_{42} &=\dfrac{7(1+6\nu)}{15(1-3\nu)}y^{3/2},\\
\delta_{41} &=\dfrac{2+507\nu}{10(1-2\nu)}y^{3/2},\\
\delta_{55} &=\dfrac{96875+857528\nu}{131250(1-2\nu)}y^{3/2}.
\end{align}
Here, following Ref.~\cite{Damour:2007yf}, we define $y\equiv (H_{\rm real}\Omega)^{2/3}$,
which gathers together relativistic corrections (like those entering the tail)
that depend on the instantaneous ADM mass of the system, namely 
$H_{\rm  real}$, rather than the total ``mechanical mass''$M$.
Concerning the last $\y^{5/2}$ 
corrections\footnote{Note that these ${\cal O}\left(\bar{y}^{5/2}\right)$ 
corrections in $\delta_{\lm}$ are the only terms in the wave amplitude that we
use here which go beyond the strict `circular limit' in that they
include contributions proportional to the (radiation-reaction-driven) 
time derivatives of the orbital radius, or of the orbital frequency.}, 
it is not clear whether they are more linked to the ADM mass or 
to the mechanical mass. This is why we use the notation $\bar{y}$, meaning that 
it could be replaced either
by $x$ or $y$ [note that in Ref.~\cite{Damour:2007yf} we 
chose $\bar{y}=x$ inside the $-24\nu\y^{5/2}$ correction 
to $\delta_{22}$, Eq.~(11) there ].
Indeed, these 2.5PN terms are not known to 1PN
fractional accuracy because we rely here on 
the available 3PN (and not 3.5PN) accurate results of~\cite{Blanchet:2008je}.

In the extreme-mass-ratio limit $\nu\to 0$, the information needed to compute
some of the higher-order PN corrections to the $\delta_{\lm}$'s is contained in the
results of Ref.~\cite{Tagoshi:1994sm}. We leave to future work the task of 
exploiting this information to complete the above $\nu$-dependent
$\delta_{\lm}$'s with higher-order $\nu=0$ corrections.
In addition we shall leave here the $\delta_{\lm}$'s in Taylor-expanded form.
We leave to future work an eventual comparison between numerically determined
phases and (possibly resummed) analytic ones.

\subsection{The fourth factor in the multiplicative decomposition 
            of the PN fractional correction $\hat{h}_{\lm}^{(\epsilon)}$ }
\label{sbsc:flm}

The fourth and last factor in the multiplicative decomposition,
Eq.~\eqref{eq:hhat}, can be computed 
as being the modulus $f_{\lm}$ of the complex ratio between 
the PN-expanded $\hat{h}_{\lm}^{(\epsilon)}$  and the 
above defined source and tail factors.
In the comparable mass case
($\nu\neq0$), the $f_{22}$ modulus correction to the leading quadrupolar
wave was computed in Ref.~\cite{Damour:2007yf} (see also
Ref.~\cite{Damour:2007xr} for the $\nu=0$ limit). 
For the  subleading partial waves, 
we compute here the other $f_{\lm}$'s to the highest 
possible PN-accuracy by starting from the currently known 
3PN-accurate $\nu$-dependent waveform~\cite{Blanchet:2008je}.
In addition, as originally proposed in Ref.~\cite{Damour:2007yf}, 
to reach greater accuracy the $f_{\lm}(x;\nu)$'s extracted from
the 3PN-accurate $\nu\neq 0$ results  are complemented by adding 
higher order contributions coming from the 
$\nu=0$ results~\cite{Tagoshi:1994sm,Tanaka:1997dj}.
In the particular $f_{22}$ case discussed 
in~\cite{Damour:2007yf}, this amounted to adding 4PN and 5PN $\nu=0$
terms. This ``hybridization'' procedure is here systematically pursued for all the other
multipoles, using the 5.5PN accurate calculation of the multipolar
decomposition of the gravitational wave energy flux done 
in Refs.~\cite{Tagoshi:1994sm,Tanaka:1997dj}.
It is worth emphasizing at this stage that 
our hybridization procedure 
is {\it not} equivalent to the
straightforward hybrid sum ansatz, 
$\tilde{h}_{\ell m}=\tilde{h}_{\ell m}^{\text{known}}(\nu)+\tilde{h}_{\ell
  m}^{\text{higher}}(\nu=0)$ (where $\tilde{h}_{\lm}\equiv h_{\lm}/\nu$)
that one may have chosen to implement.
The detailed definition of the hybridization procedure 
that we use, as well as the reasons 
why we think that our procedure is better than others, 
will be explained below.

In the even-parity case, the determination of the modulus $f_{\lm}$ is unique. 
In the odd-parity case, it depends on the choice of the source which, as
explained above, can be either connected to the effective energy  or 
to the angular momentum. We will consider both cases and 
distinguish them by adding either the label $H$ or $\J$ of 
the corresponding $f_{\lm}$.  Note, in passing, that, since 
in both cases the factorized effective source term ($H_{\rm eff}$ or $\J$) 
is a real quantity, the phases $\delta_{\lm}$'s are the same.

The above explained procedure defines the $f_{\lm}$'s as Taylor-expanded
PN series of the type
\begin{equation}
\label{scheme_flm}
f_{\lm}(x;\nu) = 1 + c_1^{f_{\lm}}(\nu) x + c_2^{f_\lm}(\nu) x^2 + c_3^{f_\lm}(\nu, \log(x))x^3 + \dots
\end{equation}
Note that one of the virtues of our factorization is to
have separated the half-integer powers of $x$ appearing in the usual 
PN-expansion of $h_{\lm}^{(\epsilon)}$ from the integer powers, the tail
factor, together with the complementary phase factor $e^{\ii\delta_{\lm}}$,
having absorbed all the half-integer powers.

We have computed all the $f_{\ell m}$'s (both for the $H$ and $\J$ choices) 
up to the highest available ($\nu$-dependent or not) PN accuracy.
In the formulas for the $f_{\lm}$'s given below we ``hybridize'' them by
adding to the known $\nu$-dependent coefficients $c_{n}^{f_{\lm}}(\nu)$
in Eq.~\eqref{scheme_flm} the $\nu=0$ value of the higher order coefficients:
$c_{n'}^{f_{\lm}}(\nu=0)$.
The 1PN-accurate $f_{\lm}$'s for  $\ell+m$ even 
and -- thanks to the new results for 1PN current multipoles obtained in Appendix A 
for arbitrary $\ell$ -- also for $\ell+m$ odd  can be written down for all $\ell$.
In Appendix~\ref{sec:app_B} we list the complete results for the $f_{\lm}$'s
that are known with an accuracy higher than 1PN.
Here, for illustrative purposes, we quote only the 
lowest  $f_\lm^{\text{even}}$ and $f_{\lm}^{\text{odd},J}$ 
up to $\ell=3$ included.
\begin{widetext}
\begin{align}
\label{eq:f22}
f_{22}(x;\nu) = 1& +\frac{1}{42} (55 \nu -86) x + 
\frac{\left(2047 \nu ^2-6745 \nu -4288\right)}{1512} x^2 \nonumber \\
  &+\left(\frac{114635 \nu ^3}{99792}-\frac{227875 \nu ^2}{33264}+
   \frac{41}{96}\pi^2\nu-\frac{34625 \nu }{3696}-\frac{856}{105}
   \text{eulerlog}_{2}(x)+\frac{21428357}{727650}\right) x^3\nonumber\\ 
  &+\left(\frac{36808}{2205}\text{eulerlog}_{2}(x)-\frac{5391582359}{198648450}\right) x^4
+\left(\frac{458816}{19845}\text{eulerlog}_{2}(x)-\frac{93684531406}{893918025}\right)x^5
+{\cal O}(x^6),
\end{align}
\begin{align}
f^J_{21}(x;\nu)=1&+\left(\frac{23 \nu}{42}-\frac{59}{28}\right) x +\left(\frac{85 \nu ^2}{252}-\frac{269\nu }{126}-\frac{5}{9}\right) x^2
+\left(\frac{88404893}{11642400}-\frac{214}{105}\text{eulerlog}_1(x)\right) x^3 \nonumber\\
&+\left(\frac{6313}{1470}\text{eulerlog}_1(x)-\frac{33998136553}{4237833600}\right)x^4+{\cal
  O}(x^5),
\end{align}
\begin{align}
f_{33}(x;\nu)= 1& +\left(2 \nu -\frac{7}{2}\right) x
+\left(\frac{887 \nu ^2}{330}-\frac{3401 \nu}{330}-\frac{443}{440}\right) x^2 
+\left(\frac{147471561}{2802800}-\frac{78}{7} \text{eulerlog}_{3}(x)\right)x^3\nonumber\\
&+\left(39\;\text{eulerlog}_{3}(x)-\frac{53641811}{457600}\right) x^4+{\cal O}(x^5), 
\end{align}
\begin{align}
f^J_{32}(x;\nu) = 1&+\frac{320 \nu ^2-1115 \nu +328}{90 (3 \nu -1)}x
+\frac{39544 \nu ^3-253768 \nu ^2+117215 \nu -20496}{11880(3\nu -1)}x^2\nonumber\\
&+\left(\frac{110842222}{4729725}-\frac{104}{21}\text{eulerlog}_{2}(x)\right)
x^3+{\cal O}(x^4),\\
\label{eq:f31}
f_{31}(x;\nu) = 1 &+ \left(-\frac{2 \nu }{3}-\frac{13}{6}\right)x+\left(-\frac{247 \nu ^2}{198}-\frac{371 \nu }{198}+\frac{1273}{792}\right) x^2\nonumber\\
              & +\left(\frac{400427563}{75675600}-\frac{26}{21}\text{eulerlog}_{1}(x)\right)x^3
                +\left(\frac{169}{63}
              \text{eulerlog}_{1}(x)-\frac{12064573043}{1816214400}\right)
              x^4+{\cal O}(x^5). 
\end{align}
\end{widetext}
For convenience and readability, we have introduced the 
following ``eulerlog'' functions $\text{eulerlog}_{m}(x)$
\begin{equation}
\label{eq:eulerlog}
\text{eulerlog}_m(x)  = \gamma_E + \log 2+\dfrac{1}{2}\log x + \log m,
\end{equation}
which explicitly reads, when $m=1,2,3$, 
\begin{align}
\label{eq:euler1}
\text{eulerlog}_1(x)  &= \gamma_E + \log 2+\dfrac{1}{2}\log x,\\
\text{eulerlog}_{2}(x)&=\gamma_E + 2\log 2+\dfrac{1}{2}\log x ,\\
\label{eq:euler3}
\text{eulerlog}_{3}(x)&=\gamma_E + \log 2+\log 3 +\dfrac{1}{2}\log x,
\end{align}
where $\gamma_{\rm E}$ is Euler's constant, $\gamma_{\rm E}=0.577215\dots$
and $\log(x)$ denotes, as everywhere else in this paper, 
the natural logarithm function.

\subsection{Resumming the modulus factor $f_{\lm}$}
\label{sec:resum_flm}

The decomposition of the total PN-correction factor
$\hat{h}_{\lm}^{(\epsilon)}$
into several factors is in itself a resummation procedure which has already
improved the convergence of the PN series one has to deal with:
indeed, one can see that the coefficients entering increasing powers of $x$ in the
$f_{\lm}$'s tend to be systematically smaller than the coefficients appearing
in the usual PN expansion of $\hat{h}_{\lm}^{(\epsilon)}$. The reason for this
is essentially twofold: (i) the factorization of $T_{\lm}$ has absorbed powers 
of $m\pi$ which contributed to making large coefficients in
$\hat{h}_{\lm}^{(\epsilon)}$,
and (ii) the factorization of either $\hat{H}_{\rm eff}$ or $\hat{j}$ has
(in the $\nu=0$ case)  removed the presence of an inverse square-root singularity
located at $x=1/3$ 
which caused the coefficient of $x^n$ in any PN-expanded quantity to grow
as $3^{n}$ as $n\to\infty$.
To prevent some potential misunderstandings, let us emphasize that we are
talking here about a singularity entering the analytic continuation (to
larger values of $x$) of a mathematical function $h(x)$ defined (for small
values of $x$) by considering the formal adiabatic circular limit. The point
is that, in the $\nu\to 0$ limit, the radius of convergence and therefore the
growth with $n$ of the PN coefficients of $h(x)$ (Taylor-expanded at $x=0$),
are linked to the singularity of the analytically continued $h(x)$ which is
nearest to $x=0$ in the complex $x$-plane.
In the $\nu\to 0$ case, the nearest singularity in the complex $x$-plane comes
from the source factor $\hat{H}_{\rm eff}(x)$ or $\hat{j}(x)$ in the waveform
and is located at the light-ring $x_{\rm LR}(\nu=0)=1/3$. In the $\nu\neq 0$
case, the EOB formalism transforms the latter (inverse square-root)
singularity into a more complicated (``branching'') singularity where
$d\hat{H}_{\rm eff}/dx$ and $d\hat{j}/dx$ have inverse square-root
singularities located at what 
is called~\cite{Buonanno:2000ef,Buonanno:2006ui,Buonanno:2007pf,Damour:2007vq,Damour:2007yf} 
the (Effective)\footnote{Beware that this ``Effective EOB-light-ring'' occurs
for a circular-orbit radius slightly larger than the purely dynamical
(circular) EOB-light-ring (where $H_{\rm eff}$ and ${\cal J}$ 
would formally become infinite). } 
``EOB-light-ring'', i.e., the (adiabatic) maximum of $\Omega$,
$x_{\rm ELR}^{\rm adiab}(\nu)\equiv 
\left(M\Omega^{\text{adiab}}_{\rm max}\right)^{2/3}\gtrsim 1/3$.

Despite this improvement, the resulting ``convergence'' of the usual
Taylor-expanded $f_{\lm}(x)$'s quoted above does not seem to be good 
enough, especially near or below the LSO, in view of the 
high-accuracy needed to define gravitational wave templates.
For this reason, Refs.~\cite{Damour:2007xr,Damour:2007yf} proposed
to further resum the $f_{22}(x)$ function via a Pad\'e (3,2) approximant,
$P^3_{2}\{f_{22}(x;\nu)\}$, so as to improve its behavior in the
strong-field-fast-motion regime. Such a resummation gave an excellent
agreement with numerically computed waveforms, near the end of the inspiral 
and during the beginning of the plunge, 
for different mass ratios~\cite{Damour:2007xr,Damour:2007vq,Damour:2008te}.
Here, however, we wish to explore a new route for resumming $f_{\lm}$,
based on replacing $f_{\lm}$ by its $\ell$-th root, say
\begin{equation}
\label{eq:lth_root}
\rho_{\lm}(x;\nu) = [f_{\lm}(x;\nu)]^{1/\ell}.
\end{equation}
Our basic motivation for replacing $f_{\ell m}$ by $\rho_{\ell m}$ 
is the following: the leading ``Newtonian-level'' contribution 
to the waveform $h^{(\epsilon)}_{\ell m}$ contains a  factor 
$\omega^\ell r_{\rm harm}^\ell v^\epsilon$ where $r_{\rm harm}$  is the
harmonic radial coordinate used in the MPM 
formalism~\cite{Blanchet:1989ki,Damour:1990ji} .
When computing the PN expansion of this factor one has to insert 
the PN expansion of the (dimensionless) harmonic radial 
coordinate $r_{\rm harm}$, $ r_{\rm harm} = x^{-1}(1+c_1 x+{\cal O }(x^2))$,
as a function of the gauge-independent
frequency parameter $x$. 
The PN re-expansion of $[r_{\rm harm}(x)]^\ell$ then generates terms of the 
type $x^{-\ell}(1 +\ell c_1 x+....)$. 
This is one (though not the only one) of the origins of 
1PN corrections in $h_{\ell m}$ and $f_{\ell m}$ 
whose coefficients grow linearly with $\ell$.
As we shall see in detail below, 
these $\ell$-growing terms are problematic for
the accuracy of the PN-expansions. 
Our replacement of $f_{\ell m}$ by $\rho_{\ell m}$ 
is a cure for this problem.

More explicitly, the investigation of $1$PN corrections to 
GW amplitudes~\cite{Blanchet:1989ki,Damour:1990ji,Kidder:2007rt} 
has shown that, in the even-parity case 
(but see also Appendix~\ref{sec:app_A} for the odd-parity case), 
\begin{align}
\label{c1lm_nu}
 c_1^{f_{\lm}^{\rm even}}(\nu)&= 
-\ell\left(1-\dfrac{\nu}{3}\right)+\dfrac{1}{2}
+\dfrac{3}{2}\dfrac{c_{\ell+2}(\nu)}{c_\ell(\nu)}
-\dfrac{b_\ell(\nu)}{c_\ell(\nu)}\nonumber\\
&-\dfrac{c_{\ell+2}(\nu)}{c_{\ell}(\nu)}\dfrac{m^2(\ell+9)}{2(\ell+1)(2\ell+3)},
\end{align}
where $c_{\ell}(\nu)$ is defined in Eq.~\eqref{eq:cl} and, 
consistently with the notation of Appendix A,
\be
b_{\ell}(\nu)\equiv X^{\ell}_2+(-)^{\ell}X_1^{\ell}.
\label{eq:bl}
\ee
As we shall see below, the $\nu$ dependence of  $c_1^{f_{\lm}}(\nu)$ is quite
mild. For simplicity, let us focus on the $\nu=0$ case, where the above result
shows that the PN expansion of $f_{\lm}$ starts as
\begin{equation}
\label{flm_1PN_even}
f_{\lm}^{\rm even}(x;0)  = 1-\ell x \left(1-\dfrac{1}{\ell}
+\dfrac{m^2(\ell+9)}{2\ell(\ell+1)(2\ell+3)}\right) + {\cal O}(x^2) .
\end{equation}
The crucial thing to note in this result is that as $\ell$ gets large (keeping
in mind that $|m|\leq \ell$), the coefficient of $x$ will be negative and will
approximately range between $-5\ell/4$ and $-\ell$. This means that when 
$\ell \geq 6$ the 1PN correction in $f_{\lm}$ would by itself make
$f_{\lm}(x)$ vanish before the ($\nu=0$) LSO $x=1/6$.
For example, for the 
$\ell=m=6$ mode, one has $f^{\rm 1PN}_{66}(x;0)=1-6x(1+11/42)\approx 1 - 6x(1+ 0.26)$
which means a correction equal to $-100\%$ at $x=1/7.57$ and
larger than $-100\%$ at the LSO, namely
$f^{\rm 1PN}_{66}(1/6;0)\approx 1 - 1.26=-0.26$.
This is definitely incompatible
with the numerical data we shall quote below. Similar results hold also for 
the odd-parity $f_{\lm}$'s,  especially in the case where we factorize the
$\J$ source (which happens to have close similarities to the $H$-factored
even-parity $f_{\lm}$). Indeed, we have extended the result of
Eq.~\eqref{flm_1PN_even} to the odd-parity case, i.e. we have computed, 
(using the comparable mass  1PN results of Ref.~\cite{Damour:1990ji}) the
1PN correction in $f_{\lm}$ and  $\rho_{\lm}$. In the $\nu\to0$ limit, 
we found that (see Appendix~\ref{sec:app_A} for more details and for 
a discussion of the comparable-mass case)
\begin{align}
f_{\lm}^J(x;0)&=1- \ell x\nonumber\\
\label{flm_1PN_odd}
&\times\left(1 + \dfrac{1}{\ell} - \dfrac{2}{\ell^2}
+ \dfrac{m^2(\ell+4)}{2\ell(\ell+2)(2\ell +3)}\right) + {\cal O}(x^2),
\end{align}
which is structurally similar to the even-parity expression quoted above. 

Let us now see how the replacement of $f_{\lm}$ by the newly defined
 $\rho_{\lm}$,
Eq.~\eqref{eq:lth_root}, cures this problem of abnormally large 1PN
corrections to the waveforms for large values of $\ell$. Indeed, 
the Taylor expansion of $\rho_{\lm}$ now starts as (say for simplicity in the
$\nu=0$, even-parity case)
\begin{equation}
\label{eq:rho1PN_even}
\rho_{\lm}^{\rm even}(x;0) = 1- x \left(1-\dfrac{1}{\ell} 
+\dfrac{m^2(\ell+9)}{2\ell(\ell+1)(2\ell+3)}\right) + {\cal O}(x^2). 
\end{equation}
Note that for large $\ell$ and arbitrary $m$ the coefficient of $x$ now
approximately ranges between $-5/4$ and $-1$.
We shall see below that the nice behavior of $\rho_{\lm}$ expected from this
1PN estimate indeed holds for the exact $\rho_{\lm}$, at least in the $\nu=0$ 
case. In addition, the same structure is found in the odd-parity $\rho^J_{\lm}$'s.
In particular, from Eq.~\eqref{flm_1PN_odd} above one finds 
\begin{align}
\rho_{\lm}^{J}(x;0) &= 1- x \nonumber\\
\label{eq:rho1PN_odd}
&\times \left(1 + \dfrac{1}{\ell}-\dfrac{2}{\ell^2}  
+\dfrac{m^2(\ell+4)}{2\ell(\ell+2)(2\ell +3)}\right)+{\cal O}(x^2),
\end{align}
where, for $\ell\gg 1$, the coefficient of $x$ again approximately ranges 
between $-5/4$ and $-1$.

We have computed all the $\rho_{\lm}$'s (both for the $H$ and $\J$ choices) 
up to the highest available ($\nu$-dependent or not) PN accuracy.
In the formulas for the $\rho_{\lm}$'s given below we ``hybridize'' them by
adding to the known $\nu$-dependent coefficients $c_{n}^{\rho_{\lm}}(\nu)$
in the Taylor expansion of $\rho_{\lm}$'s,
\begin{equation}
\rho_{\lm}(x;\nu) = 1 + c_1^{\rho_{\lm}}(\nu) x + c_2^{\rho_{\lm}}(\nu) x + c_3^{\rho_{\lm}}(\log(x);\nu) x^3+\dots ,
\end{equation}
the $\nu=0$ value of the higher order coefficients \hbox{$c_{n'}^{\rho_{\lm}}(\nu=0)$}.
Beware that this definition of an hybrid $\rho_{\lm}$ is {\it not} equivalent 
to that displayed in Eqs~\eqref{eq:f22}-\eqref{eq:f31} above of an 
analogous hybrid $f_{\lm}$ (nor is it equivalent to a straightforward
hybridization of $h_{\lm}$).
The primary hybridization procedure that we advocate (and use) in this paper
is the one based on $\rho_{\lm}$ (i.e., replacing $c_{n'}^{\rho_{\lm}}(\nu)$
by $c_{n'}^{\rho_{\lm}}(0)$ when $n'$ is beyond the maximal $\nu$-dependent
PN knowledge).
The 1PN-accurate $\rho_{\lm}$'s for $\ell+m$ even and -- thanks to the
new results for $h_{\ell m}$ for $\ell +m$ odd in Appendix A 
-- also for  $\ell+m$ odd  are explicitly known for all $\ell$. For the 1PN
coefficient of the $\rho_{\lm}$'s we explicitly have
\begin{align}
c_1^{\rho_{\lm}^{\rm even}}(\nu)&=-\left(1-\dfrac{\nu}{3}\right)+\dfrac{1}{2\ell}
+\dfrac{3}{2\ell}\dfrac{c_{\ell+2}(\nu)}{c_\ell(\nu)}
-\dfrac{1}{\ell}\dfrac{b_\ell(\nu)}{c_\ell(\nu)}\nonumber\\
&-\dfrac{m^2(\ell+9)}{2\ell
  (\ell+1)(2\ell+3)}\dfrac{c_{\ell+2}(\nu)}{c_{\ell}(\nu)},\\
\nonumber\\
\label{eq:c1Jnu}
c_1^{\rho^J_{\lm}}(\nu)&=
-\left(1-\dfrac{\nu}{3}\right)
-\frac{1}{2\ell}\left(5-\dfrac{\nu}{3}\right) - \dfrac{\nu}{2\ell^2}
 \nonumber\\
&+\frac{2\ell+3}{2\ell^2}
\frac{b_{\ell+1}(\nu)}{c_{\ell+1}(\nu)} + 2\nu\frac{\ell+1}{\ell^2}\frac{b_{\ell-1}(\nu)}{c_{\ell+1}(\nu)}\nonumber\\ 
&+\frac{1}{2}\frac{\ell+1}{\ell^2}\frac{c_{\ell+3}(\nu)}{c_{\ell+1}(\nu)}
- \frac{m^2(\ell+4)}{2\ell(\ell+2)(2\ell+3)}\frac{c_{\ell+3}(\nu)}{c_{\ell+1}(\nu)}.
\end{align}

For definiteness, we give in Appendix~\ref{sec:app_B} the complete 
results, for $\rho_{\lm}$ (even-parity) and $\rho_{\lm}^{J}$ (odd-parity), 
up to $\ell=8$ included.
Here, for illustrative purposes, we quote only some of the lowest
multipole results up to $\ell=3$ included.
\begin{widetext}
\begin{align}
\label{eq:rho22}
\rho_{22}(x;\nu)= 1 &+\left(\frac{55 \nu }{84}-\frac{43}{42}\right) x 
+\left(\frac{19583 \nu^2}{42336}-\frac{33025 \nu
}{21168}-\frac{20555}{10584}\right) x^2 \nonumber\\
&+\left(\frac{10620745 \nu ^3}{39118464}-\frac{6292061 \nu ^2}{3259872}+\frac{41 \pi
   ^2 \nu }{192}-\frac{48993925 \nu }{9779616}-\frac{428}{105}
  \text{eulerlog}_{2}(x)+\frac{1556919113}{122245200}\right) x^3 \nonumber\\
&+\left(\frac{9202}{2205}\text{eulerlog}_2(x)-\frac{387216563023}{160190110080}\right) x^4
+\left(\frac{439877}{55566}\text{eulerlog}_{2}(x)-\frac{16094530514677}{533967033600}\right)x^5+{\cal O}(x^6), 
\end{align}
\begin{align}
\rho_{21}^J(x;\nu)&=1+\left(\frac{23 \nu }{84}-\frac{59}{56}\right) x +\left(\frac{617 \nu ^2}{4704}-\frac{10993\nu }{14112}-\frac{47009}{56448}\right) x^2\nonumber\\
                    &+\left(\frac{7613184941}{2607897600}-\frac{107}{105}\text{eulerlog}_1(x)\right)x^3
                     +\left(\frac{6313}{5880}\text{eulerlog}_1(x)-\frac{1168617463883}{911303737344}\right)x^4+{\cal O}(x^5),
\end{align}
\begin{align}
\rho_{33}(x;\nu) = 1&+\left(\frac{2 \nu }{3}-\frac{7}{6}\right) x+\left(\frac{149 \nu ^2}{330}-\frac{1861 \nu }{990}-\frac{6719}{3960}\right) x^2
+\left(\frac{3203101567}{227026800}-\frac{26}{7} \text{eulerlog}_{3}(x)\right)x^3\nonumber\\
 &+\left(\frac{13}{3}\text{eulerlog}_{3}(x)-\frac{57566572157}{8562153600}\right) x^4+{\cal O}(x^5), 
\end{align}
\begin{align}
\rho_{32}^J(x;\nu)&=1 +\frac{320 \nu ^2-1115\nu +328}{270 (3 \nu -1)}x 
                      +\frac{3085640 \nu ^4-20338960 \nu ^3-4725605 \nu ^2+8050045\nu -1444528}{1603800 (1-3 \nu )^2}x^2\nonumber\\
                  &+\left(\frac{5849948554}{940355325}-\frac{104}{63}\text{eulerlog}_2(x)\right)x^3+{\cal O}(x^4),
\end{align}
\begin{align}
\label{eq:rho31}
\rho_{31}(x;\nu) &= 1+\left(-\frac{2 \nu}{9}-\frac{13}{18}\right) x 
+\left(-\frac{829 \nu ^2}{1782}-\frac{1685\nu }{1782}+\frac{101}{7128}\right) x^2
+\left(\frac{11706720301}{6129723600}-\frac{26}{63}\text{eulerlog}_1(x)\right) x^3\nonumber\\
&+\left(\frac{169}{567}
\text{eulerlog}_1(x)+\frac{2606097992581}{4854741091200}\right) x^4 + {\cal
  O}(x^5).\\
\nonumber\\
\nonumber
\end{align}
\end{widetext}
\begin{figure*}[t]
    \begin{center}
      \includegraphics[width=78 mm]{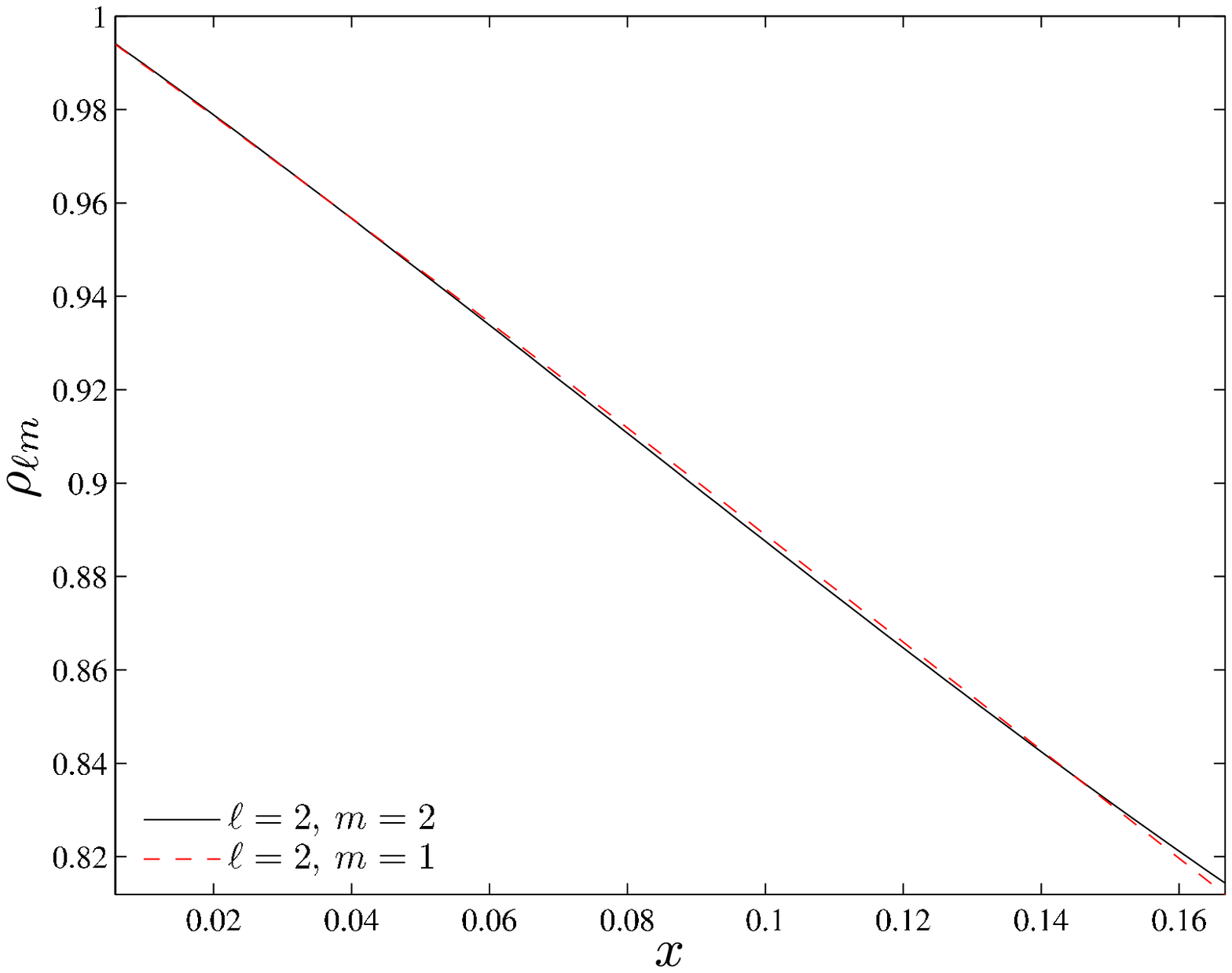}
      \hspace{3 mm}
      \includegraphics[width=78 mm]{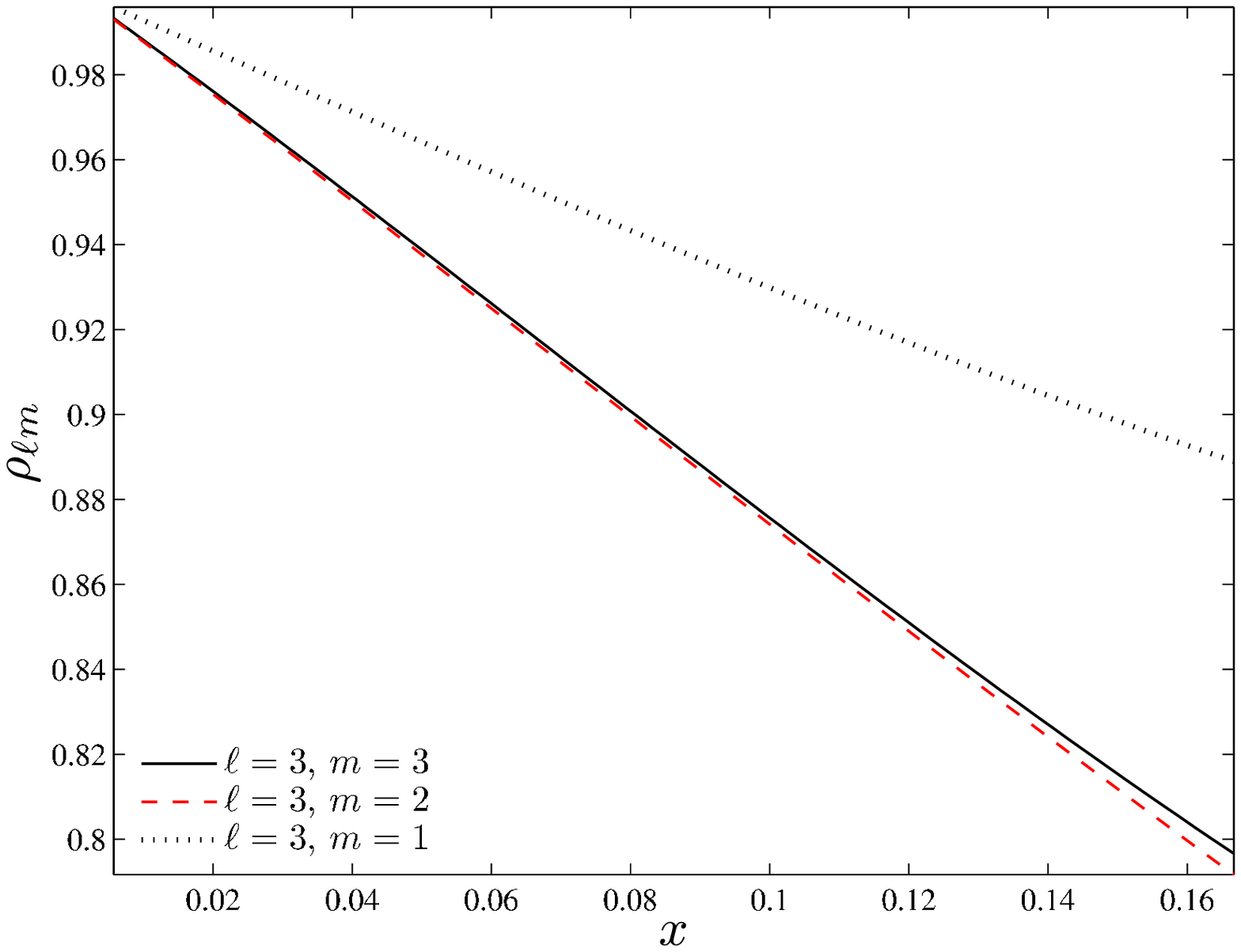}\\
      \vspace{4 mm}
      \includegraphics[width=78 mm]{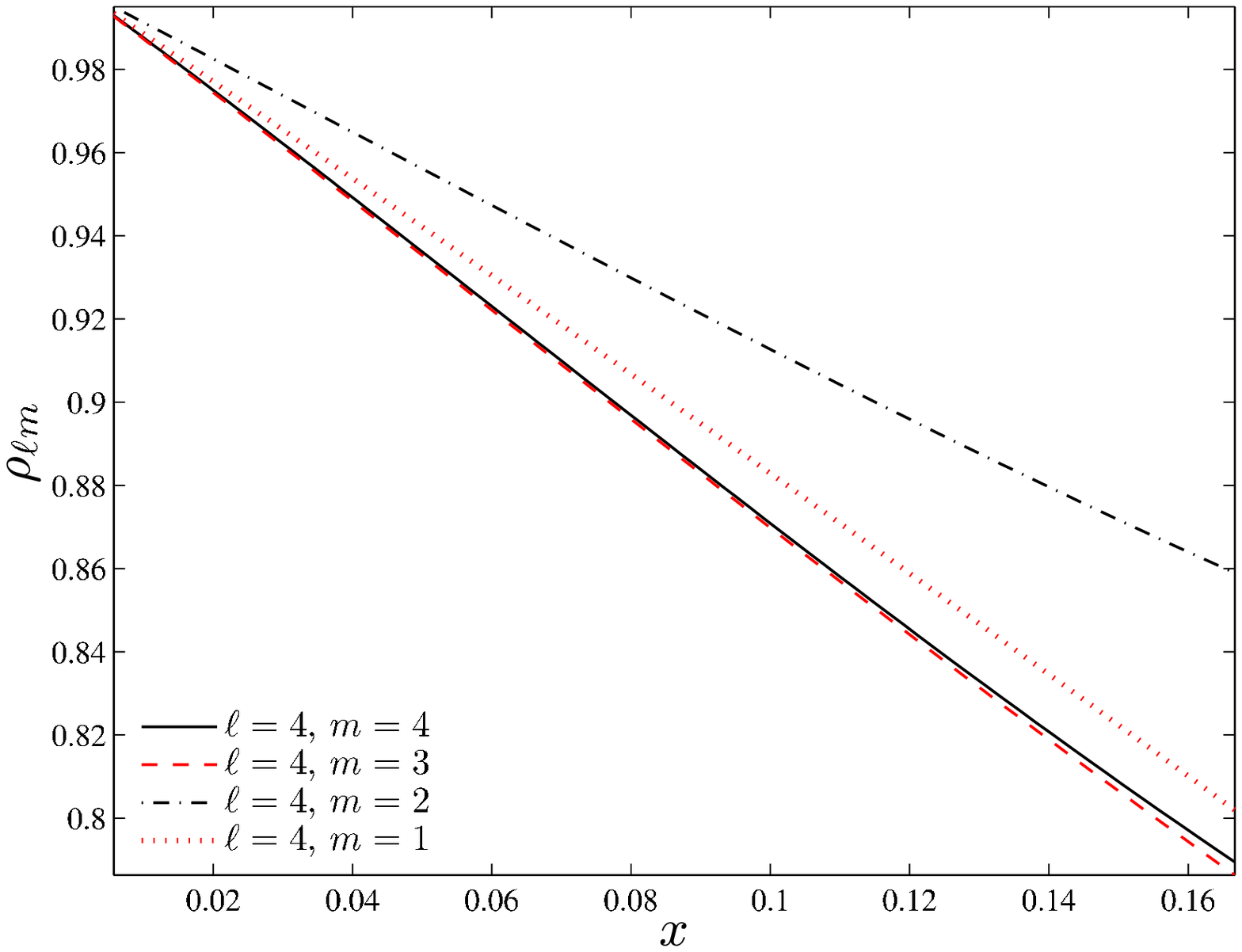}
      \hspace{3 mm}
      \includegraphics[width=78 mm]{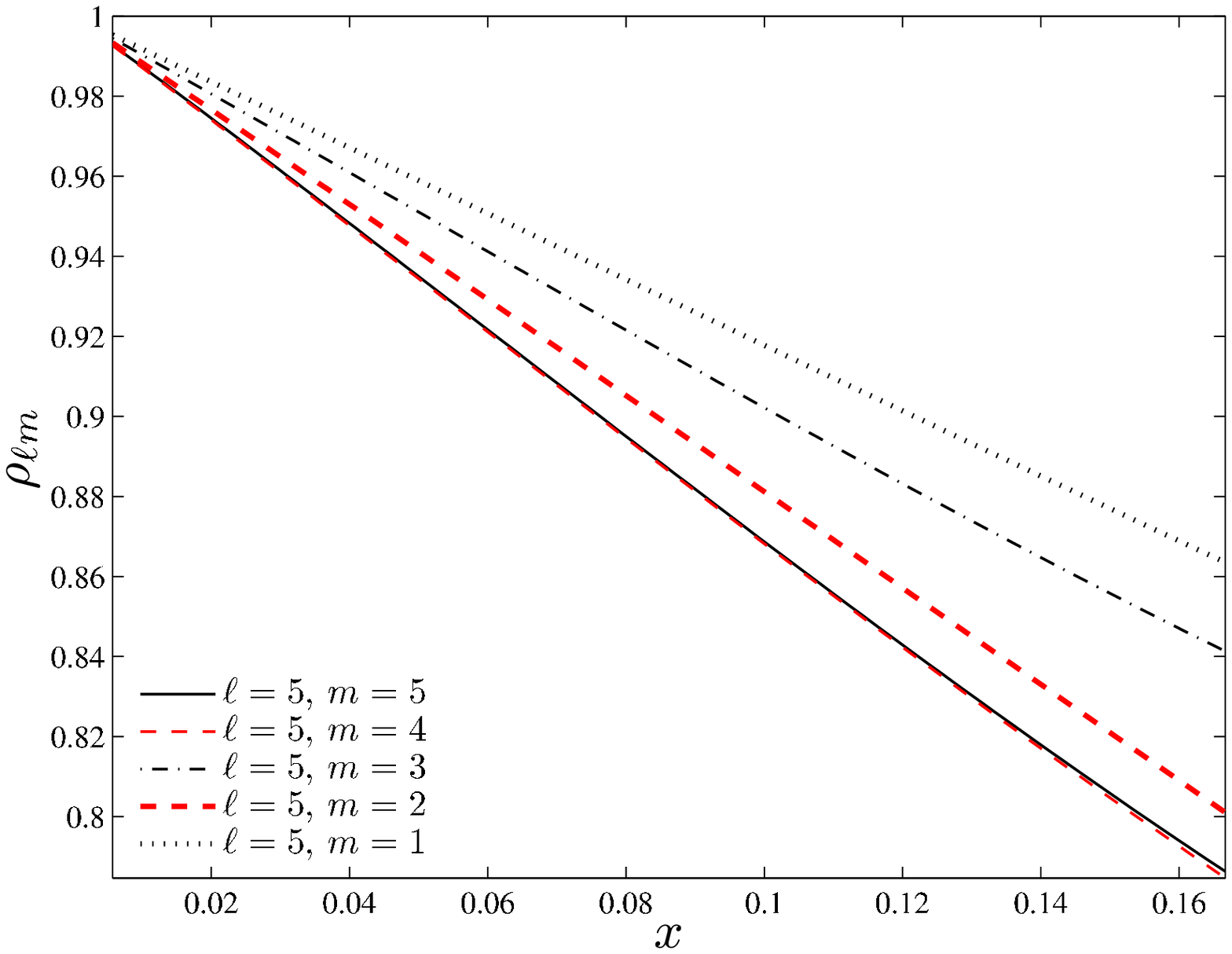}\\
      \vspace{4 mm}
      \includegraphics[width=78 mm]{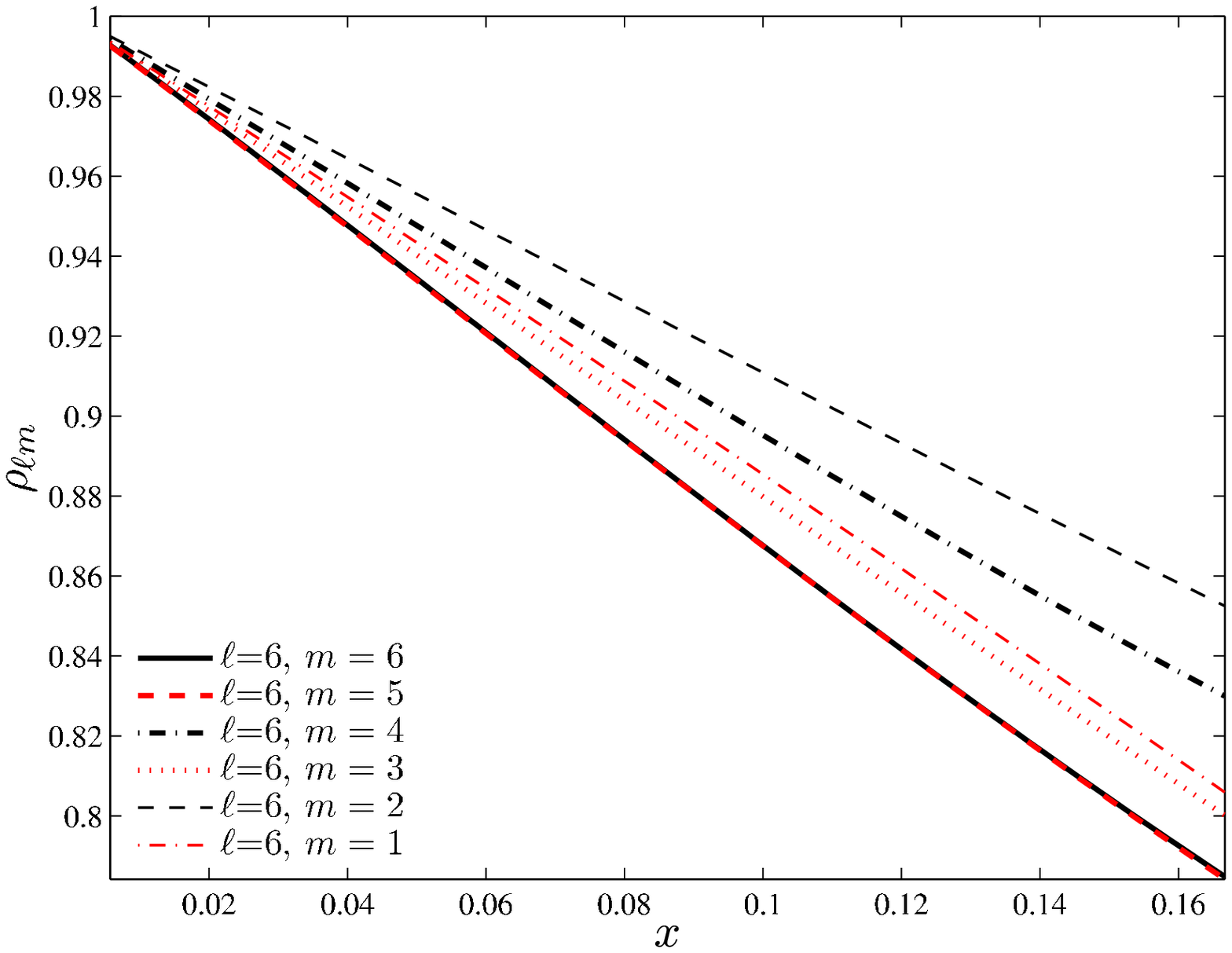}
      \vspace{2 mm}
      \caption{ \label{fig:rholm} Extreme-mass-ratio limit ($\nu=0$). 
      The ``exact'' functions $\rho_{\lm}(x)$ for $0<x<1/6$ extracted 
      from E.~Berti's numerical fluxes. Multipoles up to $\ell=6$ are
      considered. Each panel corresponds to one value of $\ell$ and
      shows the even-parity partial 
      amplitudes (black online) together with the {\it {\cal J}-normalized} 
      odd-parity ones (red online).}
  \end{center}
\end{figure*}
\section{Results for the extreme-mass-ratio case ($\nu=0$)}
\label{results_nu0}

\subsection{Extracting the $\rho_{\lm}^{\text{Exact}}$ multipoles from
      black-hole perturbation numerical data}
\label{sec:extract}

To test our new resummation procedure 
based on the $\rho_{\lm}$'s we shall compare the analytical results defined
by our multiplicative decomposition Eq.~\eqref{eq:hhat} to the
``exact'' results obtained by numerical analysis of black hole perturbation
theory. For most of the comparisons discussed below we will rely on
data kindly provided by Emanuele Berti, who
computed the multipolar decomposition of the GW flux from stable 
circular orbits above the LSO with a frequency-domain code which
solves numerically the Teukolsky equation with a point-particle source. 
(see for example Ref.~\cite{Yunes:2008tw} and references therein). 
In addition, we have complemented Fig.~\ref{fig:rho22_resum} by computing the
quadrupolar GW energy flux $F_{22}$ from a sample of unstable circular
orbits with radius between $6M$ and $3.1M$.

The result of the numerical computation is expressed in terms of 
the multipolar pieces $F_{\lm}$  of the total
``exact'' flux, Eq.~\eqref{eq:flux}. We shall only consider multipoles
up to $\ell=6$ included.
The ``exact'' $\nu\to0$ version of our new quantities $\rho_{\lm}(x;\nu)$'s
are then obtained from the ratio of the ``exact'' partial fluxes 
$F_{\lm}^{\rm Exact}$ to their Newtonian counterparts 
$F_{\lm}^{\rm Newton}$   as
\begin{align}
\rho_{\lm}^{{\rm Exact},(\epsilon)}(x;0)=
\left\{ \dfrac{\sqrt{ F_{\ell m}^{\rm Exact}/F_{\lm}^{\rm Newton}}}{|T_{\lm}|\hat{S}^{(\epsilon)}_{\rm eff}}\right\}^{1/\ell},
\end{align}
where, for $\nu=0$, we explicitly have 
\begin{align}
\label{Hnu0}
\hat{S}^{(0)}_{\rm eff}(x)&= \dfrac{1-2 x}{\sqrt{1-3x}} \ , 
\end{align}
and either
\begin{equation}
\hat{S}^{(1,H)}_{\rm eff}(x)= \dfrac{1-2 x}{\sqrt{1-3x}}, 
\end{equation}
or
\begin{equation}
\label{jnu0}
\hat{S}^{(1,J)}_{\rm eff}(x)= \dfrac{1}{\sqrt{1-3x}} \ . 
\end{equation}
Still in the $\nu\to 0$ limit, we have $H_{\rm real}\to M$, 
and, from well-known properties of the $\Gamma$ function, 
the square modulus of the tail factor $T_{\lm}$ reads
\begin{align}
|T_{\lm}|^2 = \dfrac{1}{(\ell !)^2} \dfrac{4\pi \k}{1-e^{-4\pi \k}}\prod_{s=1}^{\ell}\left[s^2 + (2 \k)^2\right] \ ,
\end{align}
where now $\k=mM\Omega=m x^{3/2}= mv^3$.

\subsection{Finding structure in the $\rho_{\lm}^{\text{Exact}}$ multipoles 
            extracted from numerical data}
\label{sec:exact}
Let us first consider the properties of the ``exact'' $\rho_{\lm}$'s.
In the odd-parity case we shall focus here on the $\J$-normalized
quantity $\rho_{\lm}^J$. We shall see that the $\rho_{\lm}$'s 
convey interesting information about the $x$ dependence of 
the multipolar GW amplitudes.
Fig.~\ref{fig:rholm} exhibits the numerical $\rho_{\lm}$ functions
for $\ell\leq 6$, versus $x=(M\Omega)^{2/3}$ (where we recall that $\Omega$ denotes 
the orbital frequency) up to the $\nu=0$ LSO, $x=1/6$.
Each panel of the figure displays, for each given $2\leq\ell\leq 6$,
the partial $\rho_{\lm}$'s for the various possible $m$'s, $1\leq m\leq\ell$
(we do not plot the negative $m$'s since they correspond to the same value
of $\rho_{\lm}$).
The even-parity ($\ell+m$ even,  black online) and odd-parity 
($\ell+m$ odd, red online) modes are shown together for comparison.

Fig.~\ref{fig:rholm} displays the following noticeable facts:
(i) to a good approximation, all the $\rho_{\lm}(x)$ are 
{\it straight  lines}\footnote{In the odd-parity case, $\ell+m$ odd, 
this quasi-linear behavior up to the LSO, is particularly clear 
for the functions $\rho_{\lm}^J(x)$'s, making us consider $\rho_{\lm}^J$
as our best default choice.  By contrast, the $H$-normalized functions 
$\rho_{\lm}^H(x)$ have a more $\ell$-dependent shape that 
the reader can figure out by noting the 
link between them: $\rho_{\lm}^H=\rho_{\lm}^J/(1-2 x)^{1/\ell}$.} 
(see below); 
(ii) for each value of $\ell$, the (negative) slopes of the dominant $m=\ell$
(even-parity), and subdominant $m=\ell-1$ (odd-parity) multipole modes 
are  very close to each other; 
and these slopes  become closer and closer as the value of $\ell$ increases
(note in particular that for $\ell=6$ $\rho_{66}(x)$ and 
$\rho_{65}(x)$ are practically coincident); 
(iii) for a given value of $\ell$, {\it and a given parity} (even or odd),  
the absolute value of the (negative) slope decreases monotonically 
as $|m|$ decreases.
This ``order'' in the ``exact'' data can be analytically understood.

The property (i) means that the {\it 1PN correction is already capturing 
most of the physical information}, which might turn out to be a useful
fact to know (see below).
We illustrate this result in Fig~\ref{fig:rho2m_1PN} which 
focuses on the 
quadrupolar ($\ell=2$) partial waves, and  exhibits the exact
$\rho_{22}$ and $\rho_{21}^J$ (solid and dashed lines) together 
with their 1PN approximations (dotted lines). 
Note, for instance, that the difference between the 1PN accurate,
$\rho^{\rm 1PN}_{22}$, that we shall denote\footnote{Here and in the
  following we shall denote the truncated $n$-PN-accurate expansion of 
any function $f(x)$ as $T_n[f(x)]\equiv f_0 + f_1 x+ \dots + f_n x^n$.}       
$T_1[\rho_{22}]\equiv\rho_{22}^{\rm 1PN}=1+c_{1}^{\rho_{\lm}}x$,
and the ``exact'' one $\rho_{22}^{\rm Exact}$ is equal, at the LSO,
to $\rho_{22}^{\rm 1PN}-\rho_{22}^{\rm Exact}=0.8294-0.8143=0.0151$,
which is only $1.9\%$ of the exact result 0.8143.
The other multipoles
exhibit a similar agreement between the exact $\rho_{\lm}$ and their
analytical 1PN representations.
To understand analytically
what underlies this agreement, let us consider the $\ell=m=2$ case. 
Numerically, from Eq.~\eqref{eq:rho22} we have, near the LSO
(for simplicity, we replace the  $\text{log}(x)$ terms present in the 
coefficients by their numerical values  at $x=1/6$)
\begin{align}
\label{rho22_exp}
\rho_{22}(x;0) &\approx 1 - 1.024 x -1.942 x^2 \nonumber\\
              & + 8.384 x^3 + 2.038 x^4 -21.690 x^5\nonumber\\
              & \approx 1 - 0.171(6x) -0.054 (6x)^2 \nonumber\\
              & + 0.039 (6x)^3 + 0.0016(6x)^4  - 0.0028(6x)^5.
\end{align}
We see that the successive coefficients of the PN-expansion of
$\rho_{22}$ are  such that, even at the LSO, the magnitudes of the
PN-corrections beyond the 1PN one are rather small. 
They are significantly smaller than the corresponding terms in the
usual PN-expansion of the total flux.
For instance, by contrast to the 
coefficient $-21.69$ which enters the 5PN-correction in $\rho_{22}$,
let us recall that the coefficient (including the $\text{log}(x)$ estimated at
the LSO) of the 5PN correction in the usual PN-expanded flux 
is $\approx -1321.402$ (see e.g.~\cite{Damour:1997ub}).
Note that the latter 5PN~contribution to the PN-expanded flux considered 
at the LSO is $-1321.40/6^5\approx -0.17$ which is as large as the 1PN
contribution to $\rho_{22}$ and about sixty times larger than the corresponding 5PN
correction to $\rho_{22}$. In addition, as the signs in Eq.~\eqref{rho22_exp}
fluctuate, there are compensations between the higher PN contributions,
as it will be clear from further results presented below.

Property (ii) can be analytically understood 
by means of the 1PN-accurate closed
formulas, Eqs.~\eqref{eq:rho1PN_even}-\eqref{eq:rho1PN_odd}. 
Indeed, it is easily checked that the difference between the coefficients
of $x$ Eq.~\eqref{eq:rho1PN_even} for $m=\ell$ and Eq.~\eqref{eq:rho1PN_odd}
for $m=\ell-1$ is of order ${\cal O}(1/\ell^2)$ when $\ell$ gets large.

Finally, property (iii) is understood by noting that the coefficients of $x$
in Eqs.~\eqref{eq:rho1PN_even}-\eqref{eq:rho1PN_odd} have the structure
$-(a(\ell)+m^2 b(\ell))$ where $a(\ell)$ and $b(\ell)$ are positive.
 
\begin{figure}[t]
   \begin{center}
     \includegraphics[width=85 mm]{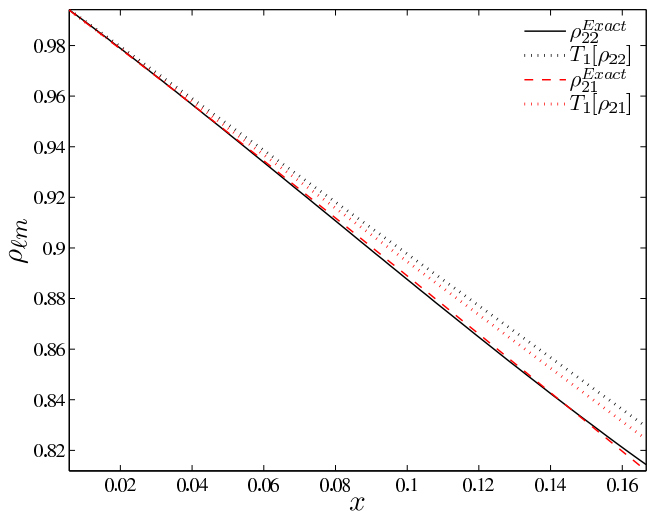}
     \caption{ \label{fig:rho2m_1PN} Extreme-mass-ratio limit ($\nu=0$). 
     Comparison between the ``exact'' leading and subleading quadrupolar 
     amplitudes $\rho_{22}$ and $\rho^J_{21}$ and the corresponding 
     1PN-accurate analytical ones.}
   \end{center}
\end{figure}
\subsection{Comparing Taylor and Pad\'e approximants of $\rho_{22}$}
\label{sec:resum_rho22}

\begin{table}[t]
\caption{\label{tab:table2} Closeness of various resummed approximants to
  $\rho_{22}$ at the LSO, $x_{\text{LSO}}=1/6$, or actually 
  $x_{\rm last}=1/6.00001$. The rightmost column lists the 
  difference $\Delta\rho_{22}$ between the resummed approximant 
  and the exact value at $x=x_{\rm last}$}
\begin{center}
  \begin{ruledtabular}
  \begin{tabular}{ccc}
    Approximant   & $\rho_{22}(x_{\rm last})$ & $\Delta\rho_{22}(x_{\text{last}})$ \\
    \hline \hline
      $\rho_{22}^{\rm Exact}$   & 0.8143372247   &  0\\
      $T_1[\rho_{22}]$          & 0.8293653638   &  0.0150281391 \\
      $T_2[\rho_{22}]$          & 0.7754188106   & -0.0389184141 \\  
      $T_3[\rho_{22}]$          & 0.8142342355   & -0.0001029892 \\ 
      $T_4[\rho_{22}]$          & 0.8158069452   &  0.0014697205 \\ 
      $T_5[\rho_{22}]$          & 0.8130176477   & -0.0013195770 \\ 
      $P^4_1\{T_5[\rho_{22}]\}$ & 0.8148012716   &  0.0004640469 \\ 
      $P^1_4\{T_5[\rho_{22}]\}$ & 0.8146954164   &  0.0003581917 \\ 
      $P^3_2\{T_5[\rho_{22}]\}$ & 0.8132320684   & -0.0011051563 \\ 
      $P^2_3\{T_5[\rho_{22}]\}$ & 0.8146954104   &  0.0003581857 \\    
  \end{tabular}
\end{ruledtabular}
\end{center}
\end{table}%

Let us now compare and contrast the ``convergence'' of various PN-approximants 
towards the ``exact'' (numerical) $\rho_{22}$. We first focus on the  values
of various approximants to $\rho_{22}(x)$ at the LSO, i.e. at $x_{\rm LSO}=1/6$
or actually the last point in the numerical data 
computed by E.~Berti, $x_{\rm last}=1/6.00001$. At the point $x=x_{\rm last}$ 
the numerical value of the Newton-normalized $\ell=m=2$ partial flux is 
$\hat{F}_{22}\equiv F_{22}/F_{22}^N=0.8927266028$. This corresponds to
$\rho_{22}^{\rm Exact}(x_{\rm last})=0.8143372247$.
In Table~\ref{tab:table2} we compare this  value to several PN-based
approximants: both Taylor approximants, 
from 1PN to 5PN ($T_1[\rho_{22}]$ to $T_5[\rho_{22}]$) and several
of the ``around the diagonal'' 5PN-accurate Pad\'e approximants, namely,
$P^4_1\{T_5[\rho_{22}]\}$, $P^1_4\{T_5[\rho_{22}]\}$, 
$P^3_2\{T_5[\rho_{22}]\}$ and $P^2_3\{T_5[\rho_{22}]\}$.
Note how the sequence of Taylor approximants to $\rho_{22}$ 
nicely approaches the exact value, especially starting with the 3PN
approximation. Probably by accident, the Taylor 3PN approximant,
$T_3[\rho_{22}]$, happens to be closer to the exact value than  the higher
order approximants $T_4$ and $T_5$. 
Besides this accidental closeness of $T_3$, the important thing to note
is the very small dispersion (within $\pm 1.8\times 10^{-3}$) of 
$T_3$, $T_4$ and $T_5$ around the correct value.
This excellent behavior of the Taylor approximants of $\rho_{22}$ 
should be contrasted with the much worse behavior of the standard 
Taylor approximants either of the 
flux or of the waveforms 
(see for example Fig.~1 in Ref.~\cite{Poisson:1995vs}, 
Fig.~3 in Ref.~\cite{Damour:1997ub} and Fig.~\ref{fig:h22} below, 
where we directly compare the usual Taylor approximants 
to the waveform to our new $\rho_{\lm}$-based approximants).
Note that when considering ``Taylor approximants to $\rho_{22}$''
we are actually speaking of a specifically {\it resummed} approximant 
to the waveform $\hat{h}_{22}$. This approximant has the factorized 
form of Eq.~\eqref{eq:hhat}, and is made of the product of several 
resummed constituents.
Even the last factor $f_{22}$ of this product is not used in Taylor-expanded 
form (which would be $T_5[f_{22}]$), but in the minimally resummed way
$f_{22}^{\text{Resummed}}=(T_5[\rho_{22}])^2$.

We have also explored several ways of further resumming $\rho_{22}$, i.e.,
of replacing its PN-expanded form $T_5[\rho_{22}]$ by various non-Taylor
approximants. In view of the good closeness of the 1PN approximation
to $\rho_{22}$ to the exact result we explored, in particular, some 
``factorized'' 
approximants (similar to those considered for the $A(u)$ function in 
Ref.~\cite{Damour:2002qh}) of the 
type $\rho_{22}(x) = (1+c_1^{\rho_{22}}x)\bar{\rho}_{22}(x)$.
We will not show our results for
these approximants here. Instead, let us discuss the use of Pad\'e
approximants for representing $T_5[\rho_{22}]$ as a rational 
function\footnote{As in Ref.~\cite{Damour:2007yf} we consider in this work the 
logarithmic terms (of the type $\text{eulerlog}_m(x)$ in Eqs.~\eqref{eq:rho22}-\eqref{eq:rho31}) as 
part of the coefficients when Padeing $\rho_{\lm}(x)$} of $x$.
As an example, we present in Table~\ref{tab:table2} the values  
of $\rho_{22}(x_{\rm last})$ predicted by
using the four ``around the diagonal'' 5PN accurate Pad\'e approximants,
namely $P^4_1\{T_5[\rho_{22}]\}$, $P^1_4\{T_5[\rho_{22}]\}$, 
$P^3_2\{T_5[\rho_{22}]\}$ and $P^2_3\{T_5[\rho_{22}]\}$.
The important thing to note is that all these approximants are both 
consistently clustered among themselves, as well as closely centered  around the 
correct numerical value (within $\pm 1.3\times 10^{-3}$). 
Note also that, apart from $T_3$, all the Pad\'e approximants 
are closer to the exact value than the $T_4$ and $T_5$. 
[Though the a priori less-accurate $T_3$ approximant 
happens to be closer to the exact value than all other approximants, we
consider that this is coincidental because the subsequent Taylor approximants
$T_4$ and $T_5$ do not exhibit such a close proximity.]
\begin{table}[t]
\caption{\label{tab:table3}Newton-normalized energy flux, and partial
  amplitudes $f_{22}$ and $\rho_{22}$, for a sample of unstable circular 
  orbits computed via the time-domain code of Ref.~\cite{Nagar:2006xv}.
  These values of $\rho_{22}$ are represented as empty circles 
  in Fig.~\ref{fig:rho22_resum}. 
  The case $r=6$ is shown here only for comparison with 
  frequency-domain-based results.}
\begin{center}
  \begin{ruledtabular}
  \begin{tabular}{ccccc}
    $r$    &       $x$ & $\hat{F}_{22}^{\rm time}$ & $f_{22}^{\rm time}$ &$\rho_{22}^{\rm time}$ \\
    \hline \hline
     6     &  0.1666   & 0.897 &  0.665 &  0.815 \\
    \hline
     5     &  0.2000   & 0.995 &  0.615 &  0.784 \\
     4     &  0.2500   & 1.378 &  0.562 &  0.750 \\
     3.5   &  0.2857   & 2.202 &  0.539 &  0.734 \\
     3.1   &  0.3226   & 6.665 &  0.513 &  0.716 \\
  \end{tabular}
\end{ruledtabular}
\end{center}
\end{table}%

\begin{figure}[t]
   \begin{center}
     \includegraphics[width=85 mm]{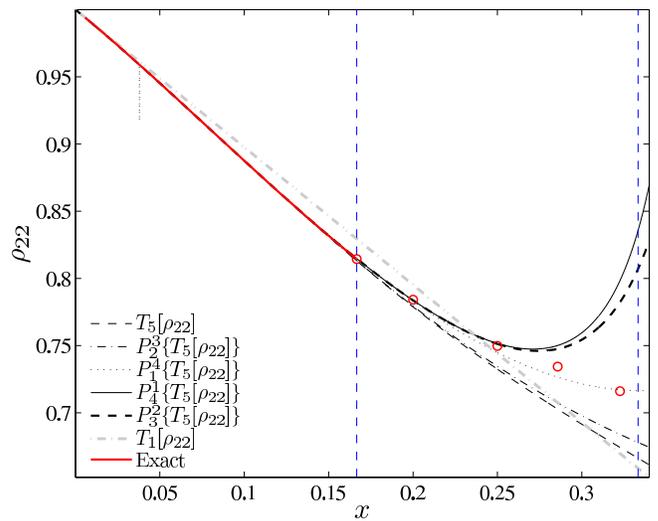}
     \caption{ \label{fig:rho22_resum} Extreme-mass-ratio limit ($\nu=0$).
       Resummation of the function $\rho_{22}(x)$ on the 
       interval $0\leq x\leq 1/3$: comparison between 
       various Taylor and Pad\'e approximants and the ``exact'' function 
       obtained from (both frequency-domain and time-domain) numerical
       calculations.
       The time-domain data points (see Table~\ref{tab:table3}) are indicated
       as empty circles.}
   \end{center}
\end{figure}

We display in Fig.~\ref{fig:rho22_resum}, 
the various 5PN-accurate approximants discussed above
(together with the 1PN-accurate $T_1[\rho_{22}]$ for comparison) 
to $\rho_{22}(x)$ over the {\it larger interval} 
$0\leq x \leq x_{\rm LR}$, where $x_{\rm LR}=1/3$ is the value 
of the frequency parameter $x$ at the ($\nu=0$) ``light-ring''.
Note that, while all the 5PN approximants stay very close to 
each other (and to the ``exact'' numerical value, red online)
up to the LSO, they start diverging from each other when $x\gtrsim 0.2$.
This motivated us to extend the numerical data of E.~Berti beyond the
LSO, i.e. to consider the GW flux emitted by unstable circular 
orbits of Schwarzschild radii $3GM\leq R\equiv GMr \leq 6GM$, corresponding
to $1/6\leq x \leq 1/3$. 
See Table~\ref{tab:table3} for results obtained for such a sample
of sub-LSO orbits (they also appear in Fig.~\ref{fig:rho22_resum}
as empty circles). These numbers have been computed with the
time-domain code described in Ref.~\cite{Nagar:2006xv}.
A resolution of $\Delta r_*=0.01$ was used. To test the accuracy 
of our numerical procedure we computed the energy flux
$\hat{F}_{22}$ at $r=6$ and compared it with the value obtained via Berti's
frequency domain code (at the very close value $r=6.00001$). 
We obtained $\hat{F}_{22}^{\rm time}=0.897342$ to be contrasted with
$\hat{F}_{22}^{\rm freq}=0.892726$, which yields a fractional difference 
$\Delta \hat{F}/\hat{F}^{\rm freq}\approx 0.005$. This gives an indication
of the accuracy of our time-domain results, though we expect, for various
numerical reasons, that the accuracy degrades as $r$ gets below $4$. 

We do not wish to give too much weight to the indication given
by our sub-LSO results on the behavior of the function $\rho_{22}(x)$ 
below the LSO. Indeed, on the one hand, the GW flux along sub-LSO 
circular orbits does represent the analytic continuation of the 
function $F(x)$ yielding the GW flux along (stable) 
super-LSO circular orbits. As such, the empty circles in Fig.~\ref{fig:rho22_resum} do
provide correct mathematical information about the analytical continuation of
the function $\rho_{22}(x)$ that we are trying to best approximate. 
On the other hand, we are evidently aware that the real dynamics 
of the ``plunge'' strongly deviates from the sequence of unstable circular
orbits below the LSO and that the GW flux emitted by a 
plunging test-mass (or effective source) will not be correctly represented
by this mathematical continuation of $F(x)$. However, we expect, especially on
the basis of the EOB formalism which has shown that plunging orbits remain
approximately quasi-circular, that, in view of the present approach where we
decompose the GW amplitude into several different factors having different
physical origin, the mathematical continuation of the $\rho_{22}$ part 
is likely to continue to capture important aspects of the nonlinear
relativistic corrections to the waveform [note that we have in mind using
our factorized waveform Eq.~\eqref{eq:hhat} along the EOB quasi-circular
plunge together with the correct instantaneous 
source $\hat{S}_{\rm eff}^{(\epsilon)}$ and tail corrections
$T_{\lm}$]. However, we are also aware that some aspects of the EOB plunge do
physically differ, near the end of the plunge, in a relevant way from
the physics included in the mathematical continuation of $\rho_{22}(x)$: namely,
the fact that the ratio $(m\Omega)^2/V_\ell(r)$ (where $V_\ell(r)$ is the
Zerilli potential) stays always small along the  real plunge, while it 
increases more along unstable circular orbits and ends up reaching values of 
order unity. In other words the part of $\rho_{22}(x)$ which takes into account
the filtering of $V_\ell(r)$ will be different in the two cases for orbits
near the light-ring.
However, with due reserve we think that the first three empty circles on
Fig.~\ref{fig:rho22_resum} do provide a guideline for selecting among the various
diverging PN approximants the ones which are likely to provide, within the
EOB formalism, a good zeroth-order approximation to the wave-amplitude 
emitted by real plunging orbits. But, we expect that
it will be necessary to correct such a zeroth-order quasi-circular
wave-amplitude by non-quasi-circular corrections of the type which has 
already been found necessary in Refs.~\cite{Damour:2007xr,Damour:2007vq} to obtain a close
agreement between EOB waveforms and numerical waveforms.

\begin{figure*}[t]
    \begin{center}
      (a)\includegraphics[width=82 mm]{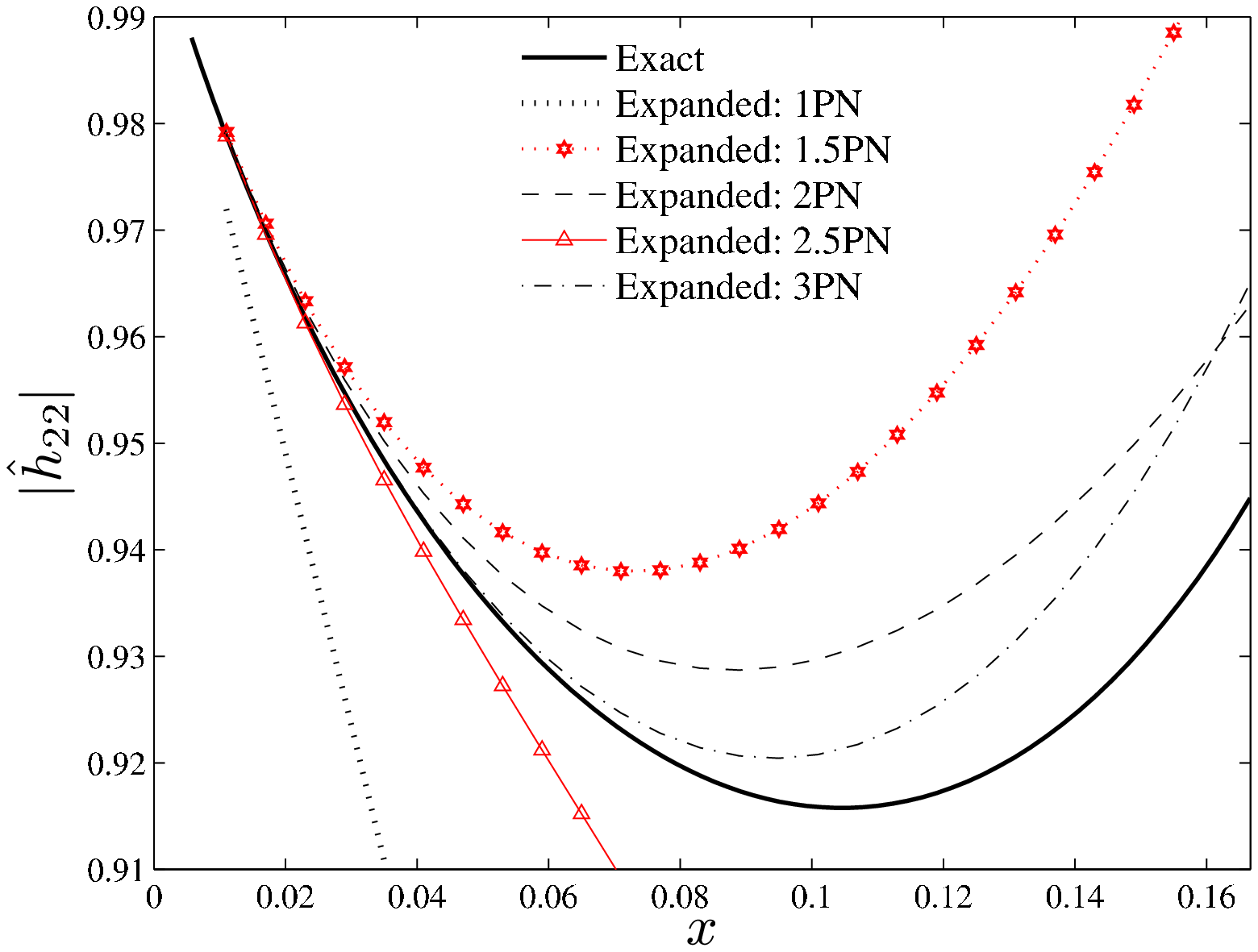}
      \hspace{2 mm}
      (b)\includegraphics[width=82 mm]{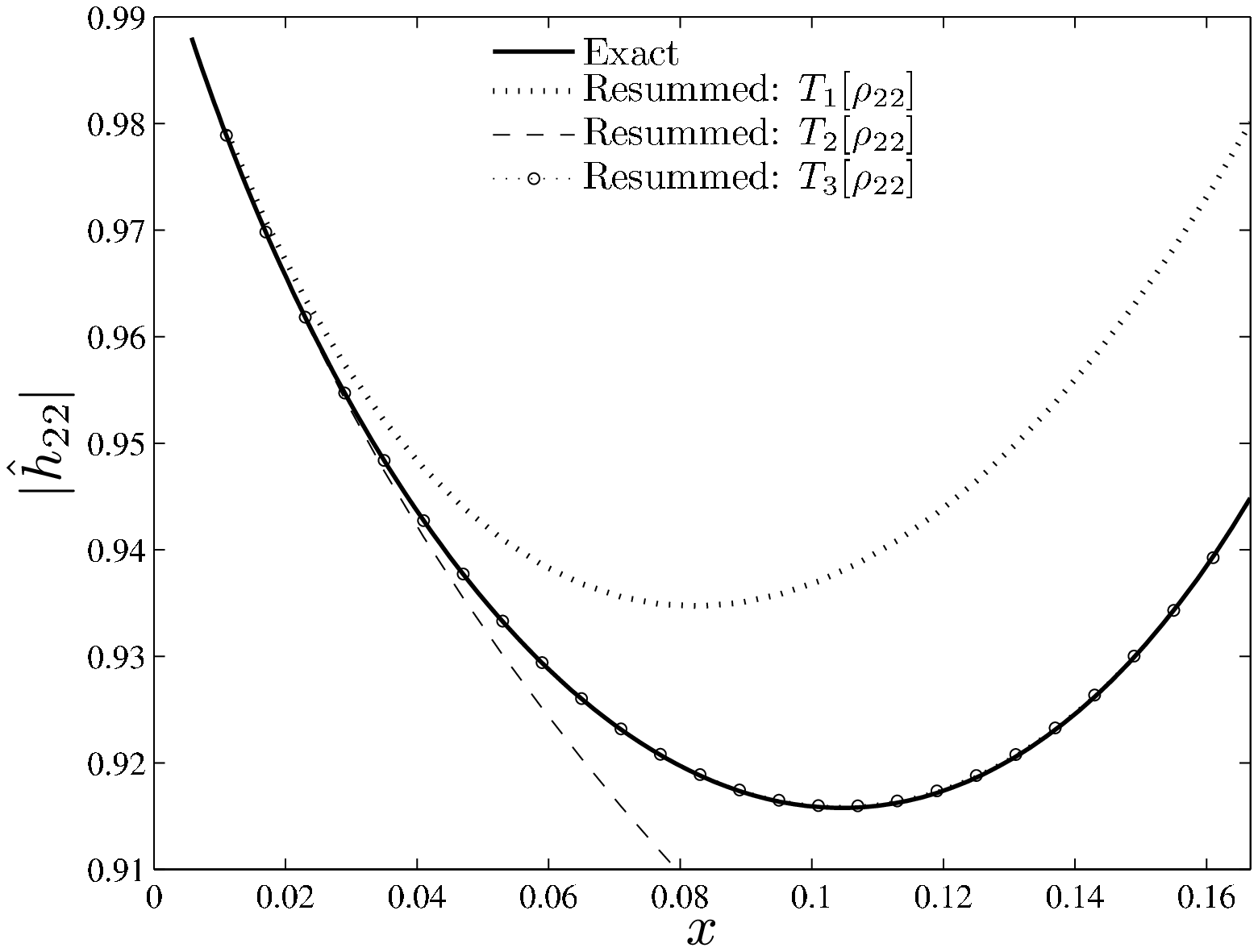}\\
      \vspace{5 mm}
      (c)\includegraphics[width=82 mm]{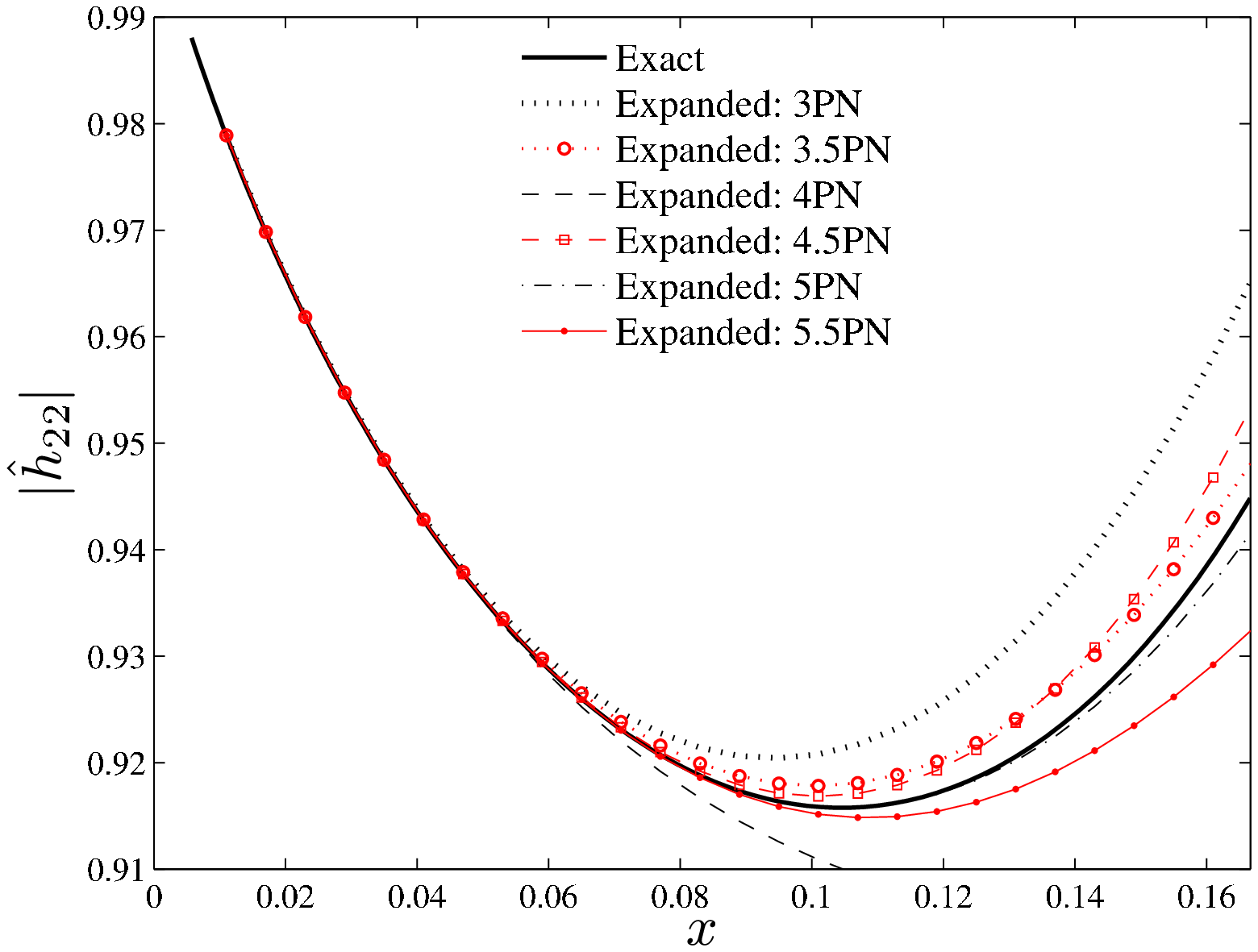}
      \hspace{2 mm}
      (d)\includegraphics[width=82 mm]{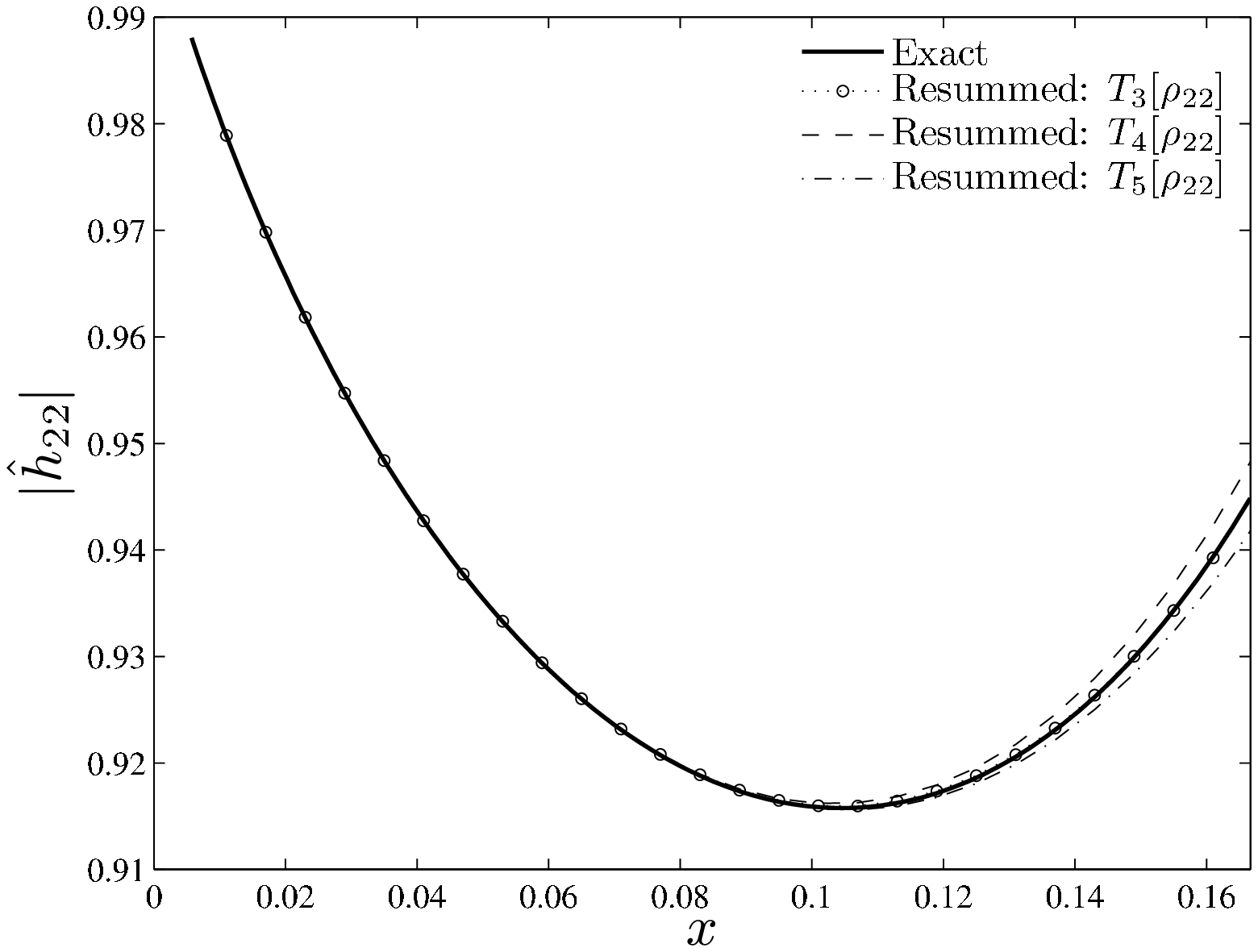}
      \vspace{3 mm}
      \caption{ \label{fig:h22} Extreme-mass-ratio limit ($\nu=0$): 
	Various representations of the 
        $|\hat{h}_{22}|$ waveform modulus. Left panels:
      standard PN-expanded amplitudes. Right panels: Various resummed
      amplitudes. See text for details.}
   \end{center}
\end{figure*}

If we use such a guideline, Fig.~\ref{fig:rho22_resum} suggests that the
best continuations of $\rho_{22}(x)$ below the LSO are given by the 
three particular Pad\'e approximants, $P^1_4$, $P^4_1$ and
$P^2_3$. However, as $P^4_1$ develops a spurious pole (which is barely visible
on the left upper corner of the figure because it is  very localized) 
at $x\approx 0.038$ we will discard it.
By contrast, the other two are robust against the presence of 
spurious poles in the useful regime  $x\lesssim 1/3$ 
(although they develop poles for higher values of $x$, namely below 
the formal ``event horizon'' value $x=1/2$).
In the following, we shall choose $P^2_3$ as our current best-bet
approximant to the $\rho_{22}$ function (notably because this is the
natural near-diagonal default Pad\'e approximant).
Note finally, in Fig.~\ref{fig:rho22_resum}, how the simple 
1PN-accurate Taylor approximant of $\rho_{22}(x)$
succeeds in providing a reasonably good representation 
of $\rho_{22}^{\rm Exact}(x)$ over a very large range of $x$  values.

\subsection{Comparing resummed waveforms to Taylor-expanded and exact ones}
\label{sec:comp_waves}

Up to now, we focussed on the ``convergence'' of various PN-based approximants
towards the numerically determined value of the fourth technical building 
block $\rho_{22}$ entering the dominant quadrupolar wave. 
 
In this subsection we shall investigate instead the ``convergence'' of various
possible PN-based approximants towards the more physically relevant
Newton-normalized GW amplitudes $\hat{h}_{\lm}$.
On the one hand, we shall consider not only the dominant 
$\ell=m=2$ wave, but also a selection of subdominant partial waves. 
On the other hand, we shall consider other PN-based approximants than those
considered above. In particular, we shall compare and 
contrast the exact moduli $|\hat{h}_{\lm}|$ both with
standard high-accuracy Taylor-expanded waveforms
($|\hat{h}_{\lm}|=1+c_1 x + c_{1.5}x^{3/2}+c_2 x^2 + \dots$),
and with our new ``resummed with Taylor$[\rho]$'' 
waveforms ( $|\hat{h}_{\lm}|=\hat{S}_{\rm eff} |T_{\lm}|\rho_{\lm}^\ell$
with $\rho_{\lm}=1+c'_1 x + c'_2 x^2 + c'_3 x^3 + \dots$).
We shall also analyze the performance of our new ``resummed with Pad\'e$[\rho]$'' 
waveforms ( $|\hat{h}_{\lm}|=\hat{S}_{\rm eff} |T_{\lm}|\rho_{\lm}^\ell$
with $\rho_{\lm}=P_p^q[1+c'_1 x + c'_2 x^2 + c'_3 x^3 + \dots]$),
at least for the $\ell=m=2$ dominant mode.\footnote{In view of the remarkable 
agreement, displayed in panel (d) of Fig.~\ref{fig:vpole},
between the exact total flux $\hat{F}(x)$ and the results 
obtained by using only ``resummed with Taylor$[\rho]$'' approximants,
we will not discuss here the
probable improvements that a further Pad\'eing of the subdominant
$\rho_{\lm}$'s might bring in.}  
For definiteness, we discuss here only, besides the dominant even-parity 
quadrupole mode $|\hat{h}_{22}|$, 
the first subdominant odd-parity mode $|\hat{h}_{21}|$, as well as the 
dominant $\ell=4$ mode $|\hat{h}_{44}|$.

Fig.~\ref{fig:h22} focusses on $|\hat{h}_{22}|$. 
The left panels, (a) and (c), display the standard Taylor-expanded 
$|\hat{h}_{22}|=1+c_1 x + c_{1.5}x^{3/2}+c_2 x^2 + \dots$.
More precisely, panel (a) considers the standard Taylor-expanded amplitudes
up to 3PN accuracy included, while panel (c) displays the standard
Taylor-expanded amplitudes from 3PN to 5.5PN accuracy.
By contrast, the right panels, (b) and (d), display our 
new ``resummed with Taylor$[\rho]$'' approximants: panel (b) exhibits the
1PN, 2PN and 3PN approximants, while panel (d) contrasts 
the 3PN, 4PN and 5PN approximants.\footnote{Note that because our tail
factor (together with $e^{\ii \delta_{\lm}}$)  has conveniently resummed 
all the half-integer powers of $x$, the left panels have to include
half-integer PN-approximants, while the right panels have only 
integer-power approximants.}
Consistently with previous studies~\cite{Poisson:1995vs,Damour:1997ub} 
(done at the level of the flux) there is evidently more scatter in
the standard Taylor-expanded amplitudes than in the resummed ones.
In particular, note that the standard 1PN-accurate Taylor approximant
gives a grossly inaccurate representation of $\hat{h}_{22}$ 
 as soon as $x\gtrsim 0.05$ (building up to $-40\%$ at the LSO), 
while our new-resummed $T_1[\rho_{22}]$-based
waveform not only captures the qualitative behavior of the exact waveform,
but also reproduces it quantitatively within $\sim 4\%$ even at the LSO.
On the other hand, for 3PN and higher accuracies
the resummed waveforms exhibit a very close agreement 
(within $\pm 1\times 10^{-3}$) with the exact
one.\footnote{As before, the fact that the resummed $T_3[\rho_{22}]$ 
approximant is closer to the exact result than 
the  $T_4[\rho_{22}]$ and $T_5[\rho_{22}]$ ones is 
probably coincidental. It is more important to note that 
panel (d) exhibits much less scatter than panel (c).}
\begin{figure}[t]
    \begin{center}
      \includegraphics[width=85 mm]{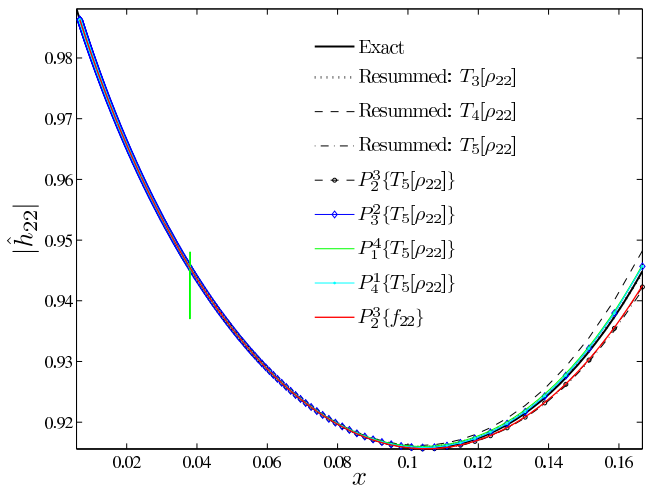}
      \caption{ \label{fig:h22_pade_nu0} Extreme-mass-ratio limit ($\nu=0$). 
      Resummation of the $|\hat{h}_{22}|$ waveform modulus: contrasting 
      ``resummed with Taylor[$\rho$]'' approximants with some
      5PN-accurate ``resummed with Pad\'e[$\rho$]'' approximants.
      See text for definitions and explanations.} 
   \end{center}
\end{figure}
In previous work~\cite{Damour:2007xr,Damour:2007yf}, we had proposed to resum $\hat{h}_{22}$
by Pad\'e ($P^3_2$) approximating $f_{22}=(\rho_{22})^2$ instead of $\rho_{22}$.
For completeness, we compare in Fig.~\ref{fig:h22_pade_nu0} our
previous best proposal to the cluster of our current best proposals
(based on various Taylor and Pad\'e approximants of $\rho_{22}$).
In first approximation, this figure shows a rather close agreement
between all these approximants. In second approximation, one can
note that some of our new approximants, namely $P^2_3$, $P^1_4$ and $P^4_1$, 
are closer to the exact numerical results.
From the pragmatic point of view, our current best-bet approximants 
are therefore our two new, pole-free, Pad\'e approximants based on 
$P^2_3\{T_5[\rho_{22}]\}$ and $P^1_4\{T_5[\rho_{22}]\}$. 
We have a slight preference for $P^2_3\{T_5[\rho_{22}]\}$ which is 
the normal sub-diagonal Pad\'e (admitting  a simple continuous 
fraction representation) and which was close to the sub-LSO numerical
results (see Fig~\ref{fig:rho22_resum}).
\begin{figure}[t]
    \begin{center}
      \includegraphics[width=75 mm]{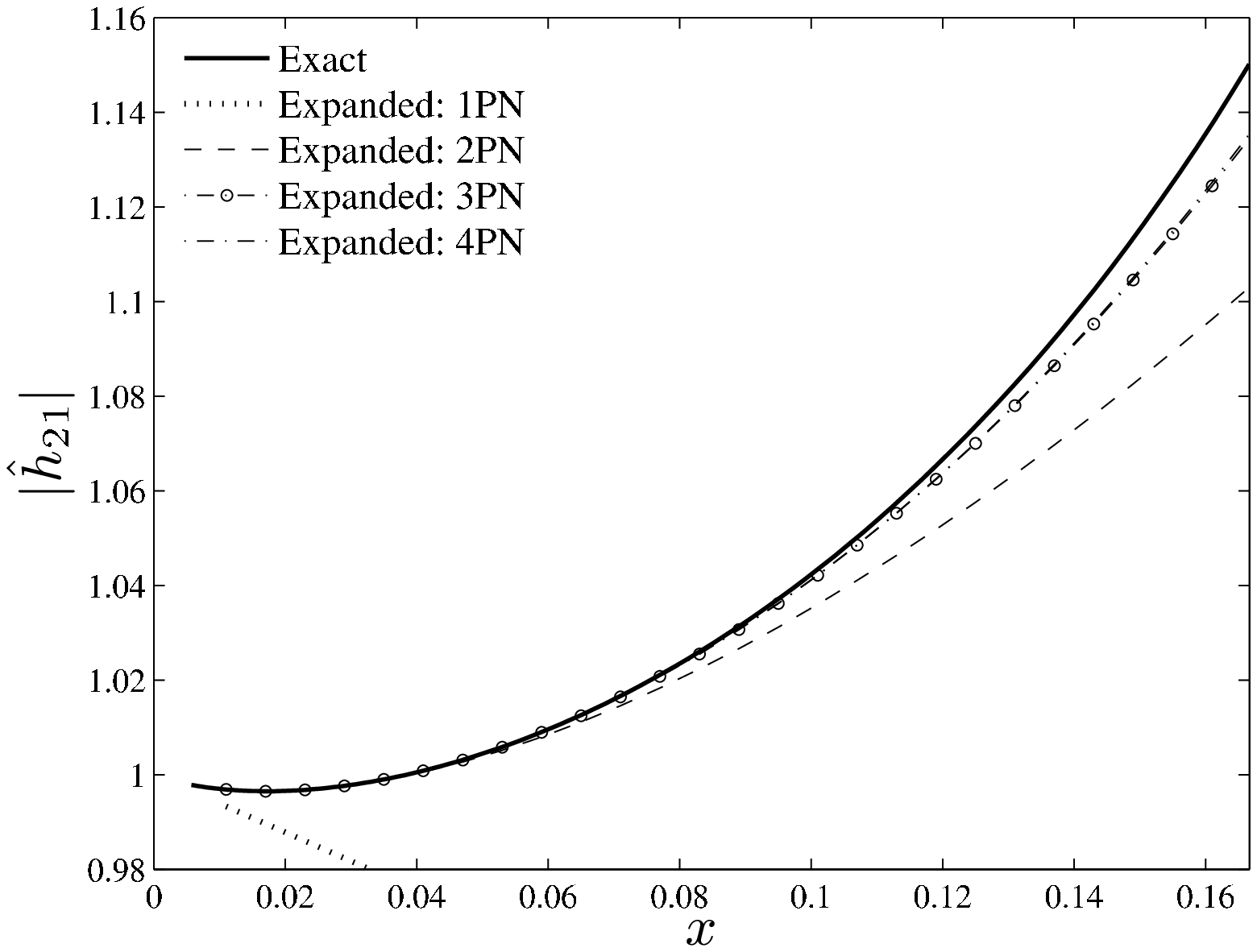}\\
      \vspace{2 mm}
      \includegraphics[width=75 mm]{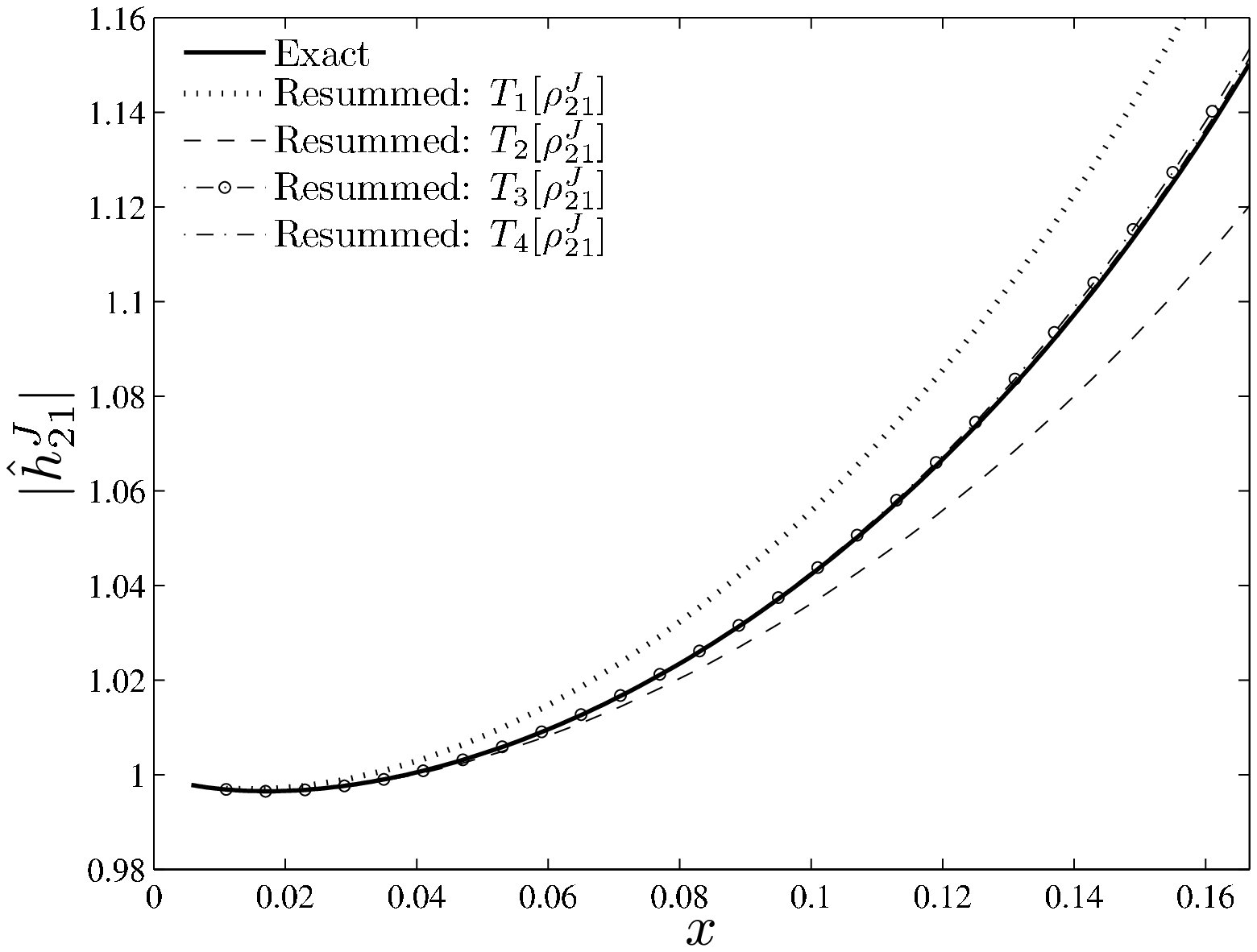}\\
      \vspace{2 mm}
      \includegraphics[width=75 mm]{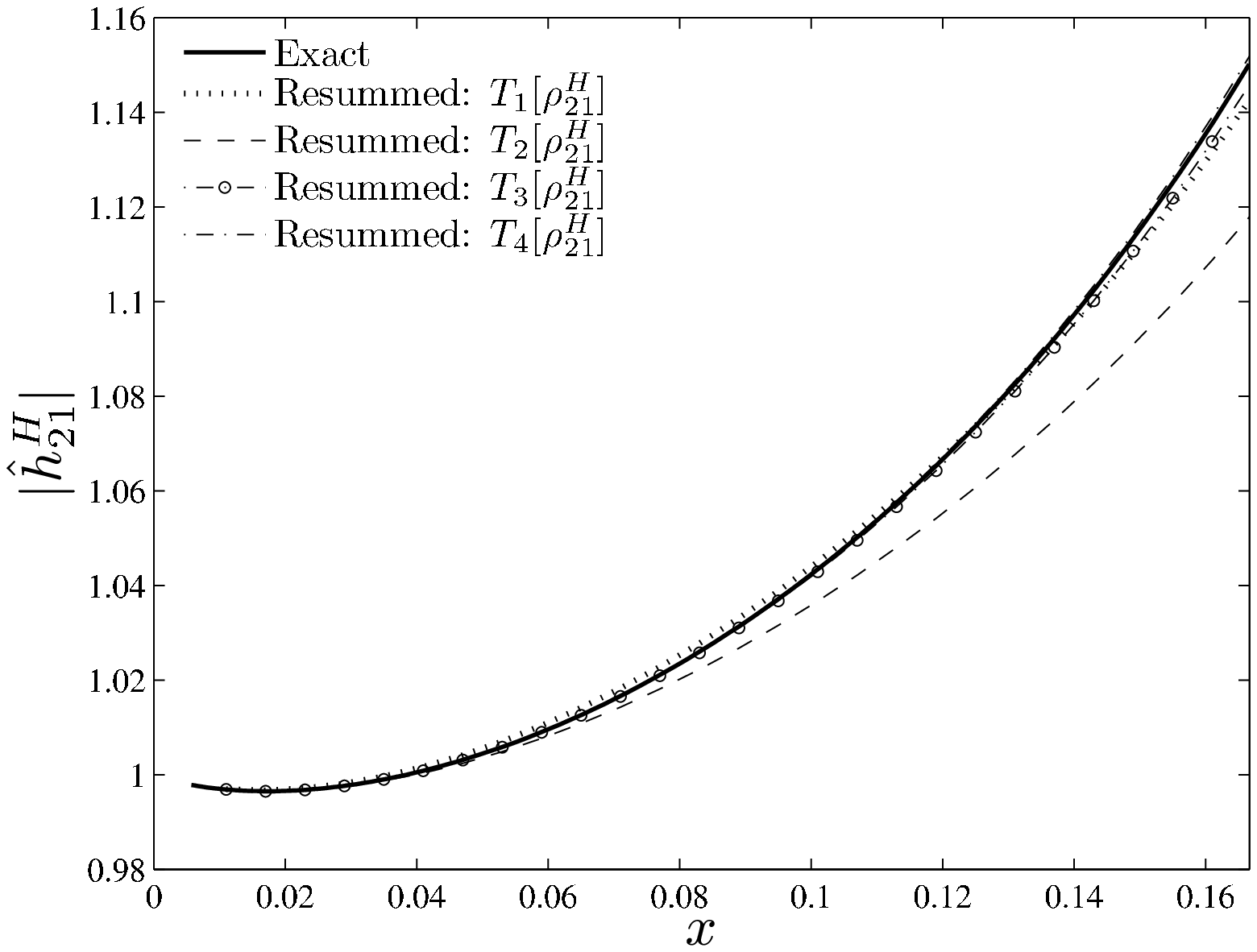}
      \caption{ \label{fig:h21}Extreme-mass-ratio limit ($\nu=0$): 
	various representations of the $|\hat{h}_{21}|$
        waveform modulus. Top panel: standard PN-expansion. Middle panel:
        resummation factoring the angular momentum $\J$. Bottom panel:
        resummation factoring the energy $H_{\rm eff}$.}
   \end{center}
\end{figure}

\begin{figure*}[t]
    \begin{center}
      \includegraphics[width=85 mm]{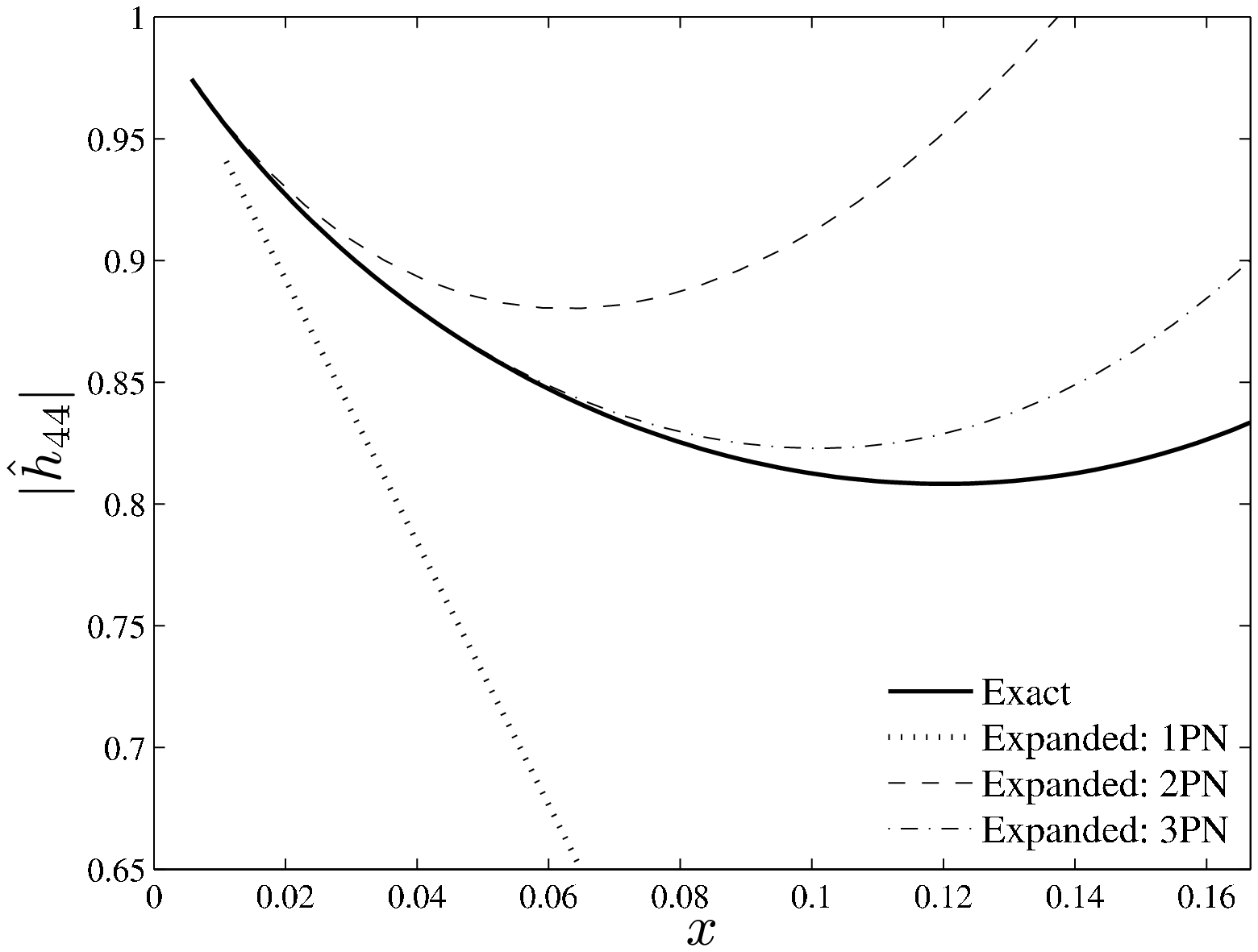}      
      \includegraphics[width=85 mm]{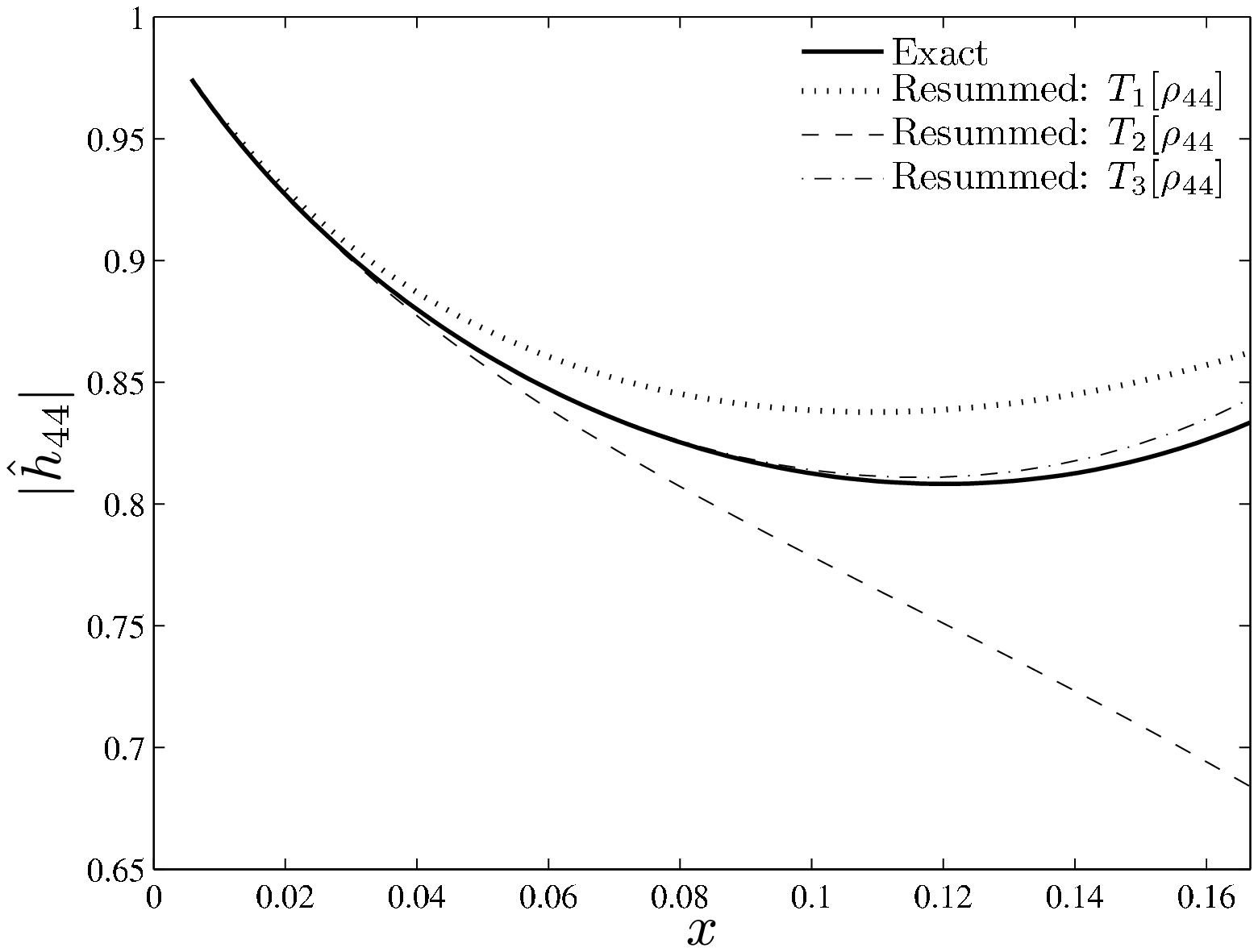}
      \caption{ \label{fig:h44}Extreme-mass-ratio limit ($\nu=0$): 
	various representation of the $|\hat{h}_{44}|$
	waveform modulus. Left panel: standard PN-expansion. 
	Right panel: our new resummation.}
   \end{center}
\end{figure*}

Fig.~\ref{fig:h21} exhibits the results for $|\hat{h}_{21}|$.
We compare and contrast: (i) standard Taylor-expanded amplitudes (top panel),
(ii) new-resummed amplitudes when factoring $\J$ (middle panel) and 
(iii) new-resummed amplitude when factoring $H_{\rm eff}$ (bottom panel). 
For brevity, the ``standard Taylor-expanded'' top panel exhibits only
the integer-order PN-approximants.
Note again, as in the case of $|\hat{h}_{22}|$ discussed above, how 
the use of a standard 1PN-accurate Taylor-expanded waveform leads
to a grossly inaccurate approximation to the exact result, building up
to $-22\%$ at the LSO. By contrast, our new-resummed $T_1[\rho_{21}^J]$-based
approximant (middle panel) or, for that matter, 
the  $T_1[\rho_{21}^H]$-based one (bottom panel), 
captures both qualitatively and quantitatively the 
correct behavior of the exact 
waveform.\footnote{The fact that the resummed $T_1[\rho^H_{21}]$-based 
approximant is extremely close to the exact result (see bottom panel)
is probably coincidental. We do not expect this coincidence to hold for 
higher-order partial waves.}
Ultimately, for 3PN and 4PN-accuracies,
both resummed waveforms exhibit a very close agreement 
(within $\sim 3\times 10^{-3}$ for the $\J$ case) 
with the exact one. The standard Taylor-expanded ones are also
close to the exact results, but visibly less close than our
new approximants.

Finally, Fig.~\ref{fig:h44} exhibits the results 
for the $|\hat{h}_{44}|$ waveforms. 
Note that $\hat{h}_{44}$ and $\hat{h}_{42}$ are the last partial 
multipoles for which the analytical $\nu=0$ 
result is known to 3PN accuracy.
The comparisons between standard Taylor-expanded and new resummed 
waveforms displayed in Fig.~\ref{fig:h44} leads to essentially the
same conclusions as above.
In particular, the standard Taylor-expanded 1PN accurate waveform
is even more grossly inaccurate\footnote{This is the consequence of
the analytical fact noted above that the 1PN correction to the waveform
 is negative and grows linearly with $\ell$. We recall that this fact
was one of our motivations for introducing the new quantities $\rho_{\lm}$.} 
than before, as  the difference builds up to about $-90\%$
at the LSO 
(i.e. $|\hat{h}_{44}^{\rm 1PN}(x_{\rm  LSO})|\approx 0.0793$ instead of 
$|\hat{h}_{44}^{\rm Exact}|(x_{\rm LSO})=0.8334$)!
Let us also emphasize that, as we could have already pointed out
for $|\hat{h}_{22}|$ and for $|\hat{h}_{21}|$, the new resummed
approximant based on the 2PN-accurate Taylor-expanded $\rho_{\lm}$'s
is systematically less good than the one based on the 
1PN-accurate Taylor-expanded $\rho_{\lm}$'s.
This suggests that for waveforms which are subdominant with respect
to $h_{44}$ and $h_{42}$ (for which one does not know the 3PN expansion
of the waveform) one will be better off, if one intends to 
use Taylor-expanded $\rho_{\lm}$'s, in employing only the 1PN accurate
$\rho_{\lm}$'s. However, as we have shown in the $\ell=m=2$ case
(see Table~\ref{tab:table2} and Fig.~\ref{fig:h22_pade_nu0}), 
we expect that a suitable Pad\'e resummation
of the highest accuracy available results will yield better agreement
than simply using the Taylor 1PN-accurate $\rho_{\lm}$'s.
In this respect, let us recall that, as exhibited in 
Eqs.~\eqref{eq:rho1PN_even}-\eqref{eq:rho1PN_odd}, 
the 1PN corrections for all even and odd-parity multipoles 
are known. 
In the $\nu\to 0$ limit they are given by
Eqs.~\eqref{eq:rho1PN_even} and~\eqref{eq:rho1PN_odd};
in the comparable-mass case the even-parity result is 
given by Eq.~\eqref{c1lm_nu} while the odd-parity result is 
given in Appendix~\ref{sec:app_A}.

\section{Results for the comparable mass case (notably the equal mass case, $\nu=0.25$)}
\label{sec:eqmass}

Let us continue to test our resummation procedure by considering the
comparable mass case $\nu\neq 0$, and notably the equal-mass case,
$\nu=1/4=0.25$.
In this case, we cannot rely on the knowledge of the ``exact'' multipolar
waveforms from comparable mass circular orbits. Indeed, though this problem
can in principle be numerically investigated for binary black hole systems
by considering the helical Killing-vector  approach (see
Ref.~\cite{Gourgoulhon:2001ec} and references therein), 
there are no presently available
results where one goes beyond the conformally flat approximation
to Einstein equations. (But see Ref.~\cite{Uryu:2005vv} for 
the case of binary neutron star systems). 
For what concerns the available numerical results on coalescing black holes,
previous work has shown that the deviations from the adiabatic-quasi-circular
approximation were far from being negligible near the LSO, so that they
cannot be directly compared to the circular waveforms investigated in
this paper. We leave to future work a comparison between suitably
non-quasi-circular corrected analytical waveforms and the results of numerical
simulations of coalescing black holes.

In the absence of exact waveforms to be compared to, we shall content ourselves
here by investigating the inner {\it consistency} and {\it robustness} of our current
best-bet analytical approximants as suggested by the $\nu=0$ results reported
above.
More precisely, we shall study the dependence of $\rho_{\lm}$ and the 
corresponding new resummed waveform $|\hat{h}_{\lm}|$ on the two
EOB deformation parameters $\nu$ and $a_5$.

\subsection{Mild dependence of $\rho_{\lm}$ on $\nu$}

To motivate the study of the $\nu$-dependence of $\rho_{\lm}$ let us start by
having a close look at the general structure of $\rho_{\lm}$, i.e.
\begin{equation}
\label{eq:rho_scheme}
\rho_{\lm}(x;\nu) = 1 + c_1^{\rho_{\lm}}(\nu) x + c_2^{\rho_{\lm}}(\nu) x + c_3^{\rho_{\lm}}(\log(x);\nu) x^3+\dots .
\end{equation}
For concreteness, let us display here $\rho_{44}(x;\nu)$ (given together with our
results in Appendix~\ref{sec:app_C}) 
\begin{widetext}
\begin{align}
\rho_{44}(x;\nu) =&1+\frac{2625 \nu ^2-5870 \nu +1614}{1320 (3 \nu -1)}x\nonumber\\
                 &+\frac{1252563795 \nu^4-6733146000 \nu^3-313857376 \nu^2+2338945704 \nu -511573572}{317116800 (1-3 \nu )^2}x^2\nonumber\\
                 &+\left(\frac{16600939332793}{1098809712000}-\frac{12568}{3465}\text{eulerlog}_4(x)\right)x^3.
\end{align}
\end{widetext}
We see on the example of $\rho_{44}$ that the $\nu$-dependence of the
coefficient $c_n^{\rho_{\lm}}(\nu)$ is not polynomial in $\nu$, but rather
given by a rational fraction. The denominator of this rational fraction in the
case of $\rho_{44}$ is proportional to some power of $1-3\nu$. 
The denominator $1-3\nu$
decreases significantly (from $1$ to $1/4$) as $\nu$ increases from the
extreme-mass-ratio case, $\nu=0$, to the equal mass case, $\nu=1/4$.
From Eq.~\eqref{eq:cl}, for the general multipole $\rho_{\lm}^{(\epsilon)}$
this denominator would be proportional to a power of 
\begin{equation}
\label{denom}
d_{\lm}(\nu)=\dfrac{c_{\ell+\epsilon}(\nu)}{X_2 + (-)^m X_1}
=\dfrac{X_2^{\ell+\epsilon-1}+(-)^m X_1^{\ell+\epsilon-1}}{X_2 + (-)^m X_1}.
\end{equation}
This ratio is expressible as a polynomial in $\nu$. 
For instance, for $\rho_{54}$, it would 
be $d_{54}=1-5\nu+5\nu^2$, which decreases from 
$1$ down to $1/16$ as $\nu$ goes from 0 to $1/4$.
More generally, $d_{\lm}(\nu)$ decreases,
as $\nu$ varies from 0 to $1/4$,
from $1$ down to
$1/2^{\ell+\epsilon-2}$  when $\pi(m)=0$ and 
to $(\ell+\epsilon-1)/2^{\ell+\epsilon-2}$, when $\pi(m)=1$.
The presence of such ``small denominators'' raises the issue 
of a possible large increase of the 
coefficients $c_n^{\rho_{\lm}}(\nu)$ as $\nu$ increases from $0$ to $1/4$.
If that were true, this would undermine the applicability to the comparable
mass case of the conclusions that we
have drawn above from the $\nu\to 0$ limit.
\begin{table}[t]
\caption{\label{tab:table4} Analysis of the fractional 
variation \hbox{$\bar{\Delta} c^{\rho_{\lm}}_n(\nu) = c_n^{\rho_{\lm}}(\nu)/c_n^{\rho_{\lm}}(0)-1$ }
of the coefficients $c^{\rho_{\lm}}_n(\nu)$ in Eq.~\eqref{eq:rho_scheme} for
a selected sample of values of $(\ell,m)$.}
\begin{center}
  \begin{ruledtabular}
  \begin{tabular}{cccc}
    $(\ell,m)$    &   $\bar{\Delta} c_1^{\rho_{\lm}}(1/4)$ & $\bar{\Delta} c_2^{\rho_{\lm}}(1/4)$ &
    $\bar{\Delta} c_3^{\rho_{\lm}}(1/4,\log(1/6)) $ \\
    \hline \hline
     (2,2)     &-0.159884     &0.185947       &-0.100421         \\
     (4,4)     &-0.230328     &0.46265       &  \dots  \\
     (5,4)     &-0.176295     & \dots &  \dots   
  \end{tabular}
\end{ruledtabular}
\end{center}
\end{table}%
Therefore, we have studied the $\nu$-dependence of the known 
coefficients $c^{\rho_{\lm}}_n(\nu)$ to check whether the presence of
these ``small denominators'' might cause them to 
grow uncontrollably when $\nu$ increases. 
In Table~\ref{tab:table4} we list the fractional differences 
$\bar{\Delta} c^{\rho_{\lm}}_n(\nu) = c_n^{\rho_{\lm}}(\nu)/c_n^{\rho_{\lm}}(0)-1$ 
at $\nu=1/4$ (and at $\log(x)=\log(1/6)$ for the logarithms 
contained in the higher coefficients) for a sample of 
the $\rho_{\lm}$'s whose $\nu$-dependence is 
analytically known. The good news is  that Table~\ref{tab:table4} 
indicates that the fractional variation of the 
coefficients $c_n^{\rho_{22}}(\nu)$ when going from the extreme-mass 
ratio case to the equal-mass ratio one is typically of the order of $20\%$.

\begin{figure}[t]
    \begin{center}
      \includegraphics[width=85 mm]{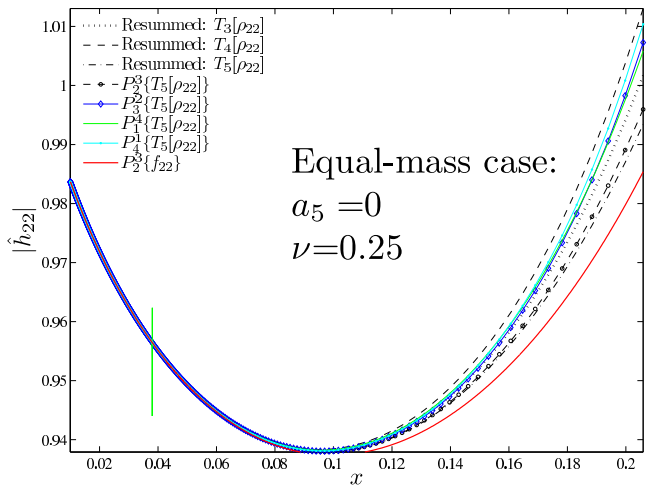}
      \caption{ \label{fig:h22_nu025_a50} Equal-mass case ($\nu=1/4$):
	contrasting various methods for resumming the waveform
	modulus $|\hat{h}_{22}|$ for $a_5=0$.
        Note the presence of a localized 
	spurious pole in $P^4_1\{T_5[\rho_{22}]\}$ at $x\approx 0.038$.}
   \end{center}
\end{figure}
This mild dependence of the coefficients $c_n^{\rho_{\lm}}(\nu)$ on $\nu$ is
the basis of the proposal~\cite{Damour:2007yf} of improving the accuracy 
of known $\nu$-dependent
$\rho_{\lm}$'s by adding the $\nu\to0$ limit of 
higher order PN-corrections (``hybridization'').
[For instance, in the case of $\rho_{22}(x;\nu)$, where the $\nu$-dependent terms
are known up to 3PN, we have added the 4PN and 5PN $\nu=0$ corrections].
Indeed, this procedure consists in using, for some higher corrections, the 
{\it approximation}\footnote{Note that our results on the mild
  $\nu$-dependence of $c_n^{\rho_{\lm}}(\nu)$ 
show that, {\it a contrario}, a naive hybridization of the waveform of the
type 
$\tilde{h}_{\lm}^{\text{hybrid}}=\tilde{h}^{\rm
  known}_{\lm}(\nu)+\tilde{h}_{\lm}^{\rm higher}(\nu=0)$ would probably be
rather unreliable, especially for $\ell\geq 3$, because it would 
not incorporate the overall strong
decrease approximately proportional to 
the ``small denominator'' $d_{\lm}(\nu)$, Eq.~\eqref{denom}.} 
$c_n^{\rho_{\lm}}(\nu)\approx c_n^{\rho_{\lm}}(0)$.

We have validated this approximate completion of known 
$\nu$-dependent terms in the following way.
In view of the results of Table~\ref{tab:table4}, 
we have tested our procedure by modifying
the 4PN coefficient for $\nu=0$, $c_{4}^{\rho_{22}}(0)$, 
by multiplying it by the factor $(1+0.8\nu)$, in order to mimic a 
possible $20\%$ increase of this coefficient when $\nu$ increases 
up to $1/4$.
We then found that such a modification of the 4PN coefficient yielded
a corresponding modification of $T_5[\rho_{22}(x;\nu)]$ equal to
$T_5[\rho_{22}(x;\nu)]^{\rm modified}/T_5[\rho_{22}(x;\nu)]=1.00038$ when
evaluated for $\nu=0.25$ at $x=1/6$. Even at $x=1/3$ we find that such
a modification yields 
$T_5[\rho_{22}(x;\nu)]^{\rm modified}/T_5[\rho_{22}(x;\nu)]=1.013$.
In the $\rho_{44}$ case, where the $\nu$-dependent corrections are 
known only up to 2PN accuracy, a similar modification of 
the 3PN term for $\nu=0$ by a factor $(1+0.8\nu)$ yields 
a corresponding fractional change of $\rho_{44}$ between 
$\nu=0$ and $\nu=0.25$ equal to 1.0099
at $x=1/6$ and $1.079$ at $x=1/3$.
These results confirm the reliability of the hybridization procedure 
adopted here, and give us an idea of the related small uncertainty.
For instance, for the dominant quadrupolar wave, we can anticipate 
that our hybridization procedure introduces an uncertainty in the
waveform $h_{22}(x)\propto \left(\rho_{22}(x)\right)^2$ 
of order $8\times 10^{-4}$ at
the LSO. This level of uncertainty is comparable to the fractional
difference between our best-bet quadrupolar amplitude based on 
$P^2_{3}[\rho_{22}(x,\nu=0)]$ and the exact result (see Table~\ref{tab:table2}).
\begin{figure}[t]
    \begin{center}
      \includegraphics[width=85 mm]{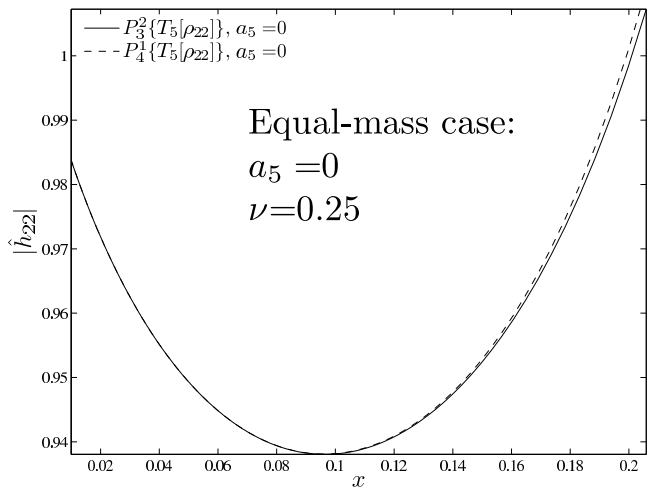}
      \caption{ \label{fig:h22_only}Equal-mass case ($\nu=1/4$): Same as 
	Fig.~\ref{fig:h22_nu025_a50}, but focussing on only 
	the ``best'' ($3^{+2}$PN-accurate) Pad\'e approximants 
	to the waveform.}
   \end{center}
\end{figure}

\subsection{Mild sensitivity of $|\hat{h}_{\lm}|$ to $\nu$}

In Fig.~\ref{fig:rho22_resum} we had put together, in the extreme-mass-ratio case, 
the predictions for $|\hat{h}_{22}|$ made by all the higher-order 
approximants within our new resummation method.
Let us now ``deform'' the results of Fig.~\ref{fig:rho22_resum} by turning on $\nu$
and increasing it up to $\nu=1/4$. 
Fig.~\ref{fig:h22_nu025_a50} is the ``$\nu=1/4$-deformed''
version of Fig.~\ref{fig:rho22_resum}.
In constructing this figure we have used the value $a_5=0$ for the 4PN
EOB parameter entering Eq.~\eqref{eq:Au}, and we have defined the EOB
radial potential $A(u)$ as being $P^1_4[A^{\rm Taylor}(u;a_5)]$.
The horizontal axis has been extended up to the location 
of $x_{\rm LSO}(a_5,\nu)$ as predicted by the corresponding 
adiabatic EOB dynamics, namely $x_{\rm LSO}(0,1/4)=0.2112$.
Some of the lessons we might draw from comparing 
the $\nu=1/4$-deformed Fig~\ref{fig:h22_nu025_a50} to its
$\nu=0$ counterpart, Fig.~\ref{fig:rho22_resum},
are the following: (i) apart from $P^4_1\{T_5[\rho_{22}]\}$ 
(which still has a spurious pole) and our old $P^3_2\{f_{22}\}$, 
the {\it relative stacking order} of {\it all} the other approximants
is maintained in the deformation between $\nu=0$ and $\nu=1/4$; 
(ii) our old prescription~\cite{Damour:2007xr,Damour:2007yf} 
based on $P^{3}_{2}\{f_{22}\}$, which in the $\nu=0$ case
was clustered together with the other approximants 
(as well as with the exact  curve), seems now to have drifted
apart from the cluster of the other ones; (iii) indeed, all the new
approximants are rather well clustered together, with a dispersion
which reaches only about $2\%$ at $x_{\rm LSO}(0,1/4)=0.2112$.

One of the results of the $\nu=0$ study above, particularly in the
dominant $\ell=m=2$ case, was to select, among the array of new
approximants, a small sample of ``best approximants''.
This sample was made of the approximants based 
on $P^2_3\{T_5[\rho_{22}]\}$ and $P^1_4\{T_5[\rho_{22}]\}$. 
In Fig.~\ref{fig:h22_only} we extracted from the previous figure 
the $\nu=1/4$-deformed version of only these two ``best approximants''.
Remembering that in the $\nu=0$ case these two curves were both 
extremely close (within $6\times 10^{-9}$ at $x=1/6$!) 
to each other, as well as being very close to the correct answer,
we note that their $\nu=1/4$-deformed versions are still very close
to each other (within $3\times 10^{-3}$ at $x_{\rm LSO}(0,1/4)=0.2112$ ). 
We therefore expect that this doublet of curves is a good indication 
of where the currently unknown (circular, adiabatic) correct $\nu=1/4$ curve might lie.

\subsection{Weak dependence of $|\hat{h}_{\lm}(x)|$ on  $a_5$.}
\begin{figure}[t]
    \begin{center}
      \includegraphics[width=85 mm]{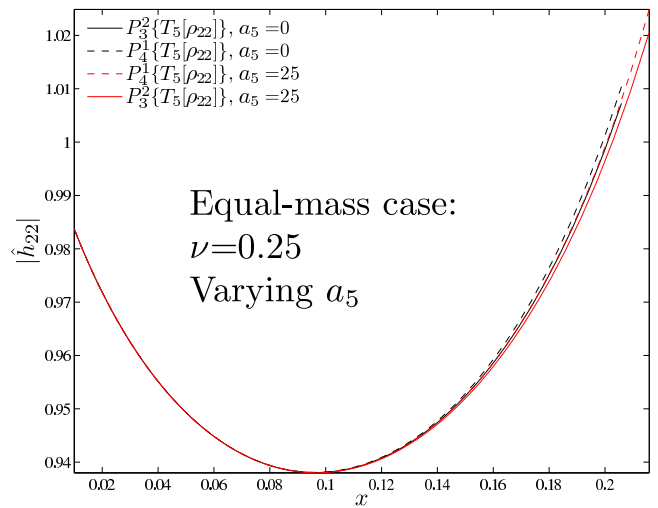}
      \caption{ \label{fig:h22_nu025_a5_25} Equal-mass case ($\nu=1/4$): 
      Effect of varying $a_5$ (between $0$ and $25$) 
      on the ``best'' Pad\'e approximants displayed in Fig.~\ref{fig:h22_only}.}
   \end{center}
\end{figure}

Finally, we study in Fig.~\ref{fig:h22_nu025_a5_25} the sensitivity 
of our new resummed {\it circular waveform} to the 4PN EOB parameter
$a_5$. This sensitivity comes from several sources.
Both the source term $S^{(\epsilon)}_{\rm eff}$ in Eq.~\eqref{eq:hlm}
and the tail term $T_{\lm}$ depend on the EOB dynamical
quantities $H$ and $\J$. Therefore, when expressing the waveform as
a function of the frequency parameter $x$, obtained by solving
Eq.~\eqref{eq:Omega} above, the $a_5$-dependent radial potential $A(u)$ 
comes in at several different places.

For concreteness, we shall study the ``deformation'' of our two best
approximants when $a_5$ increases from $0$ to $25$ (such a range is
motivated by 
recent work ~\cite{Buonanno:2007pf,Damour:2007yf,Damour:2007vq,Damour:2008te} ).
The $a_5$-deformed version of Fig.~\ref{fig:h22_only} is plotted as 
Fig.~\ref{fig:h22_nu025_a5_25}. This figure compares two doublets of
curves: our two best Pad\'e approximants ($P^2_3\{T_5[\rho_{22}]\}$, 
$P^1_4\{T_5[\rho_{22}]\}$) for
$a_5=0$ versus the same Pad\'e approximants when $a_5=25$.

The main thing to note is that the $a_5$ deformation is 
continuous and monotonic. The displacement of each curve is 
only of order $2\times 10^{-3}$ at $x=0.2112=x_{\rm LSO}(0,1/4)$ 
(the horizontal axis of the figure has been extended up 
to $x_{\rm LSO}(25,1/4)=0.2236$). In addition, the separation of 
the $a_5$-deformed doublet of Pad\'e curves is about the same 
as it was before deformation.

\section{Conclusions}
\label{conclusions}

In this paper we have explored the properties of a new resummation method 
of post-Newtonian multipolar waveforms from circular nonspinning compact binaries.
The two characteristic features of this method are: (i) the 
{\it  multiplicative} decomposition of the (complex) $h_{\lm}^{(\epsilon)}$ 
waveform into the product of several factors corresponding to various
physical effects, and (ii) the replacement of the last (real) 
factor, $f_{\lm}$, in this decomposition, by 
its {\it$\ell$-th root} $\rho_{\lm}(x)=\left(f_{\lm}(x)\right)^{1/\ell}$.

To test this resummation method we have first considered 
the extreme-mass-ratio limit ($\nu\to 0$), for which 
``exact'' results for the waveform can be obtained by 
numerical analysis of black hole perturbation theory.
We first noted (see Fig.~\ref{fig:rholm}) that the new quantity 
that we introduced, $\rho_{\lm}(x)$, has a remarkably simple
quasi-linear behavior as a function of the orbital frequency 
parameter $x=(GM\Omega/c^3)^{2/3}$. [In the odd-parity case, this
quasi-linear behavior is especially pronounced when factoring out
the angular momentum ${\cal J}$ from the wave amplitude. This leads
us to consider $\rho_{\lm}^J$ as our ``best-bet'' default choice.]
We related the simple properties of the function 
$\rho_{\lm}(x)$ (including those concerning its dependence on $(\ell,m)$),
to analytical results on the 1PN corrections to
multipole moments.
In this regard, we explicitly computed 
new expressions for the 1PN source current multipoles 
for arbitrary $\ell$ and in consequence
the coefficient of the 1PN correction in 
the {\it odd-parity} waveform (and $\rho_{\lm}$).
The quasi-linear behavior of the functions $\rho_{\lm}(x)$ 
also means that 2PN and higher-order corrections to them
are smaller than analogous corrections in 
usual quantities, like the waveform.

We have shown that, even if one uses only (without any further resummation) 
the successive Taylor approximants to $\rho_{\lm}$, this defines
a sequence of {\it  new resummed approximants} to the waveform
which ``converges'' towards the exact waveform  much less erratically 
than the standard PN approximants. Moreover, for all the waveforms
for which 3PN corrections are known (at least when $\nu\to 0$), our
results show that the new resummed waveform nearly coincide with the
exact results starting with the 3PN 
approximation (see Figs.~\ref{fig:h22},~\ref{fig:h21},~\ref{fig:h44}).
We have also shown that we can further improve the quality of our
new approximants by suitably Pad\'e-resumming the function $\rho_{\lm}(x)$
before using it to construct the 
waveform $h_{\lm}^{(\epsilon)}(x)\propto\left(\rho_{\lm}(x)\right)^\ell$.
In particular, two Pad\'e approximants to $\rho_{22}$, 
namely $P^2_3\{T_5[\rho_{22}]\}$ and
$P^1_4\{T_5[\rho_{22}]\}$, stand out as 
defining the most accurate representation of the
exact waveform (see Fig.~\ref{fig:h22_pade_nu0}).

We have finally explored the robustness of our approximants when
considering a finite mass ratio. We have checked that the $\nu$-dependence
of the coefficients entering the Taylor expansion of the 
function $\rho_{\lm}(x;\nu)$ is rather mild in spite of  the presence
of $\nu$-dependent denominators that decrease very significantly as $\nu$
increases from $0$ to $1/4$.
This justifies the proposal
of completing the known $\nu$-dependent $\rho_{\lm}$'s by adding the 
$\nu\to0$ limit of higher order PN corrections.
We have also shown that the relative stacking order of all the best approximants is
maintained in the ``$\nu$-deformation'' between $\nu=0$ and $\nu=1/4$.
In addition, our new approximants are rather well clustered together,
with a dispersion which reaches only about $2\%$ at the Last Stable Orbit.

Let us finally note that we have compared in the four panels 
of  Fig.~\ref{fig:vpole} four different resummation approaches to
the total (Newton-normalized) GW energy 
flux $\hat{F}(x)$ (for $\nu\to 0$): (a) the standard 
post-Newtonian (Taylor) expansion, (b) the Pad\'e resummation advocated 
long ago~\cite{Damour:1997ub}, (c) the improved $\vp$-tuned Pad\'e resummation 
advocated in~\cite{Damour:2007yf}, and (d) our present new resummation 
method (using only Taylor expanded $\rho_{\lm}$'s). 
The $\vp$-flexed technique, panel (c), is clearly superior to the results of 
the first two techniques, panels (a) and (b). It has however the disadvantage
that it needs to rely
on some external knowledge (such as the exact value of the flux
at the LSO) to determine the optimal value of $\vp$.
On the other hand, our new resummation procedure not only 
stands out, among all other proposals, as yielding the best agreement 
with the exact flux (when $\nu=0$), but it has also the further advantage 
of being parameter-free.

\medskip

\section*{Acknowledgements}
We are grateful to E.~Berti for kindly providing us with his frequency-domain 
numerical data. The activity of A.~Nagar at Institut des Hautes Etudes 
Scientifiques (IHES) is supported by INFN. B.~R.~Iyer thanks IHES and 
the Institut d'Astrophysique de Paris for 
hospitality and support during different stages of this work.

\appendix
\section{Results for Odd-parity (current) 1PN-accurate multipoles}
\label{sec:app_A}
The 1PN-accurate results for the source current (i.e., odd-parity) moments (of any multipolar
order $\ell$) were obtained long ago in Eqs.~(5.18) and/or (5.21) of Ref.~\cite{Damour:1990ji}.
Alternatively, we can use as starting point
for the explicit determination of
the 1PN-accurate source current multipole 
moment Eqs.~(4.3) and (4.4) in~\cite{Blanchet:2001aw}. 
Recalling  the notation $\gamma= GM/Rc^2$ and 
(consistently with Eq.~\eqref{eq:cl}) using the notation
\begin{align}
\label{app_bl}
b_\ell(\nu)\equiv\;& X_2^{\ell}+(-)^\ell X_1^{\ell}\\
\label{app_cl}
c_\ell(\nu)\equiv\;& X_2^{\ell-1}+(-)^\ell X_1^{\ell-1}\\
 {}_cY^{L}_b(\mathbf{y}_1,\mathbf{y}_2)\equiv\;& \frac{\partial}{\partial y_1^c}
\frac{\partial}{\partial y_2^b}Y^{L}(\mathbf{y}_1,\mathbf{y}_2)\\
Y^{L}(\mathbf{y}_1,\mathbf{y}_2)\equiv\;& \frac{r_{12}}{\ell+1}\sum^{\ell}_{p=0} y_1^{\langle L-P}y_2^{P\rangle},
\end{align}
at 1PN accuracy the ``compact'' terms for the source 
current multipole moment $J_L$ can be explicitly 
evaluated, in the circular orbit case, for a general 
value of $\ell$ (as in the $\ell +m$ even case).
They read 
\begin{widetext}
\begin{align}
J_L^{\rm compact}&=STF_{L}\;\; \nu M\epsilon_{abi_\ell}x_av_b\bigg\{x^{L-1}
\bigg[c_{\ell+1}(\nu)+\nonumber\\
  \gamma&\left(b_{\ell+1}(\nu)+2\nu b_{\ell-1}(\nu)
      +\left(\frac{1}{2}-\frac{(\ell-1)(\ell+4)}{2(\ell+2)(2\ell+3)}\right)c_{\ell+3}(\nu)\right)\bigg]
+\frac{r^2}{c^2}x_{L-3}v_{i_{\ell-2}}v_{i_{\ell-1}}
\frac{(\ell-1)(\ell-2)(\ell+4)}{2(\ell+2)(2\ell+3)}c_{\ell+3}(\nu)
\bigg\}\,.
\end{align}
\end{widetext}
\begin{table*}[t]
\caption{\label{tab:tableblcl} 
List of the $b_\ell(\nu)$ and $c_\ell(\nu)$ functions that appear in the text
for some values of $\ell$.
In the following formulas, we have introduced the 
notation $X_{12}\equiv X_1-X_2=\text{sign}(m_1-m_2)\sqrt{1-4\nu}$.  }
\begin{center}
  \begin{ruledtabular}
  \begin{tabular}{lll}
    $\ell$    &   $b_\ell(\nu)$ &$ c_\ell(\nu)$ \\
    \hline \hline
     1     & $-X_{12}$     & 0             \\
     2     & $1-2\nu$     & 1             \\
     3     & $-X_{12}(1-\nu)$     & $-X_{12}$             \\
     4     & $1-4\nu+2\nu^2$     & $1-3\nu $            \\
     5     & $-X_{12}(1-3\nu+\nu^2)$     & $-X_{12}(1-2\nu)$             \\
     6     & $1-6\nu+9\nu^2-2\nu^3$     & $1-5\nu+5\nu^2 $            \\
     7     & $-X_{12}(1-5\nu+6\nu^2-\nu^3)$     & $-X_{12}(1-4\nu+3\nu^2)$             \\
     8     & $1-8\nu+20\nu^2-16\nu^3+2\nu^4$     & $1-7\nu+14\nu^2-7\nu^3 $            \\
     9     & $-X_{12}(1-7\nu+15\nu^2-10\nu^3+\nu^4)$     & $-X_{12}(1-6\nu+10\nu^2-4\nu^3)$             \\
     10     & $1-10\nu+35\nu^2-50\nu^3+25\nu^4-2\nu^5$     & $1-9\nu+27\nu^2-30\nu^3+9\nu^4 $            \\
     11     & $-X_{12}(1-9\nu+28\nu^2-35\nu^3+15\nu^4-\nu^5)$     & $-X_{12}(1-8\nu+21\nu^2-20\nu^3+5\nu^4)$             \\
  \end{tabular}
\end{ruledtabular}
\end{center}
\end{table*}
In addition to the above ``compact terms''(generated by compact-support terms
in the effective stress-energy tensor $\tau^{\mu\nu}$), there exist
three ``non-compact'' contributions that make the 1PN 
current moments more involved than the corresponding 1PN
mass moments. These noncompact contributions can  be
expressed in terms of the $Y^L$ objects introduced in~\cite{Damour:1990ji},
so as  to obtain,
\begin{align}
J_L^{\rm noncompact}&=STF_{L}\;\; \nu M\epsilon_{abi_\ell}
\frac{GM}{c^2}\nonumber\\
&\times \bigg[2X_1  v^c\, {}_cY^{L-1a}_b +\frac{3}{2}X_2 v^c\, {}_cY^{L-1a}_b\nonumber\\
&-\frac{2\ell+1}{2(\ell+2)(2\ell+3)}\frac{d}{dt}\left( {}_aY^{L-1cb}_c\right)
+1\leftrightarrow 2
\bigg]
\end{align}
More explicit expressions for  these non-compact contributions  
can be provided for a general value of $\ell$ by straightforward 
but slightly long computations. 
For circular orbits one can check that the last term does not 
contribute   and  the final result for the other two terms
can be simply re-expressed in terms of the
polynomials $b_{\ell}(\nu)$ and $c_{\ell}(\nu)$,
Eqs.~\eqref{app_bl}-\eqref{app_cl}, as for 
the compact terms. 
The final result (for circular orbits) is given by
\begin{align}
J_L^{\rm noncomp}&= STF_{L}\;\;\;  \nu M
\gamma\epsilon_{abi_\ell}x_av_b x^{L-1}\nonumber\\
&\times\left[ \frac{c_{\ell+3}(\nu)+3b_{\ell+1}(\nu)}{2\ell}+\nu\frac{4b_{\ell-1}(\nu)-c_{\ell+1}(\nu)}{2\ell}\right].
\end{align}
In the test-mass limit ($\nu\to0$) this expression reduces to
\begin{equation}
J_L^{\rm noncomp}= STF_{L}\;\;\; 2 \nu M \gamma\epsilon_{abi_\ell}x_av_b
x^{L-1}\frac{(-1)}{\ell}^{\ell+1}.    
\end{equation}
Thus, in the circular orbit case,
the 1PN-accurate current multipole for 
a general value of $\ell$ finally reads:
\begin{widetext}
\begin{align}
J_L=&STF_{L}\;\; \nu M\epsilon_{abi_\ell}x_av_b\bigg\{x^{L-1}
\bigg[c_{\ell+1}(\nu)+
\gamma\left(-\frac{\nu}{2\ell}c_{\ell+1}(\nu)+\frac{2\ell+3}{2\ell}b_{\ell+1}(\nu)+2\nu\frac{\ell+1}{\ell}
 b_{\ell-1}(\nu)+\right.
\nonumber\\ &     \left.
      \frac{1}{2}\left(\frac{\ell+1}{\ell}-\frac{(\ell-1)(\ell+4)}{(\ell+2)(2\ell+3)}\right)c_{\ell+3}(\nu)\right)\bigg]
+\frac{r^2}{c^2}x_{L-3}v_{i_{\ell-2}}v_{i_{\ell-1}}
\frac{(\ell-1)(\ell-2)(\ell+4)}{2(\ell+2)(2\ell+3)}c_{\ell+3}(\nu)
\bigg\}\,.
\end{align}
\end{widetext}
Adapting the reasoning line of Ref.~\cite{Kidder:2007rt},  
recalling the additional velocity dependence of 
the current moments that leads to $(v/c)^{\ell+1}$ 
and noting that $\gamma =x$ to this order of accuracy,
one can finally show that, for circular orbits,
the 1PN-accurate odd-parity $\hat{h}_{\lm}^{(1)}$'s read

\begin{align}
&\hat{h}_{\lm}^{(1)}(x;\nu)=
1-x\bigg\{
(\ell+1)\left(1-\dfrac{\nu}{3}\right)+\frac{\nu}{2\ell}\nonumber\\
&-\frac{2\ell+3}{2\ell}
\frac{b_{\ell+1}(\nu)}{c_{\ell+1}(\nu)}
-2\nu\frac{\ell+1}{\ell}\frac{b_{\ell-1}(\nu)}{c_{\ell+1}(\nu)} \\
& - \frac{1}{2}\frac{\ell+1}{\ell}\frac{c_{\ell+3}(\nu)}{c_{\ell+1}(\nu)}
 + \frac{m^2(\ell+4)}{2(\ell+2)(2\ell+3)}
\frac{c_{\ell+3}(\nu)}{c_{\ell+1}(\nu)}\bigg\}+{\cal O}(x^2)\nonumber,
\end{align}
where we have not simplified on purpose in order to allow the reader
to explicitly track the origin of each single contribution.
In the extreme-mass-ratio limit, $M\equiv m_1\gg \mu\equiv m_2$ ($\nu\equiv\mu/M\to 0$),
one has $c_\ell(0)=b_\ell(0)=(-1)^{\ell}$, and so this equation simply
reduces to
\be
\hat{h}_{\lm}^{(1)}(x;0)=1-x\left(\ell-\frac{1}{2}-\frac{2}{\ell}
+\frac{m^2(\ell+4)}{2(\ell+2)(2\ell+3)}\right)+{\cal O}(x^2).
\ee
When computing the amplitude $f_{\lm}^J(x;\nu)$ 
(where $\hat{S}^{(1,J)}_{\rm eff}\equiv \hat{j}$ is factorized), 
an additional contribution of $-(3/2+\nu/6)x$ 
(see Eq.~\eqref{j_expansion}) comes in, so that the 
1PN-accurate $f_{\lm}^J(x;\nu)$'s read
\begin{widetext}
\begin{align}
f_{\lm}^J(x;\nu)= &
1-x\left\{
(\ell+1)\left(1-\dfrac{\nu}{3}\right)+\frac{3}{2}+\frac{\nu}{6}+
\frac{\nu}{2\ell}-\frac{2\ell+3}{2\ell}
\frac{b_{\ell+1}(\nu)}{c_{\ell+1}(\nu)}-2\nu\frac{\ell+1}{\ell}
 \frac{b_{\ell-1}(\nu)}{c_{\ell+1}(\nu)}-\right.
\nonumber\\ &     \left.
      \frac{1}{2}\frac{\ell+1}{\ell}\frac{c_{\ell+3}(\nu)}{c_{\ell+1}(\nu)}
+ \frac{m^2(\ell+4)}{2(\ell+2)(2\ell+3)}
\frac{c_{\ell+3}(\nu)}{c_{\ell+1}(\nu)}\right\}+{\cal O}(x^2).
\end{align}
\end{widetext}
In the test-mass limit, this equation becomes
\be
f_{\lm}^J(x;0)=1-x\left(\ell+1-\frac{2}{\ell}
+\frac{m^2(\ell+4)}{2(\ell+2)(2\ell+3)}\right)+{\cal O}(x^2).
\ee
These results lead to the 1PN-accurate $\rho_{\lm}^J$'s, 
Eqs.~\eqref{eq:rho1PN_odd} and~\eqref{eq:c1Jnu}, that we have used in this paper.

For completeness, we conclude this Appendix by quoting the
$\nu$-dependent, 1PN-accurate $f_{\lm}$'s for $\ell+m$ even
and a ready-reckoner of the $b_\ell(\nu)$ and $c_\ell(\nu)$ 
functions for $\ell$-values relevant for this work, Table~\ref{tab:tableblcl}.
From Eq.~(C5) of Ref.~\cite{Kidder:2007rt}, the general
expression of $\hat{h}_{\lm}^{(0)}$ at 1PN reads
\begin{widetext}
\begin{equation}
\hat{h}_{\lm}^{(0) }(x;\nu)= 1-x
\left\{\ell\left(1-\dfrac{\nu}{3}\right)-\dfrac{3}{2}\dfrac{c_{\ell+2}(\nu)}{c_\ell(\nu)}
+
\dfrac{b_\ell(\nu)}{c_\ell(\nu)}+\dfrac{c_{\ell+2}(\nu)}{c_{\ell}(\nu)}\dfrac{m^2(\ell+9)}{2(\ell+1)(2\ell+3)}\right\}+{\cal
  O}(x^2).
\end{equation}
From this expression, the even-parity $f_{\lm}$'s follow as
\begin{equation}
f_{\lm}(x;\nu)= 1-x
\left\{\ell\left(1-\dfrac{\nu}{3}\right)-\dfrac{1}{2}-\dfrac{3}{2}\dfrac{c_{\ell+2}(\nu)}{c_\ell(\nu)}
+
\dfrac{b_\ell(\nu)}{c_\ell(\nu)}+\dfrac{c_{\ell+2}(\nu)}{c_{\ell}(\nu)}\dfrac{m^2(\ell+9)}{2(\ell+1)(2\ell+3)}\right\}+{\cal
  O}(x^2).
\end{equation}
\end{widetext}
which reduces to Eq.~\eqref{flm_1PN_even} in the test-mass limit.

\section{Explicit form of  the $f_{\lm}$'s with higher PN accuracy }
\label{sec:app_B}
In this Appendix we complete the information given in the text by 
explicitly listing the $f_{\lm}$'s  that are known at an accuracy higher 
than 1PN. This means considering multipoles up to $\ell=5$ for even-parity
modes ($\ell+m$ even) and $\ell=4$ for odd-parity modes ($\ell+m$ odd).
We consider separately the even-parity $f_{\lm}$'s and the odd-parity
$f_{\lm}^J$'s and $f_{\lm}^H$'s.
\subsection{Even-parity $f_{\lm}$'s}
\label{sec:even_all_flm}
The even-parity $f_{\lm}$'s (with $\nu\neq 0$ and $\nu= 0$ contributions) 
are given by 
\begin{widetext}
\begin{align}
f_{22}(x;\nu) = 1& +\frac{1}{42} (55 \nu -86) x + 
\frac{2047 \nu ^2-6745 \nu -4288}{1512} x^2 \nonumber \\
  &+\left(\frac{114635 \nu ^3}{99792}-\frac{227875 \nu ^2}{33264}+
   \frac{41}{96}\pi^2\nu-\frac{34625 \nu }{3696}-\frac{856}{105}
   \text{eulerlog}_{2}(x)+\frac{21428357}{727650}\right) x^3\nonumber\\ 
  &+\left(\frac{36808}{2205}\text{eulerlog}_{2}(x)-\frac{5391582359}{198648450}\right) x^4
+\left(\frac{458816}{19845}\text{eulerlog}_{2}(x)-\frac{93684531406}{893918025}\right)x^5
+{\cal O}(x^6),\\
\nonumber\\
f_{33}(x;\nu)= 1& +\left(2 \nu -\frac{7}{2}\right) x
+\left(\frac{887 \nu ^2}{330}-\frac{3401 \nu}{330}-\frac{443}{440}\right) x^2 
+\left(\frac{147471561}{2802800}-\frac{78}{7} \text{eulerlog}_{3}(x)\right)x^3\nonumber\\
&+\left(39\;\text{eulerlog}_{3}(x)-\frac{53641811}{457600}\right) x^4+{\cal
  O}(x^5), \\
\nonumber\\
f_{31}(x;\nu) = 1 &+ \left(-\frac{2 \nu }{3}-\frac{13}{6}\right)x+\left(-\frac{247 \nu ^2}{198}-\frac{371 \nu }{198}+\frac{1273}{792}\right) x^2\nonumber\\
              & +\left(\frac{400427563}{75675600}-\frac{26}{21}\text{eulerlog}_{1}(x)\right)x^3
                +\left(\frac{169}{63}
              \text{eulerlog}_{1}(x)-\frac{12064573043}{1816214400}\right)
              x^4+{\cal O}(x^5),
\end{align}
\end{widetext}
\begin{widetext}
\begin{align}
f_{44}(x;\nu) = 1 &+\frac{2625 \nu ^2-5870 \nu +1614}{330 (3 \nu -1)} x
+\frac{23740185 \nu ^3-106831480 \nu ^2+50799672 \nu -4536144}{1801800 (3 \nu -1)} x^2\nonumber\\
&- 2\left(\frac{1132251120}{156080925}
   \text{eulerlog}_{4}(x)-\frac{5992751383}{156080925}\right) x^3+{\cal O}(x^4),\\
f_{42}(x;\nu)  = 1&+\frac{285 \nu ^2-3530 \nu +1146}{330 (3 \nu -1)}x
-\frac{2707215 \nu ^3+28154560 \nu ^2-26861688 \nu +5538096}{1801800(3 \nu -1)}x^2\nonumber\\
&-\left( \frac{1132251120}{312161850}
\text{eulerlog}_{2}(x)-\dfrac{5180369659}{312161850}\right)x^3 + {\cal O}(x^4),
\end{align}
\end{widetext}
\begin{equation}
f_{55}(x;\nu) = 1 +\frac{512 \nu ^2-1298 \nu  +487}{78 (2 \nu
  -1)}x +\frac{50569}{6552} x^2+{\cal O}(x^3),
\end{equation}
\begin{equation}
f_{53}(x;\nu) = 1 +\frac{176 \nu ^2-850 \nu +375}{78 (2 \nu
  -1)}x+\frac{69359}{10920}x^2 + {\cal O}(x^3), 
\end{equation}
\be
f_{51}(x;\nu) = 1+\frac{8 \nu ^2-626 \nu +319}{78 (2 \nu
  -1)}x+\frac{28859}{4680}x^2+{\cal O}(x^3),
\ee%
%
\subsection{Odd-parity $f_{\lm}^J$'s}
\label{sec:odd_all_flm}
Let us focus now on the odd-parity case and list the $f_{\lm}^J$
in which the Newton-normalized angular momentum $\hat{j}$ has
been factorized as an effective source. We have
\begin{widetext}
\begin{align}
f^J_{21}(x;\nu)=1&+\left(\frac{23 \nu}{42}-\frac{59}{28}\right) x +\left(\frac{85 \nu ^2}{252}-\frac{269\nu }{126}-\frac{5}{9}\right) x^2
+\left(\frac{88404893}{11642400}-\frac{214}{105}\text{eulerlog}_1(x)\right) x^3 \nonumber\\
&+\left(\frac{6313}{1470}\text{eulerlog}_1(x)-\frac{33998136553}{4237833600}\right)x^4+{\cal
  O}(x^5),\\
\nonumber\\
f^J_{32}(x;\nu) = 1&+\frac{320 \nu ^2-1115 \nu +328}{90 (3 \nu -1)}x
+\frac{39544 \nu ^3-253768 \nu ^2+117215 \nu -20496}{11880
  (3\nu -1)}x^2\nonumber\\
&+\left(\frac{110842222}{4729725}-\frac{104}{21}\text{eulerlog}_{2}(x)\right)
x^3+{\cal O}(x^4),\\
\nonumber\\
f^J_{43}(x;\nu)=1&+\frac{\left(160\nu^2-547\nu +222\right)}{44(2\nu-1)}x
+\frac{225543}{40040}x^2+{\cal O}(x^3),\\
f^J_{41}(x;\nu)=1&+\frac{\left(288\nu^2-1385 \nu+602\right)}{132(2\nu
  -1)}x+\frac{760181}{120120}x^2+{\cal O}(x^3).
\end{align}
\end{widetext}

\subsection{Odd-parity $f_{\lm}^H$'s}
We finally list the  odd-parity $f_{\lm}^H$
in which the effective energy $\hat{H}_{\rm eff}$ has 
been factorized as an effective source. We have
\begin{widetext}
\begin{align}
f_{21}^H(x;\nu) = 1&+\left(\frac{5\nu }{7}-\frac{3}{28}\right) x+\left(\frac{79 \nu ^2}{168}-\frac{485 \nu }{126}-\frac{97}{126}\right) x^2
                    +\left(\frac{70479293}{11642400}-\frac{214}{105}\text{eulerlog}_1(x)\right) x^3\nonumber\\
                   &+\left(\frac{107}{490}\text{eulerlog}_1(x)+\frac{9301790917}{1412611200}\right)x^4 + {\cal O}(x^5),\\
f_{32}^H(x;\nu) = 1&+\frac{365 \nu ^2-590 \nu +148}{90 (3 \nu -1)}x
+\frac{16023 \nu ^3-93976 \nu ^2+612 \nu+6192}{3960 (3 \nu -1)}x^2\nonumber\\
&+\left(\frac{96051082}{4729725}-\frac{104}{21}
\text{eulerlog}_2(x)\right)x^3+ {\cal O}(x^4),\\
f_{43}^H(x;\nu)=1&+\frac{524 \nu ^2-1135 \nu
  +402}{132(2\nu-1)}x-\frac{1667}{3640}x^2 + {\cal O}(x^3) ,\\ 
f_{41}^H(x;\nu)=1&+\frac{332 \nu ^2-879 \nu +338}{132 (2 \nu
  -1)}x+\frac{145021}{120120}x^2 + {\cal O}(x^3).
\end{align}
\end{widetext}

\section{Complete expressions of the
         $\rho_{\lm}$'s for $2\leq \ell\leq 8$ }
\label{sec:app_C}
We finally list the ``hybridized'' expressions of all the even- and odd-parity $\rho_{\lm}$'s 
obtained from the corresponding $f_{\lm}$'s [with the proviso explained above
that the $\ell$-th power of the ``hybridized'' $\rho_{\lm}$ presented 
here would generate some specific $\nu$-dependent higher-order 
coefficients $c_{n'}^{f_{\lm}}(\nu)$ which differ from the
$c_{n'}^{f_{\lm}}(\nu=0)$
listed in, e.g. the equations of Appendix~\ref{sec:app_B}.
In the odd-parity case, we only list the ${\cal J}$-normalized
$\rho^J_{\lm}$'s obtained from the
$f_{\lm}^J$'s.
For completeness and future reference we present the $\rho_{\lm}$'s explicitly
up to $\ell=8$ included.
\begin{widetext}
\begin{align}
\rho_{22}(x;\nu)&= 1 +\left(\frac{55 \nu }{84}-\frac{43}{42}\right) x 
+\left(\frac{19583 \nu^2}{42336}-\frac{33025 \nu
}{21168}-\frac{20555}{10584}\right) x^2 \nonumber\\
&+\left(\frac{10620745 \nu ^3}{39118464}-\frac{6292061 \nu ^2}{3259872}+\frac{41 \pi
   ^2 \nu }{192}-\frac{48993925 \nu }{9779616}-\frac{428}{105}
  \text{eulerlog}_{2}(x)+\frac{1556919113}{122245200}\right) x^3 \nonumber\\
&+\left(\frac{9202}{2205}\text{eulerlog}_2(x)-\frac{387216563023}{160190110080}\right) x^4
+\left(\frac{439877}{55566}\text{eulerlog}_{2}(x)-\frac{16094530514677}{533967033600}\right)x^5+{\cal O}(x^6), \\
\nonumber\\
\rho_{21}^J(x;\nu)&=1+\left(\frac{23 \nu }{84}-\frac{59}{56}\right) x +\left(\frac{617 \nu ^2}{4704}-\frac{10993\nu }{14112}-\frac{47009}{56448}\right) x^2\nonumber\\
                    &+\left(\frac{7613184941}{2607897600}-\frac{107}{105}\text{eulerlog}_1(x)\right)x^3
                     +\left(\frac{6313}{5880}\text{eulerlog}_1(x)-\frac{1168617463883}{911303737344}\right)x^4+{\cal O}(x^5),\\
\nonumber\\
\rho_{33}(x;\nu) &= 1+\left(\frac{2 \nu }{3}-\frac{7}{6}\right) x+\left(\frac{149 \nu ^2}{330}-\frac{1861 \nu }{990}-\frac{6719}{3960}\right) x^2
+\left(\frac{3203101567}{227026800}-\frac{26}{7} \text{eulerlog}_{3}(x)\right)x^3\nonumber\\
 &+\left(\frac{13}{3}\text{eulerlog}_{3}(x)-\frac{57566572157}{8562153600}\right) x^4+{\cal O}(x^5), \\
\nonumber\\
\rho_{32}^J(x;\nu)&=1 +\frac{320 \nu ^2-1115\nu +328}{270 (3 \nu -1)}x 
                      +\frac{3085640 \nu ^4-20338960 \nu ^3-4725605 \nu ^2+8050045\nu -1444528}{1603800 (1-3 \nu )^2}x^2\nonumber\\
                  &+\left(\frac{5849948554}{940355325}-\frac{104}{63}\text{eulerlog}_2(x)\right)x^3+{\cal O}(x^4),\\
\nonumber\\
\rho_{31}(x;\nu) &= 1+\left(-\frac{2 \nu}{9}-\frac{13}{18}\right) x 
+\left(-\frac{829 \nu ^2}{1782}-\frac{1685\nu }{1782}+\frac{101}{7128}\right) x^2
+\left(\frac{11706720301}{6129723600}-\frac{26}{63}\text{eulerlog}_1(x)\right) x^3\nonumber\\
&+\left(\frac{169}{567}
\text{eulerlog}_1(x)+\frac{2606097992581}{4854741091200}\right) x^4 + {\cal O}(x^5),\\
\nonumber\\
\rho_{44}(x;\nu) &=1+\frac{2625 \nu ^2-5870 \nu +1614}{1320 (3 \nu -1)}x\nonumber\\
                 &+\frac{1252563795 \nu^4-6733146000 \nu^3-313857376 \nu^2+2338945704 \nu -511573572}{317116800 (1-3 \nu )^2}x^2\nonumber\\
                 &+\left(\frac{16600939332793}{1098809712000}-\frac{12568}{3465}\text{eulerlog}_4(x)\right)x^3
                 +{\cal O}(x^4),\\
\nonumber\\
\rho_{43}^J(x;\nu)&=1+ \frac{160 \nu ^2-547 \nu +222}{176 (2\nu-1)}x
                    -\frac{6894273}{7047040}x^2+{\cal O}(x^3),\\ 
\nonumber\\
\rho_{42}(x;\nu)&=1 +\frac{285 \nu ^2-3530 \nu +1146}{1320 (3 \nu -1)}x\nonumber\\
                 &  +\frac{-379526805 \nu ^4-3047981160 \nu ^3+1204388696 \nu ^2+295834536 \nu -114859044}{317116800 (1-3 \nu )^2}x^2\nonumber\\
                 &
                 +\left(\frac{848238724511}{219761942400}-\frac{3142}{3465}\text{eulerlog}_2(x)\right)
                 x^3+{\cal O}(x^4),\\
\nonumber\\
\rho_{41}^J(x;\nu)&= 1 +\frac{ 288 \nu^2 -1385\nu + 602 }{528 (2 \nu -1)}x
                    -\frac{7775491}{21141120}x^2+{\cal O}(x^3),
\end{align}
\begin{align}
\rho_{55}(x;\nu)&=1 + \dfrac{512 \nu ^2-1298 \nu +487}{390(2\nu-1)} x -
\dfrac{3353747}{2129400}x^2+{\cal O}(x^3),\\
\nonumber\\
\rho_{54}^J(x;\nu)&= 1 + \dfrac{33320\nu^3 - 127610\nu^2 + 96019\nu - 17448 }{13650(5\nu^2 - 5\nu + 1 )}x+{\cal O}(x^2),\\
\nonumber\\
\rho_{53}(x;\nu)&= 1 + \dfrac{176 \nu ^2-850 \nu +375}{390(2\nu-1)}x   - \dfrac{410833}{709800}x^2+{\cal O}(x^3),\\
\nonumber\\
\rho_{52}^J(x;\nu)&=1 + \dfrac{21980\nu^3 - 104930\nu^2 + 84679\nu - 15828}{13650(5\nu^2 -5\nu +1)}x+{\cal O}(x^2),\\
\nonumber\\
\rho_{51}(x;\nu)&= 1 + \dfrac{8\nu^2-626\nu+319}{390(2\nu-1)}x - \dfrac{31877}{304200}x^2+{\cal O}(x^3),\\
\nonumber\\
\rho_{66}(x;\nu)&=1+\frac{273 \nu ^3-861 \nu ^2+602 \nu
  -106}{84\left(5 \nu ^2-5 \nu +1\right)}x+{\cal O}(x^2),\\
\nonumber\\
\rho_{65}^J(x;\nu)&=1+\frac{220 \nu ^3-910 \nu ^2+838 \nu -185}{144 \left(3
  \nu ^2-4 \nu +1\right)}x+{\cal O}(x^2),\\
\nonumber\\
\rho_{64}(x;\nu)&=1+\frac{133 \nu ^3-581 \nu ^2+462 \nu -86}{84
  \left(5 \nu ^2-5 \nu +1\right)}x+{\cal O}(x^2),\\
\nonumber\\
\rho_{63}^J(x;\nu)&=1+\frac{156 \nu ^3-750 \nu ^2+742 \nu -169}{144 \left(3
  \nu ^2-4 \nu +1\right)}x+{\cal O}(x^2) \\
\nonumber\\
\rho_{62}(x;\nu)&=1+\frac{49 \nu ^3-413 \nu ^2+378 \nu
  -74}{84\left(5 \nu ^2-5 \nu +1\right)}x+{\cal O}(x^2),\\
\nonumber\\
\rho_{61}^J(x;\nu)&=1+\frac{124 \nu ^3-670 \nu ^2+694 \nu -161}{144 \left(3 \nu ^2-4 \nu +1\right)}x+{\cal O}(x^2), \\
\nonumber\\
\rho_{77}(x;\nu)&=1+\frac{1380 \nu ^3-4963 \nu ^2+4246 \nu -906}{714 \left(3 \nu ^2-4 \nu +1\right)}x+{\cal O}(x^2),\\
\nonumber\\
\rho_{76}^J(x;\nu)&=1+
\frac{6104 \nu ^4-29351 \nu ^3+37828 \nu ^2-16185 \nu +2144}{1666 \left(7 \nu
  ^3-14 \nu ^2+7 \nu -1\right)}x + {\cal O}(x^2),\\
\nonumber\\
\rho_{75}(x;\nu)&=1+\frac{804 \nu ^3-3523 \nu ^2+3382 \nu
  -762}{714 \left(3 \nu ^2-4 \nu +1\right)}x+{\cal O}(x^2),\\
\nonumber\\
\rho_{74}^J(x;\nu)&=1+\frac{41076 \nu ^4-217959 \nu ^3+298872 \nu ^2-131805 \nu +17756}{14994 \left(7 \nu ^3-14 \nu ^2+7 \nu -1\right)}x+{\cal O}(x^2),
\end{align}
\begin{align}
\rho_{73}(x;\nu)&=1+\frac{420 \nu ^3-2563 \nu ^2+2806 \nu
  -666}{714 \left(3 \nu ^2-4 \nu +1\right)}x+{\cal O}(x^2),\\
\nonumber\\
\rho_{72}^J(x;\nu)&=1+\frac{32760 \nu ^4-190239 \nu ^3+273924 \nu ^2-123489
  \nu +16832}{14994 \left(7 \nu ^3-14 \nu ^2+7 \nu -1\right)}x + {\cal O}(x^2),\\
\nonumber\\
\rho_{71}(x;\nu)&=1+\frac{228 \nu ^3-2083 \nu ^2+2518 \nu
  -618}{714 \left(3 \nu ^2-4 \nu +1\right)}x+{\cal O}(x^2),
\end{align}
\begin{align}
\rho_{88}(x;\nu)&=1+\frac{12243 \nu^4-53445 \nu^3+64659 \nu^2-26778 \nu
  +3482}{2736 \left(7 \nu^3-14 \nu^2+7\nu -1\right)}x+{\cal O}(x^2),\\
\nonumber\\
\rho_{87}^J(x;\nu)&=1+\frac{38920 \nu ^4-207550 \nu ^3+309498 \nu ^2-154099
  \nu +23478}{18240 \left(4 \nu ^3-10 \nu ^2+6 \nu -1\right)}x+{\cal O}(x^2), \\
\nonumber\\
\rho_{86}(x;\nu)&= 1 +\frac{2653 \nu^4-13055 \nu^3+17269 \nu^2-7498 \nu
  +1002}{912\left(7 \nu^3-14\nu^2+7\nu-1\right)}x + {\cal O}(x^2),\\
\nonumber\\
\rho_{85}^J(x;\nu)&=1+\frac{6056 \nu ^4-34598 \nu ^3+54642 \nu ^2-28055 \nu
  +4350}{3648 \left(4 \nu ^3-10 \nu ^2+6 \nu -1\right)}x + {\cal O}(x^2),\\
\nonumber\\
\rho_{84}(x;\nu)&=1+\frac{4899 \nu ^4-28965 \nu ^3+42627 \nu ^2-19434 \nu
  +2666}{2736 \left(7 \nu ^3-14 \nu ^2+7 \nu -1\right)}x+{\cal O}(x^2),\\
\nonumber\\
\rho_{83}^J(x;\nu)&=1+\frac{24520 \nu ^4-149950 \nu ^3+249018 \nu ^2-131059
  \nu +20598}{18240 \left(4 \nu ^3-10 \nu ^2+6 \nu -1\right)}x + {\cal O}(x^2),\\
\nonumber\\
\rho_{82}(x;\nu)&=1+\frac{3063 \nu^4-22845 \nu^3+37119 \nu ^2-17598 \nu
+2462}{2736 \left(7 \nu^3-14 \nu^2+7 \nu -1\right)}x + {\cal O}(x^2),\\
\nonumber\\
\rho_{81}^J(x;\nu)&=1+\frac{21640 \nu ^4-138430 \nu ^3+236922 \nu ^2-126451
  \nu +20022}{18240 \left(4 \nu ^3-10 \nu ^2+6 \nu -1\right)}x + {\cal O}(x^2).
\end{align}
\end{widetext}

\bibliography{bibliography}


\end{document}